\documentclass[final,3p,times,twocolumn]{elsarticle}

\usepackage{graphicx}

\usepackage{amssymb}
\usepackage[utf8]{inputenc}
\usepackage{units}

\journal{Physics Reports}

\newcommand{\xmax}{X$_{\mathrm{max}}$ }

\begin{document}

\begin{frontmatter}

\title{Radio detection of cosmic ray air showers in the digital era}

\author[KITIK]{Tim Huege}

\address[KITIK]{Institut f\"ur Kernphysik, Karlsruher Institut f\"ur Technologie - Campus Nord, Postfach 3640, 76021 Karlsruhe, Germany}

\cortext[cor]{Email: tim.huege@kit.edu}

\begin{abstract}

In 1965 it was discovered that cosmic ray air showers emit impulsive 
radio signals at frequencies below 100~MHz. After a period of intense 
research in the 1960s and 1970s, however, interest in the detection 
technique faded almost completely. With the availability of powerful digital signal 
processing techniques, new attempts at measuring cosmic ray air 
showers via their radio emission were started at the beginning 
of the new millennium. Starting with modest, small-scale digital prototype setups, the 
field has evolved, matured and grown very significantly in the past 
decade. Today's second-generation digital radio detection 
experiments consist of up to hundreds of radio antennas or cover
areas of up to 17 km$^{2}$. We understand the physics of the radio emission in extensive air showers in 
detail and have developed analysis strategies to accurately derive 
from radio signals parameters which are related to the astrophysics of 
the primary cosmic ray particles, in particular their energy, arrival direction and estimators 
for their mass. In parallel to these successes, limitations inherent in 
the physics of the radio signals have also become increasingly clear. In 
this article, we review the progress of the past decade and the current state of the field, discuss 
the current paradigm of the radio emission physics and present the 
experimental evidence supporting it. Finally, we discuss the potential 
for future applications of the radio detection technique to advance the field of cosmic ray physics.

\end{abstract}


\end{frontmatter}




\section{Introduction}

\begin{figure*}[!htb]
  \centering
  \includegraphics[width=0.7\textwidth]{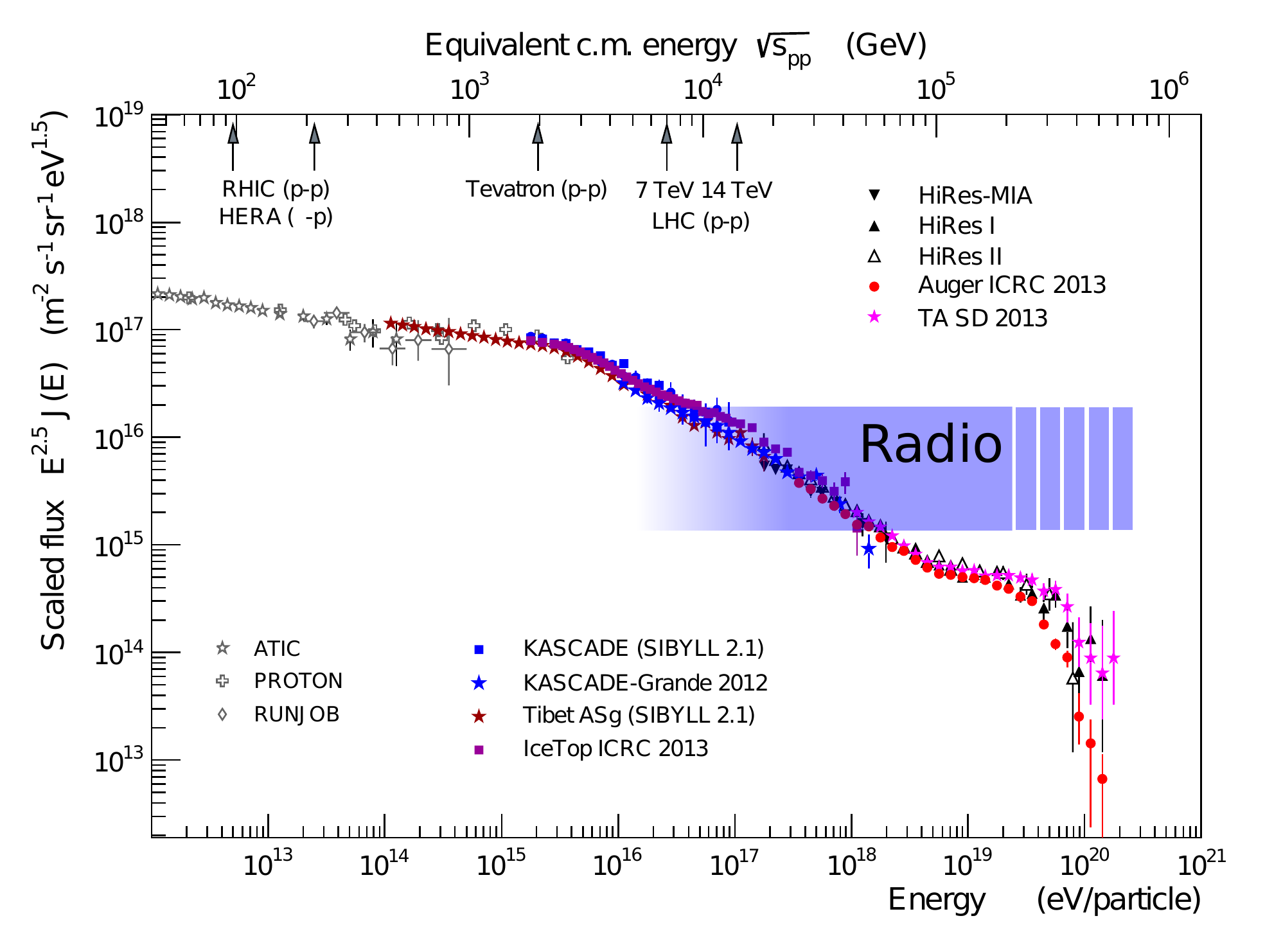}
  \caption{The energy spectrum of the highest energy cosmic rays 
  measured by various experiments. The energy range accessible to 
  radio measurements is indicated. At low particle energies, radio signals 
  become weak and are overwhelmed by background. At high energies, concepts 
  to cover very large effective detection areas have yet to be 
  developed. Diagram updated and adapted from \citep{Engel:2011zzb}.}
  \label{fig:crspectrum}
 \end{figure*}

Even though more than 100 years have passed since the discovery of cosmic rays, 
many questions about their origin, the physics of their acceleration and 
their hadronic interactions in the atmosphere are still unanswered 
\citep{Bluemer2009293}. To tackle the complexity of the problem, two ingredients 
are very important: First, cosmic rays have to be measured 
with sufficient statistics, a difficult task at the highest 
energies where the particle flux becomes as small as one particle per 
km$^2$ per century, see Fig.\ \ref{fig:crspectrum}. Second, the measurement quality has to 
be as good as possible to provide enough information, in particular, to
identify the mass of the primary particles, an essential piece of 
information in testing hypotheses for particle acceleration and propagation. Techniques such as large-scale particle 
detection with ground-based arrays and fluorescence detection of air 
showers with optical telescopes have been employed with great success 
over many decades \citep{Kampert2012660}. These approaches detect ``extensive 
air showers'', cascades of secondary particles initiated by the 
primary cosmic ray in the atmosphere \citep{Engel:2011zzb}. 
However, the established detection methods all have their drawbacks, and the community is 
constantly looking for ways to improve on the established techniques.
A prime example of such an endeavor is the proposed AugerPrime \citep{AugerPrimeIcrc2015} upgrade of the 
Pierre Auger Observatory, which strives to achieve sensitivity to the 
mass composition of cosmic rays at the highest energies via separate
measurements of the electromagnetic and muonic air shower components 
using an additional layer of scintillators deployed on top of the
existing water-Cherenkov detectors.

\begin{figure*}[!htb]
  \centering
  \includegraphics[angle=90,width=0.7\textwidth,clip=true,trim= 0cm 9cm 20cm 0cm]{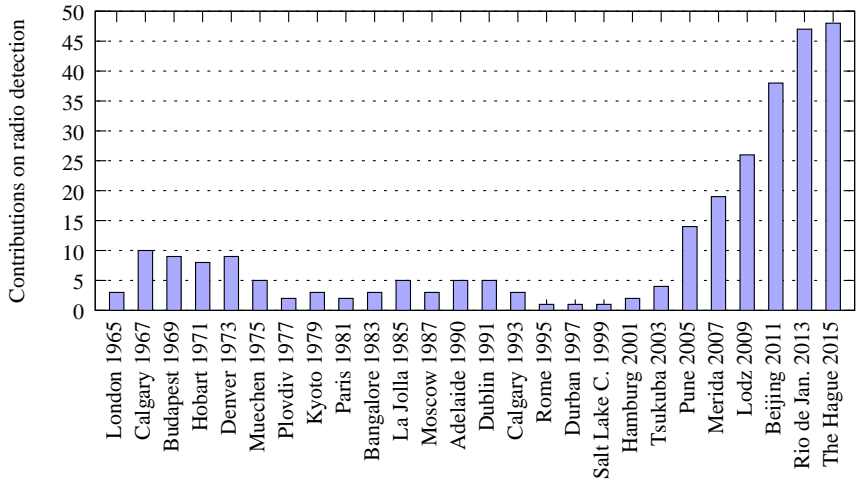}
  \caption{Number of contributions related to radio detection of 
  cosmic rays or neutrinos at the bi-yearly International Cosmic Ray
  Conferences. Data up to 2007 were taken from \citep{Nichol2011}.}
  \label{fig:icrcstats}
 \end{figure*}

In the past decade, the field of radio detection of cosmic ray air 
showers has undergone an impressive renaissance. Building on the 
knowledge gathered from historical radio detection experiments in the 1960s 
and 1970s, innovative projects were started in the early 2000s, driven 
by high expectations \citep{FalckeGorhamProtheroe2004}. The goal of these 
projects was to first provide a proof of principle for the detection of air showers using digital radio techniques, and 
then to evolve these approaches into a new technology for large-scale air shower 
measurements. Having met with great success, these activities steadily gained
in momentum, as is illustrated in Fig.\ \ref{fig:icrcstats}. Today's 
experiments have matured well beyond the prototyping phase. They are aimed either at 
covering large areas with a minimum number of antennas or at measuring
individual air showers with hundreds of radio antennas at a time. Radio signals are expected to be measurable above 
background at energies $\lesssim 10^{17}$~eV, and probably down to 
energies as low as $\gtrsim 10^{16}$~eV when applying interferometric 
analysis techniques, see Fig.\ \ref{fig:crspectrum}.

In parallel to the experimental activities, models for the physics of 
the radio emission emanating from extensive air showers
have matured to a degree that the emission mechanisms are now 
generally assumed to be well-understood. As it turns out, there is a 
large overlap between the physics of radio emission from air showers 
and the physics of radio emission from particle showers in dense media. 
We will mention these parallels where appropriate. However, we 
deliberately focus this review on the case of air showers and the methods to detect them with 
radio techniques.

After a short introduction of the starting point for the modern-day 
experiments, including an overview of the merits warranting the 
investigation of radio detection of cosmic rays, we will set the 
scene with a review of the current paradigm of air shower radio 
emission physics and the most important characteristics of the emission.
Next, we will discuss the evolution of modelling efforts which, 
in conjunction with results from various experiments, 
led to this paradigm. Afterwards, we will describe the experimental projects 
which were developed over the past decade and highlight their goals 
and technological choices, before discussing some analysis-related 
aspects and then moving on to a detailed description of the important 
experimental results achieved to date and how they compare to theoretical 
predictions. Finally, we close with an outlook to possible future 
directions of the field of air shower radio detection.


\section{The starting point for digital radio detection of air showers}

Modern radio experiments built on knowledge gained 50 
years ago, which provided a valuable starting point. Here, we quickly 
discuss the most relevant information available from the historical 
works and then outline the promises of the radio detection technique 
which led to renewed interest and sparked the new projects.

\subsection{The knowledge from historical experiments}

Radio detection of cosmic rays {\it per se} is not a new technique. In fact, the 
experimental proof that air showers emit impulsive radio signals was 
made as early as 1965 \citep{JelleyFruinPorter1965}. As a consequence, several groups engaged in
experimental and theoretical work to study the details of the radio 
emission. It is not the goal of this article to review these 
historical works, and we kindly refer the reader to the excellent 
article of Allan \citep{Allan1971} for such a review.

However, let us briefly discuss the most relevant pieces of 
information which were available from the historical works at the time 
that the community rediscovered its interest in radio detection of 
cosmic rays. These were:

\begin{itemize}
\item Air showers initiated by cosmic rays emit impulsive radio emission. 
The emission was originally discovered at a frequency of 44~MHz, but successful 
detections from as low as 2~MHz up to 500~MHz followed.
\item The radio signal, at least at frequencies below 100~MHz, is 
coherent. In other words, the received power generally scales quadratically with the 
number of emitting particles, and thus with the energy of the primary 
cosmic ray (with the exception of showers being truncated when hitting 
the ground).
\item The emission is dominated by a geomagnetic effect, since a clear 
correlation of the radio signal strength is seen with the angle between the 
air shower axis (the axis determined by the arrival direction of the 
primary cosmic ray particle) and the geomagnetic field axis. Due to 
the geomagnetic nature of the emission, the signal is generally 
expected to be strongest in antennas measuring the polarisation 
component aligned with the Lorentz force. For detection sites at 
geographic mid-latitudes, this corresponds to the east-west polarisation 
component.
\item The signal strength measured by an antenna depends on the 
lateral distance from the air shower axis and can be fitted with an 
exponential lateral distribution function (LDF).\footnote{Keep in 
mind, however, that the historical experiments only measured the radio LDF 
averaged over many different air showers, not the radio LDF of an 
individual air shower itself.}
\end{itemize}

The gist of this knowledge can be summarized in one formula, often 
referred to as the ``Allan-formula'' (eq.\ (84) in Ref.\ \citep{Allan1971}), as follows:
\begin{eqnarray}
\epsilon_{\nu} &=& 20\ \mathrm{\mu V\ m}^{-1}\ \mathrm{MHz}^{-1}
\left(\frac{E_{\mathrm{p}}}{10^{17}\ \mathrm{eV}}\right)\ \nonumber \\
 &\times & 
\sin{\alpha}\ \cos{\theta}\ \exp\left(-\frac{R}{R_{0}(\nu, \theta)}\right),
\end{eqnarray}
in which $\epsilon_{\nu}$ denotes the peak total amplitude (modulus) of the electric field 
vector divided by the effective bandwidth of the measurement, 
$E_{\mathrm{p}}$ is the energy of the primary cosmic ray, $\alpha$ is 
the so-called ``geomagnetic angle'', i.e.\ the angle between the air 
shower axis and the geomagnetic field axis, $\theta$ is the ``zenith 
angle'', i.e.\ the angle wrt.\ vertical incidence of the cosmic ray 
primary, and $R$ denotes the lateral distance perpendicular from the air shower 
axis, often denoted ``axis distance''. The scale factor $R_{0}$ depends 
on frequency and zenith angle, but there were no particularly quantitative results 
available at the time.

In spite of this significant knowledge, a number of important questions were open:

\begin{itemize}
\item Several secondary emission mechanisms had been investigated on a 
theoretical basis, and there were some experimental results suggesting 
that the geomagnetic emission was not the only mechanism. However, 
these were indications at best, and they were far from being accepted 
in the community. The relevance of the atmospheric refractive index 
gradient had also been studied to some extent \citep{AllanRefractive1971,HoughXmaxLDF}.
\item There were theoretical investigations on the sensitivity of the 
radio signal, in particular its LDF, on the longitudinal evolution of 
the air shower \citep{AllanRefractive1971,HoughXmaxLDF}, and thus the mass of the primary particle,
but no experimental tests or quantitative studies were available.
\item The absolute strength of the radio emission was reported very 
differently, up to factors of 100 in amplitude, by different groups 
\citep{AtrashkevichVedeneevAllan1978}. The assumption was that this was due to difficulties in 
providing an absolute calibration for the measurements.
\item It was unclear how important the influence of electric fields in the 
atmosphere (say in thunderclouds \citep{Mandolesi19741431}) could be and whether an effect on the 
radio signal was to be expected even for fair weather. If the latter were 
true, it would make the technique unreliable for any quantitative 
measurements, as the atmospheric electric field at altitudes of several 
kilometers is hard to monitor.
\end{itemize}

In the mid-1970s, the activities on radio detection of cosmic rays 
ceased almost completely, as is evident also from Fig.\ 
\ref{fig:icrcstats}. This was due to a number of reasons, for example 
there were problems in associating the radio measurements 
reliably with the relevant air shower characteristics, which was 
sometimes attributed to the effects of unknown atmospheric electric 
fields. Also, the fluorescence imaging technique pioneered in the Fly's Eye experiment 
\citep{Bergeson:1975nh} seemed more promising and made good progress, which
shifted the interest of the cosmic ray community. It took 30 years for the 
interest in radio detection to renew. We will discuss the reasons for 
this renewal in the next subsection.

\subsection{The promises of radio detection}

As mentioned in the introduction, existing techniques using particle 
detector arrays and optical fluorescence detectors (or optical 
Cherenkov light detectors) have been very successful in studying cosmic rays 
over a very wide range of energies. However, they do have their 
shortcomings.

\begin{figure}[!htb]
\centering
\includegraphics[width=0.5\textwidth,clip=true,trim=2cm 2.5cm 10cm 9cm]{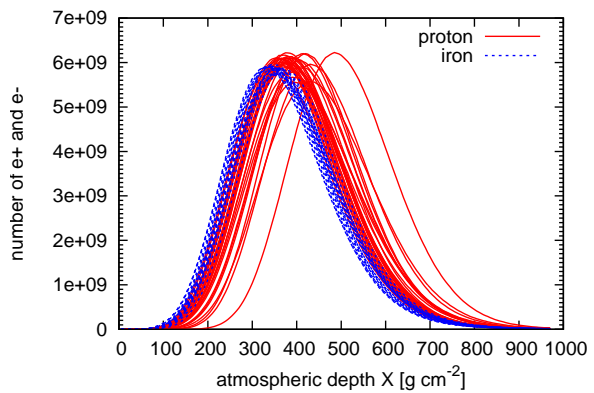}
\caption{Longitudinal evolution profiles of the electromagnetic 
components of extensive air showers initiated by proton and iron primaries with an energy of $10^{19}$~eV, 
as simulated with CORSIKA \citep{HeckKnappCapdevielle1998}. The depth of the maximum of the air shower 
evolution, $X_{\mathrm{max}}$, provides valuable information about the mass of the primary 
particle. More massive particles on average have a lower depth of 
shower maximum, and their distribution of \xmax values scatters less than for lighter 
particles.\label{fig:showerevolution}}
\end{figure}

All detection techniques at energies beyond $\gtrsim 10^{14}$~eV rely 
on the measurement of the extensive air shower cascade initiated by a 
primary cosmic ray in the atmosphere. This air shower is dominated by 
the electromagnetic component (electrons, positrons and photons) and 
has a characteristic evolution with atmospheric depth, shown in Fig.\ 
\ref{fig:showerevolution}. The shower first grows, then 
reaches a maximum, and afterwards dies out.

Particle detectors only measure a momentary snapshot of the secondary 
particles in an air shower reaching the ground, and typically only 
sample a small fraction of these particles due to their limited area coverage. This yields very 
indirect information about the original primary cosmic ray particle. 
In particular, uncertainties in the hadronic interactions at energies 
well beyond those accessible in collider experiments introduce 
significant systematic uncertainties in the reconstruction of the 
characteristics of the primary particle from the ground-based particle 
detector measurements \citep{Engel:2011zzb}. An important problem unsolved to date is the 
discrepancy in the number of muons predicted by simulations, which at 
10$^{19}$~eV is at least 30\% lower than the one derived from hybrid 
measurements with fluorescence telescopes and particle detectors at
the Pierre Auger Observatory \citep{AugerMuonExcess}. The most likely 
explanation for this discrepancy is in the extrapolation of hadronic 
interaction physics well beyond the scale probed with measurements at 
particle accelerators, which also measure in a very different regime 
than the extreme forward interactions in air showers. While particle detectors can be seen 
as the ``work-horse'' of cosmic ray detection, since in particular 
they can measure with 100\% duty cycle, information from other 
detectors is thus needed to exploit them to their full potential, 
especially with regard to a reliable reconstruction of the absolute 
energy scale of cosmic-rays. The Pierre Auger Observatory in particular has committed to such a 
``hybrid approach'' by combining particle detectors with optical 
fluorescence telescopes which are used to calibrate the 
energy scale of the particle detectors. It remains very difficult, 
however, to determine the mass of the primary particle from 
ground-based particle detectors. This requires at least the 
separate measurement of the electromagnetic and muonic components of the air 
shower, as is the goal of the AugerPrime upgrade \citep{AugerPrimeIcrc2015}, and still might suffer from systematic uncertainties in 
hadronic interaction models.

The most-used optical detection technique is the detection of 
ultraviolet light emitted by excited air molecules in the air shower. 
Using pixelated UV cameras, the longitudinal evolution of an air 
shower as shown in Fig.\ \ref{fig:showerevolution} can be imaged. The integral of the longitudinal evolution 
profile yields a calorimetric energy measurement of the air shower. 
As the emission of fluorescence light is vastly dominated by the electromagnetic component of the air 
shower, it is much less affected by hadronic interaction 
uncertainties. Also, the atmospheric depth at which the air shower reaches its 
maximum particle number, the ``shower maximum'' or $X_{\mathrm{max}}$ 
(in g~cm$^{-2}$) can be read off from the profile and yields precious 
information about the mass of the primary particles, again visible in Fig.\ 
\ref{fig:showerevolution}. Fluorescence 
detectors thus yield very high quality data --- however, they do it 
effectively in less than 10-15\% of the time. Their ``duty cycle'' is 
limited to such small numbers because the technique relies on clear, 
moon-less nights. Fiducial volume cuts further limit the 
fraction of data usable for analyses. At the very high energies, the 
loss in statistics of more than a factor of 10 is an important 
drawback. Also, the air quality at the site has to be good and has to 
be monitored very closely \citep{AugerNIM2014}. 

In light of these limitations, interest in the radio detection 
technique re-emerged in the early 2000s; for a review of the field at 
that time see \citep{FalckeGorhamProtheroe2004}. Radio detection was expected 
to have the following advantages:

\begin{itemize}
\item The radio emission is caused by the electromagnetic component of 
the air shower. As such it does not suffer strongly from uncertainties in the 
hadronic interaction models.
\item The radio signal is integrated over the full shower evolution. 
(There is no relevant damping in the atmosphere at VHF frequencies.) 
It thus represents a calorimetric energy measurement.
\item Radio measurements can be performed with essentially 100\% duty 
cycle.\footnote{As we will see, only thunderstorms lower this value, 
but it typically remains at a level of more than 95\% at most sites.}
\item The radio signal should be sensitive to the longitudinal shower 
evolution and thus $X_{\mathrm{max}}$.
\item Radio antennas can be built comparably cheaply. Possibly, very large areas could be 
instrumented economically with radio detectors to detect cosmic rays 
at the highest energies.
\item In contrast to the 1960s and 1970s, powerful digital signal 
processing is available today. (No more photographing of oscilloscope traces!) 
While digital electronics do come with a price tag, radio detections 
directly profits from Moore's law. Digital 
electronics get exponentially cheaper in time.
\end{itemize}

In the early 2000s, the idea arose to apply digital radio detection to the problem of 
cosmic ray physics \citep{FalckeGorham2003}. One vision driving these 
activities was the hope that a combination of particle 
detectors and radio antennas could yield similar information as the 
combination of particle and fluorescence detectors --- but with 100\% 
duty cycle instead of 10\% duty cycle.

We will discuss in the course of this review which of these promises 
can be kept and which cannot. Before we go into any more detail of the 
evolution of the radio emission experiments, let us set the scene with a 
summary of what we know about the nature of the radio emission from 
air showers today.


\section{The physics of radio emission from extensive air showers} 
\label{sec:emissionparadigm}

Before reviewing the progress of the last decade on both the theoretical and the 
experimental side in detail, let us first go through a concise summary 
of the radio emission physics as we understand it today. We will keep 
the discussion mostly non-technical, readers interested in the details 
are encouraged to study the original publications referenced in the 
text.

\subsection{Geomagnetic emission}

The main emission mechanism for radio pulses from cosmic ray showers 
is associated with the geomagnetic field: Secondary electrons and 
positrons in the air shower are accelerated in the magnetic field. One 
idea that was followed was that this acceleration directly leads to 
the radio emission as in synchrotron emission, hence the term ``geosynchrotron emission'' was 
coined \citep{FalckeGorham2003}. However, this view does not correctly describe the emission 
physics in air showers. The reason is that electrons and positrons do 
not propagate unimpeded on long, let alone periodic orbits. Instead, they interact 
continuously with air molecules. The situation is comparable to the one of electrons in a conductor to which a 
voltage is applied. In the equilibrium of acceleration by 
the magnetic field and deceleration in interactions with air 
molecules a net drift of the electrons and positrons arises in opposite directions as governed by the 
Lorentz force
\begin{equation}
\vec{F}= q\ \vec{v} \times \vec{B}.
\end{equation}
where $q$ denotes the particle charge, $\vec{v}$ is its velocity 
vector and $\vec{B}$ is the magnetic field vector. For particles 
originally moving along the shower axis, the resulting current will be 
perpendicular to the shower axis, i.e.\ we can refer to them as 
``transverse currents''.

\begin{figure}[!htb]
\centering
\includegraphics[width=0.4\textwidth]{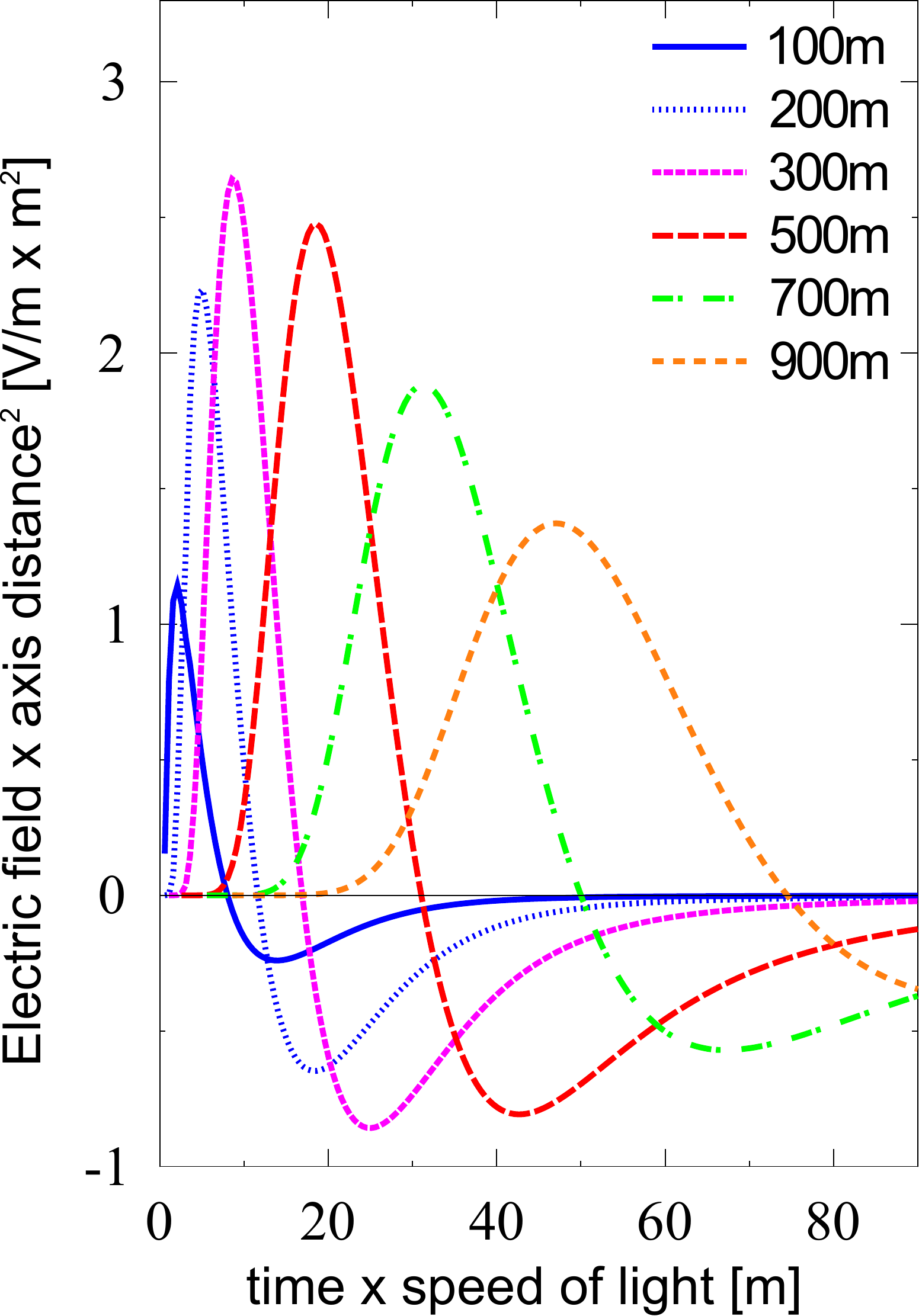}
\includegraphics[width=0.41\textwidth]{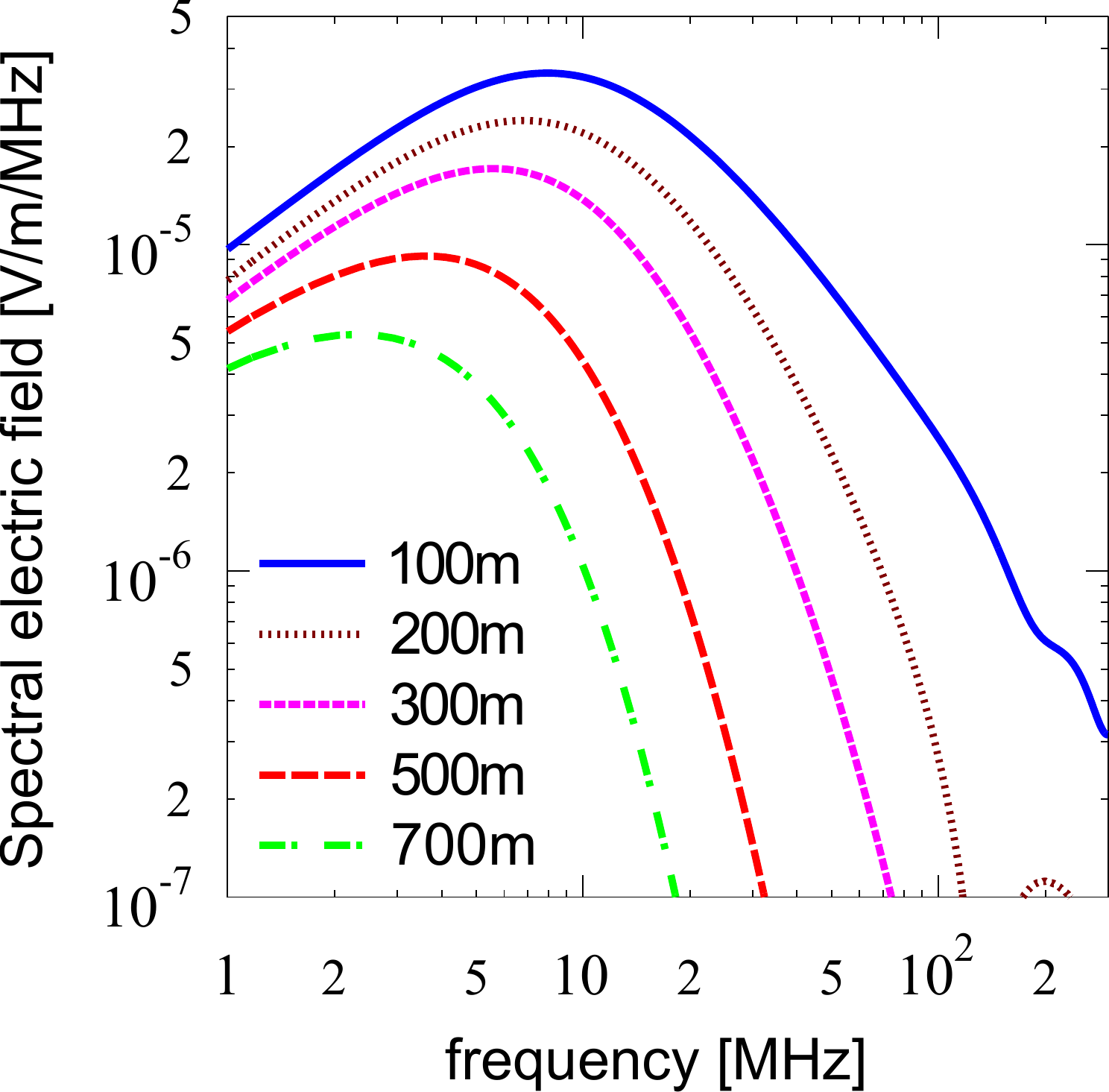}
\caption{Radio pulses (top) arising from the time-variation of the 
geomagnetically induced transverse currents in a $10^{17}$~eV air 
shower as observed at various observer distances from the shower 
axis and their corresponding frequency spectra (bottom). Refractive 
index effects are not included. Adapted from \citep{ScholtenWernerRusydi2008}.\label{fig:radiopulses}}
\end{figure}

There is one more important ingredient: These transverse currents vary 
as the air shower evolves and the number of secondary particles first 
grows, then reaches a maximum, and then declines as the shower dies 
out (cf.\ Fig.\ \ref{fig:showerevolution}). It is this {\em time-variation} of the transverse 
currents which leads to electromagnetic radiation. Due to the 
relativistic speed of the emitting particles, the emission is 
compressed in short pulses in the forward direction, along the shower axis (Fig.\ 
\ref{fig:radiopulses}, top). Correspondingly, the emission has broad-band frequency spectra
(Fig.\ \ref{fig:radiopulses}, bottom). For geometrical reasons, the pulses get broader and the 
frequency spectra cut off at lower frequencies as the observer moves 
away from the shower axis. Interpreted in a 
microscopic way, the time-variation of the transverse currents can be 
associated with acceleration of individual particles, i.e., it is in 
fact acceleration of charged particles that produces the radiation.

\begin{figure*}[!htb]
  \centering
  \includegraphics[width=0.23\textwidth,clip=true,trim=0cm 0cm 0cm 12cm]{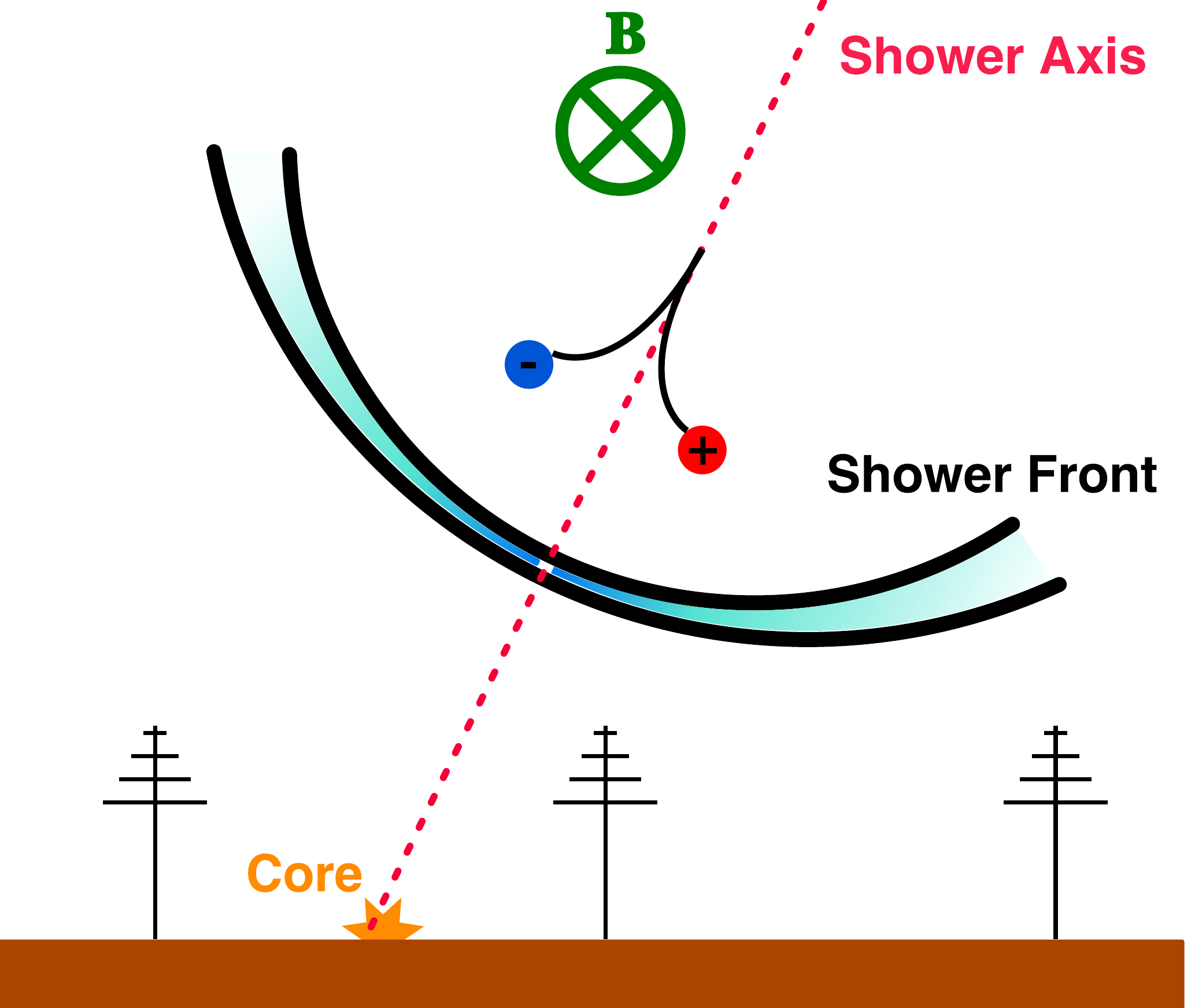}
  \includegraphics[width=0.21\textwidth]{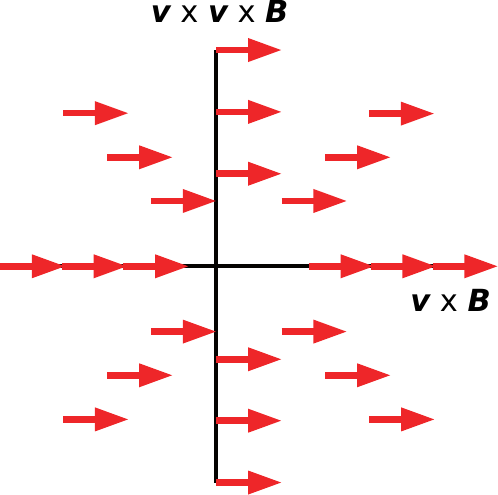}
  \hspace{0.06\textwidth}
  \includegraphics[width=0.23\textwidth,clip=true,trim=0cm 0cm 0cm 12cm]{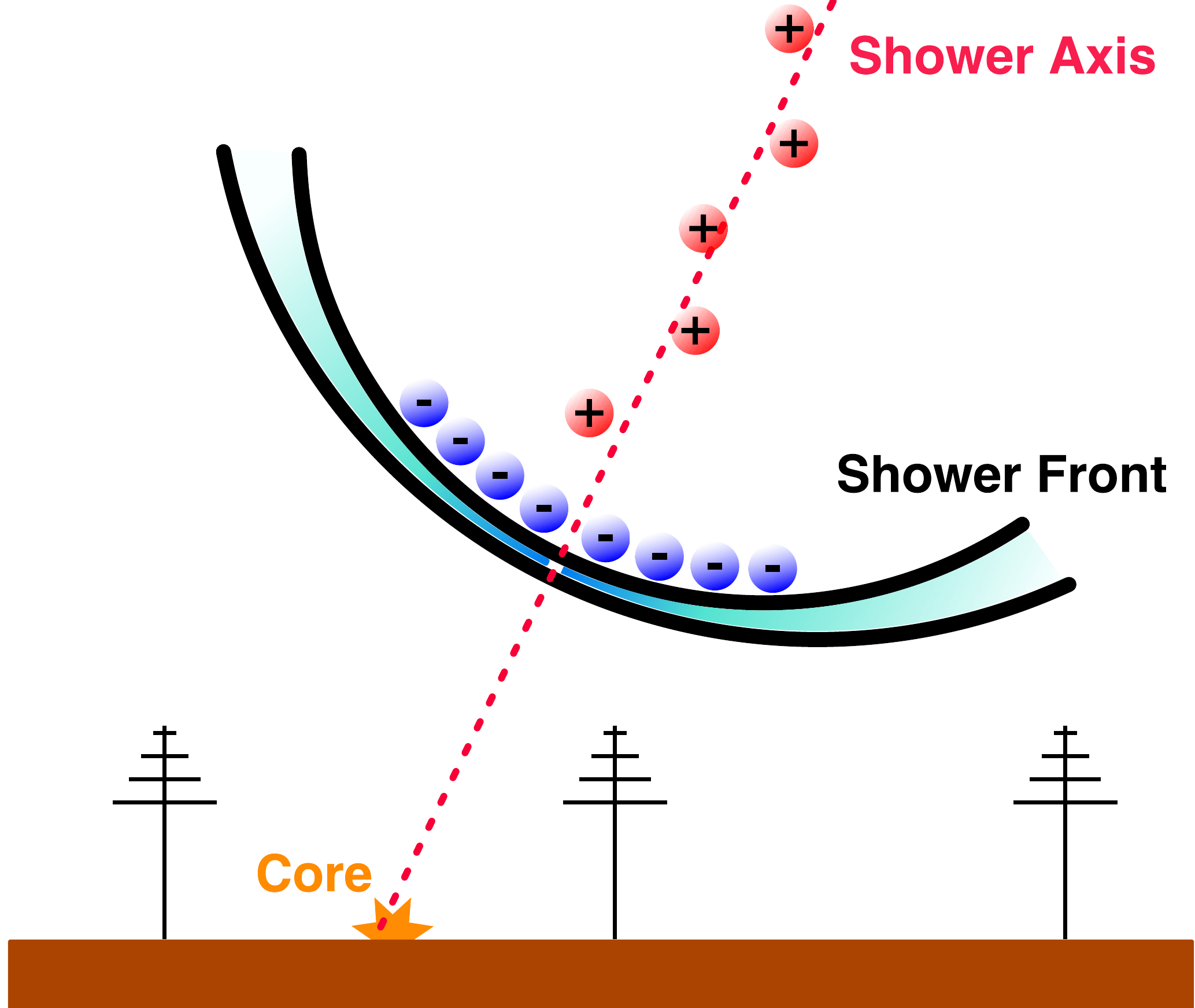}
  \includegraphics[width=0.21\textwidth]{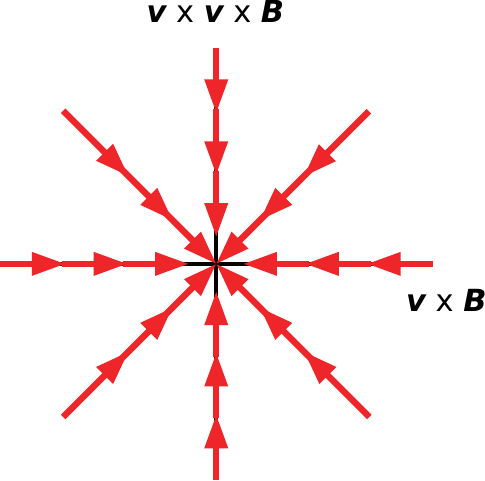}
  \caption{Left: Illustration of the geomagnetic radiation mechanism. 
  The arrows denote the direction of linear polarisation in the plane 
  perpendicular to the air shower axis. Irrespective of the observer 
  position, the emission is linearly polarised along the direction 
  given by the Lorentz force, $\vec{v} \times \vec{B}$ (east-west for 
  vertical air showers). Right: 
  Illustration of the charge excess (Askaryan) emission. The arrows 
  illustrate the linear polarisation with electric field vectors oriented
  radially with respect to the shower axis. Diagrams have been adapted from 
  \citep{SchoorlemmerThesis2012} and \citep{deVries2012S175}.}
  \label{fig:mechanisms}
 \end{figure*}

The polarisation of the radiation by the time-varying transverse 
currents is linear with the electric field vector aligned with the 
Lorentz force, i.e.\ along the $\vec{v} \times \vec{B}$ direction, 
where the propagation direction of the particles $\vec{v}$
can be approximated with the shower axis. This 
is illustrated in the left panel of Fig.\ \ref{fig:mechanisms}.

In principle, any charged particle undergoes the processes described 
here. However, only electrons and positrons contribute significantly 
to the radio signal as they have by far the highest charge/mass 
ratio. Already muons are much too heavy to make a significant 
contribution.

This emission physics has already been 
described by Kahn \& Lerche \citep{KahnLerche1966}. A modern 
formulation was developed by Scholten, Werner and Rusydi 
\citep{ScholtenWernerRusydi2008}.

\subsection{Charge excess emission (Askaryan effect)}

In addition to the dominating geomagnetic contribution\footnote{Obviously, the 
geomagnetic contribution vanishes for air showers arriving parallel to the 
magnetic field. For air showers with a small geomagnetic angle thus 
the charge excess emission can actually become dominant.} a secondary effect 
exists. It is well known that there is a negative charge excess of $\approx 
10-20$\% in air showers, which is caused mostly by the fact that 
the ambient medium is ionized by the air shower particles and the ionization electrons are swept 
with the cascade, while the much heavier positive ions stay behind. As 
the shower evolves, the absolute negative charge present in the moving 
cascade grows, reaches a maximum and finally decreases when the shower dies 
out. Hence, again there is a {\em time-varying} charge excess, and 
this leads to pulses of electromagnetic radiation.

This radiation also has linear polarisation. However, the electric 
field vector is oriented radially with respect to the shower axis. In 
other words, the orientation of the electric field vector depends on 
the location of an observer (radio antenna) with respect to the shower 
axis, as is illustrated in the right panel of Fig.\ \ref{fig:mechanisms}.

The mechanism described here, together with Cherenkov-like effects 
that will be described in the next section, is essentially the 
Askaryan-effect \citep{Askaryan1962a,Askaryan1965}. It usually plays a sub-dominant role in 
air shower physics, however it is the sole relevant emission mechanism 
in particle showers in dense media and has been investigated in 
considerable depth in the context of neutrino detection via radio 
emission in ice and the lunar regolith (see, e.g., \citep{AlvarezMunizAskaryan}). Since the length scales of particle 
showers in dense media are much smaller, the resulting radiation is strongest at 
GHz frequencies. The underlying physics, however, is 
the same as in air.

\subsection{Superposition of the contributions and signal asymmetries}

\begin{figure}[!htb]
  \vspace{2mm}
  \centering
  \includegraphics[width=0.49\textwidth]{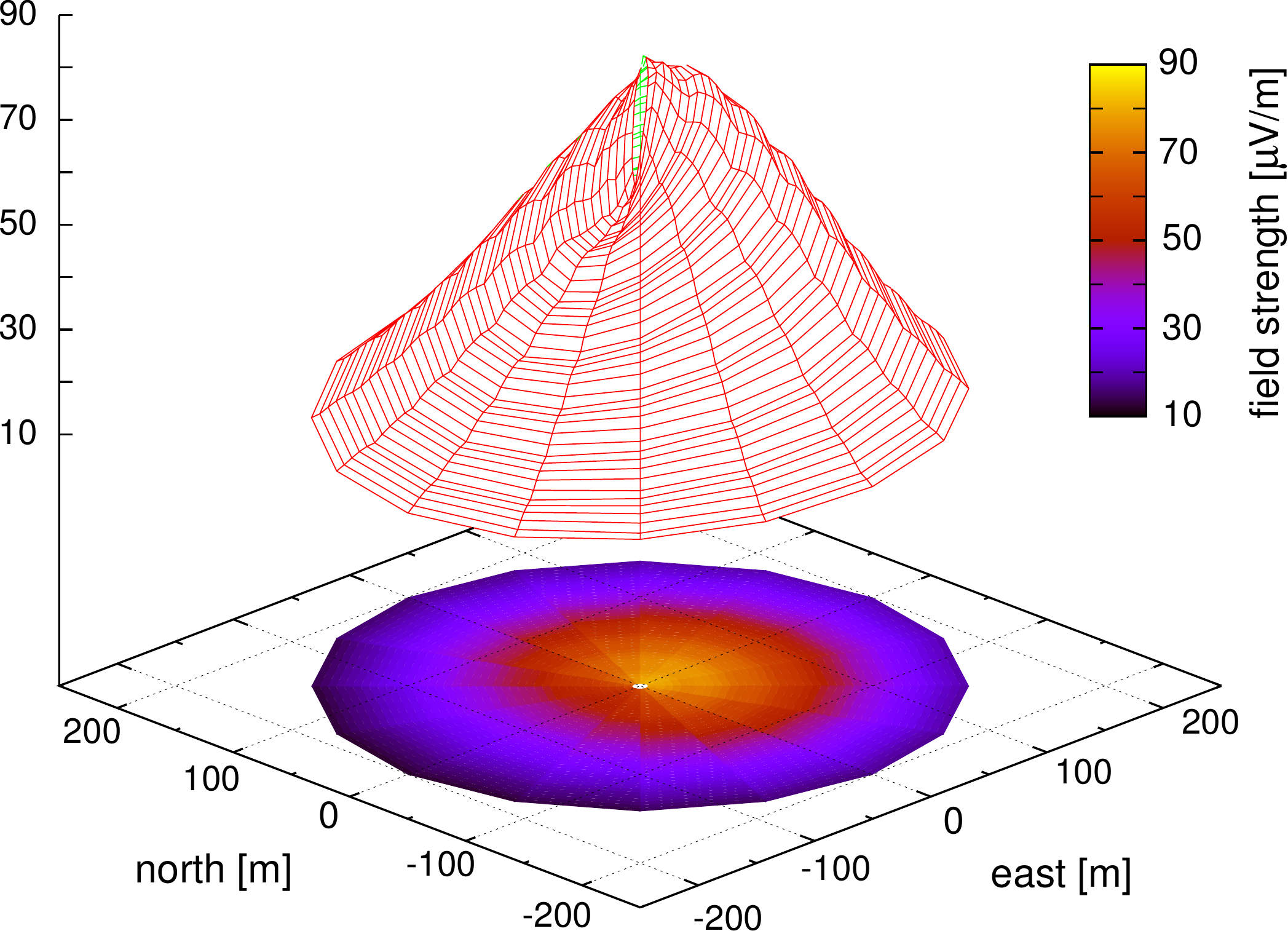}
  \caption{Simulation of the total electric field amplitude in the 
  40-80~MHz band for a vertical cosmic ray air shower at the site of 
  the LOPES experiment. The asymmetry arises 
  from the superposition of the geomagnetic and charge-excess emission 
  contributions. Refractive index effects are included. Adapted from 
  \citep{HuegeARENA2012a}.}
  \label{fig:footprint}
 \end{figure}

When the electric field vectors associated with the two emission mechanisms
are superposed, complex asymmetries in the radio signal arise, as 
depending on the observer location, the two contributions can add 
constructively or destructively. The arising asymmetry, specifically 
along the direction denoted by $\vec{v} \times \vec{B}$ (east-west for 
vertical air showers) is illustrated by the visualization of the ``radio footprint'' 
(two-dimensional radio LDF) depicted in Fig.\ \ref{fig:footprint}. The 
degree of asymmetry depends on the relative strength of the 
geomagnetic and charge excess contributions, and thus in particular on 
the geomagnetic angle of a given air shower as well as the 
strength of the local geomagnetic field.

\begin{figure*}[h!t]
  \includegraphics[height=0.75\textwidth]{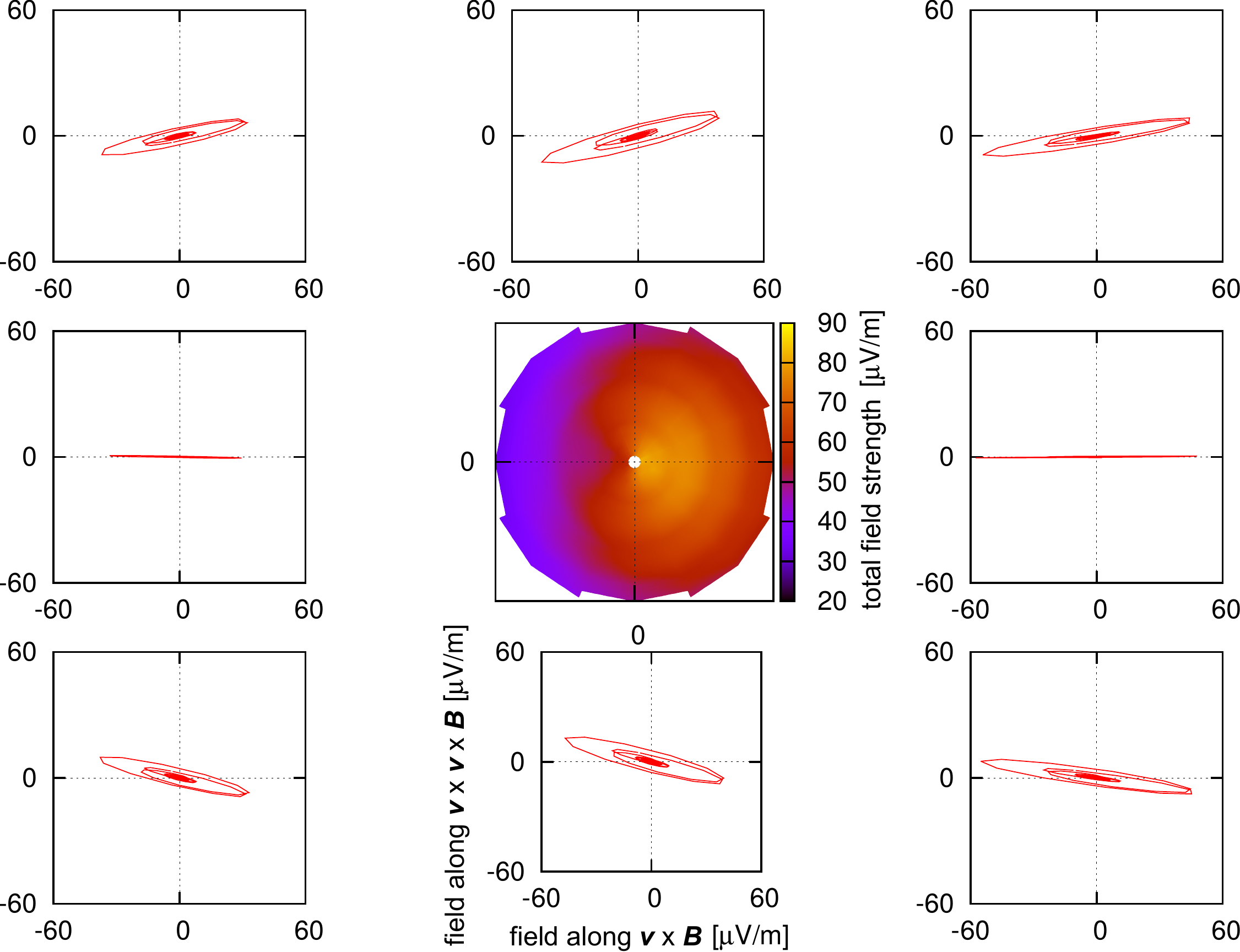}
  \caption{Illustration of the time-evolution of the electric field 
  vector in the 40-80~MHz frequency band for antennas at 100~m lateral 
  distance to the north, north-east, east, south-east, south, 
  south-west, west, and north-west (counting clockwise from the top 
  panel) of the impact point of a vertical $10^{17}$~eV air shower 
  at the LOPES site. The field is decomposed in components along the
  direction of the Lorentz force and perpendicular to that.
  The map in the center illustrates the total 
  amplitude footprint. Refractive index effects are included.
  Adapted from \citep{HuegeARENA2012a}.}
  \label{fig:lopespolarisation}
\end{figure*}

While the footprint shown in Fig.\ \ref{fig:footprint} illustrates the 
peak amplitude measured at various observer locations, a closer look 
at the time-evolution of the impulsive radio emission is shown in Fig.\ 
\ref{fig:lopespolarisation}. The pulses associated with the two 
emission mechanisms are not perfectly synchronized, a sign that the 
time-variation of the transverse currents induced by geomagnetic 
effects and the time-variation of the net charge excess are slightly offset over the 
course of the longitudinal evolution of the extensive air shower.  Therefore, the 
electric field vector does not generally trace a line in the plane 
perpendicular to the shower axis; instead, it generally traces an ellipse. In other words, the radio 
emission from cosmic ray air showers is generally of elliptical 
polarisation, i.e.\ an admixture of linear and circular polarisation. 
This effect remains to be proven experimentally.

\subsection{Forward beaming, coherence and Cherenkov-like effects} 
\label{sec:cherenkov}

An important factor in the emission physics is coherence. As long as 
radiation at a given frequency from different particles acquires negligible relative phase shifts 
during its propagation to the observer, the vectorial electric fields 
add up coherently. This means that the electric field amplitude scales linearly with particle number and thus 
(approximately) with the energy of the primary particle. The linearity 
of this dependence is a very useful feature for energy measurements
with radio techniques. Equivalently, the received power scales 
quadratically with the energy of the primary particle.

Obviously, coherence is frequency-dependent and more pronounced 
at low frequencies. Coherence is influenced by the spatial particle 
distribution (mainly the thickness of the air shower disk, but for 
observers at large lateral distances also its lateral extent) as 
well as geometrical effects and propagation physics.

Due to the relativistic motion of the radiating particles along the air shower 
axis, the radiation is strongly forward-beamed, requiring antennas 
placed within a constrained ``illuminated area''. 
Furthermore, the refractive index of the atmosphere is not unity. At 
sea level a typical value is $n = 1.000292$, and it scales  
proportionally with the air density to higher altitudes.

\begin{figure}[!htb]
\centering
\includegraphics[width=0.5\textwidth]{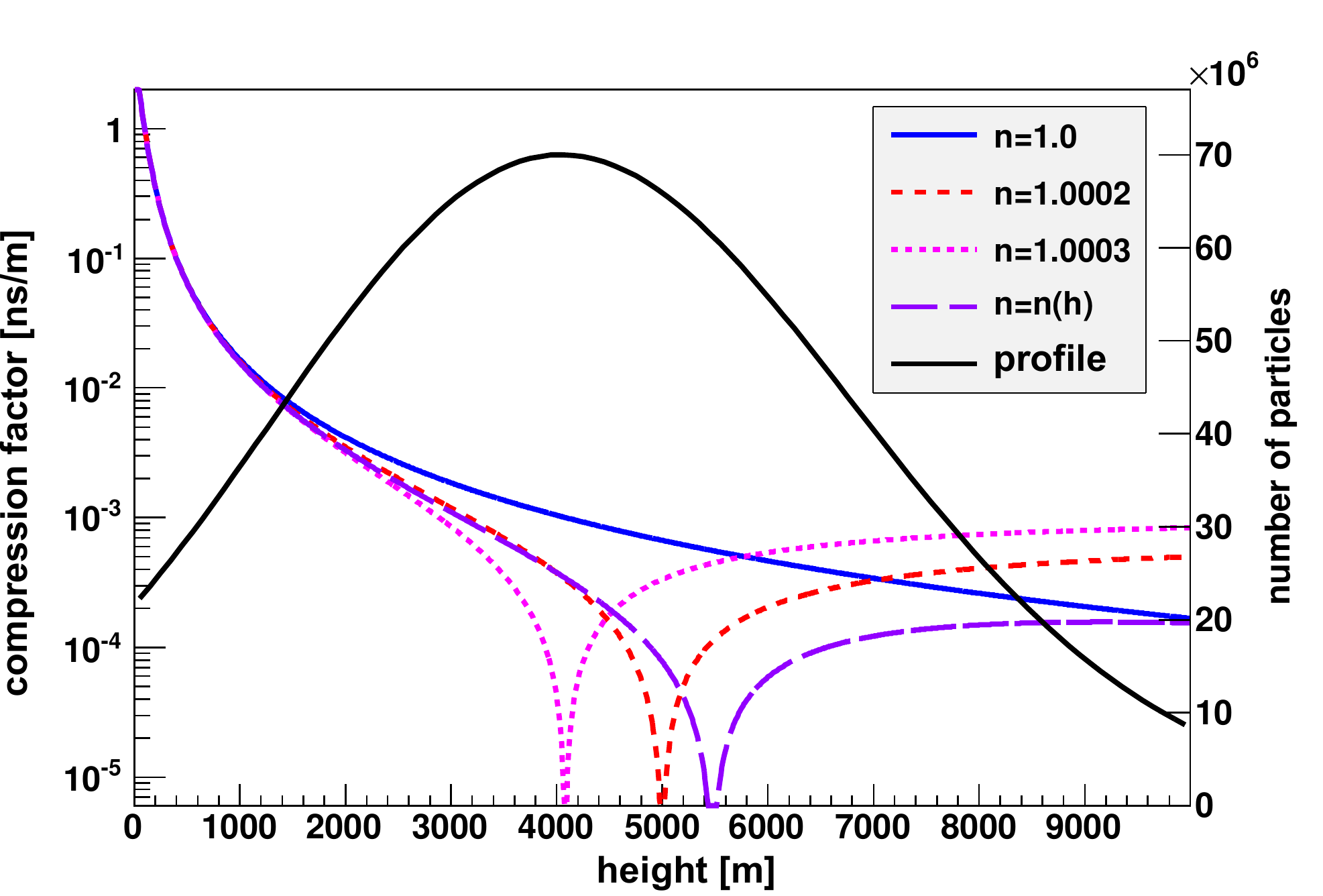} 
\caption{For different models of the atmospheric refractive index $n$, 
emission emanating from certain heights $h$ can be strongly compressed 
in time, as indicated by the compression factor on the left vertical 
axis. If the strongly compressed emission region coincides with those parts of the 
longitudinal shower evolution profile (black line and right vertical 
axis) at which the shower changes rapidly, strong compressed radio pulses occur. Adapted from 
\citep{AlvarezMunizCarvalhoZas2012}.\label{fig:compressionfactor}}
\end{figure}

\begin{figure}[!htb]
\centering
\includegraphics[width=0.48\textwidth]{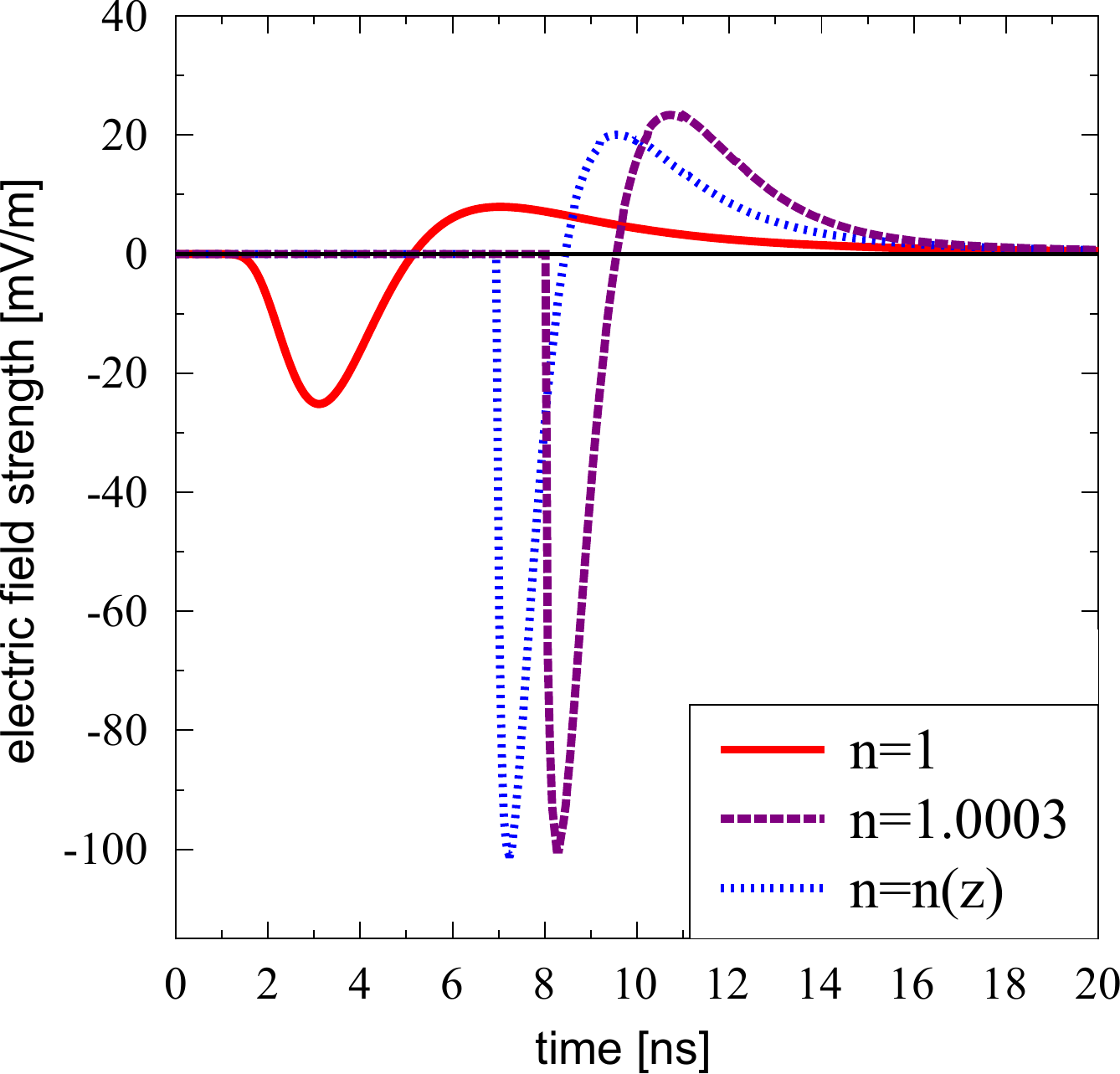}
\caption{Simulated radio pulses for an air shower with a primary 
energy of $5 \times 10^{17}$~eV. The observer is located at an axis 
distance of 100~m. The refractive index $n$ has been adopted as unity (vacuum),
1.0003 (sea level) and a realistic gradient in the 
atmosphere $n(z)$, illustrating the ensuing time compression of the radio 
pulses. The particle distribution is approximated to have no lateral extent.
Adapted from \citep{DeVriesBergScholten2011}.\label{fig:cherenkovcompression}}
\end{figure}

\begin{figure}[!htb]
  \includegraphics[width=0.48\textwidth]{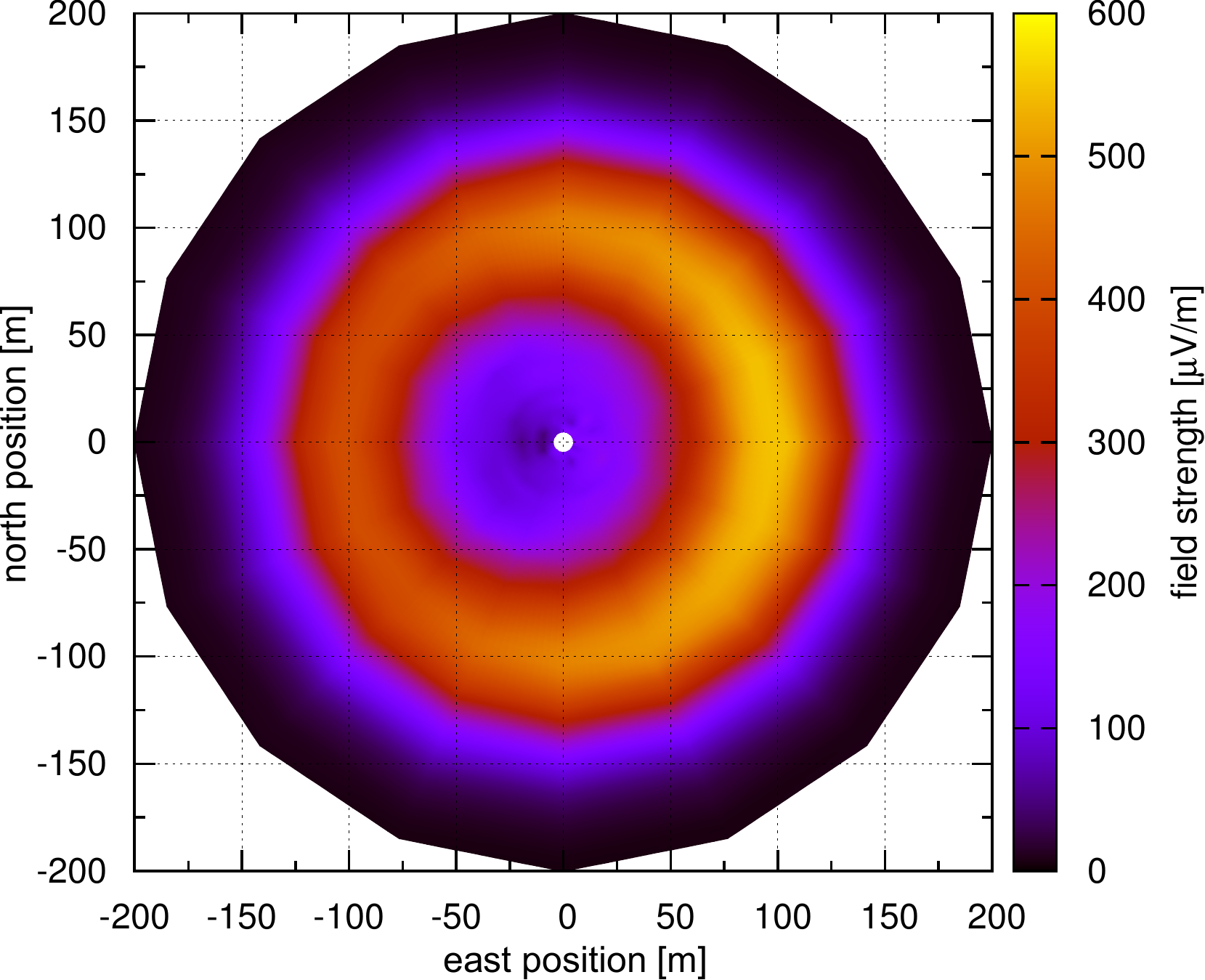}
  \caption{Radio-emission footprint of the total field strength of a vertical 
  $10^{17}$~eV air shower induced by an iron 
  primary at the LOPES site as seen in the frequency range from 300 to 
  1,200~MHz. Adapted from \citep{HuegeARENA2012a}. 
  \label{fig:anitafreqs}}
\end{figure}

This refractive index gradient has important consequences for the 
resulting radiation pattern. If for a given emission region along the shower axis an observer is 
located at the corresponding Cherenkov angle\footnote{Due to the 
refractive index gradient in the atmosphere, there is of course strictly 
speaking no well-defined, unique Cherenkov angle.}, radiation emitted from all 
along this region arrives simultaneously at the observer. In other 
words, pulses are compressed in time and can thus become very short, 
as is shown in Figs.\ \ref{fig:compressionfactor} and \ref{fig:cherenkovcompression}. This can lead to 
coherent emission up to GHz frequencies for observers 
located on a ``Cherenkov ring'' with typical ring radii for vertical 
$10^{17}$~eV air showers of order 100~m, as is shown in Fig.\ \ref{fig:anitafreqs}.

At lateral distances which fall inside the Cherenkov ring, the pulses are 
stretched by the refractive index effects and the transition frequency 
from coherent to incoherent emission is decreased. Also, the time-ordering of signals is 
reversed: signals emitted in the early stages of the air shower arrive 
later than those emitted in late stages.

At large lateral distances, outside the Cherenkov ring, the refractive 
index effects are negligible, as the pulse-widths are dominated by 
geometrical effects. The larger the lateral distance, the broader the 
received pulses and the lower is the transition frequency from 
coherent to incoherent emission.

We stress here that the refractive index gradient influences the 
radiation emitted by the time-varying transverse currents and 
time-varying charge excess by compressing (or stretching) it in time. 
It is valid to term this ``Cherenkov-like'' effects. However, the 
reader should not confuse this with ``Cherenkov radiation'' in the 
sense of a (constant) net charge moving through a medium with a 
velocity which is higher than the medium-speed-of-light \citep{JamesFalckeHuege2012}. Such a 
contribution by ``Cherenkov radiation'' must certainly be present, but to 
our knowledge it is completely negligible for the case of air showers 
at radio frequencies.

\subsection{Source distance effects} \label{sec:distanceeffects}

As the radio emission is strongly forward-beamed, into a cone of 
a few degrees opening angle, the distance of the radio source 
from the observer has a strong influence on the size of the 
illuminated area. It should be noted that for the radio emission, 
geometrical distance scales, in particular the distance from source to 
observer, matters. This is in contrast to the air shower evolution which 
is governed by the amount of matter traversed (atmospheric depth).

\begin{figure*}[tbh]
\centering
\includegraphics[width=0.95\textwidth,clip=true,trim=0cm 25cm 0cm 40cm]{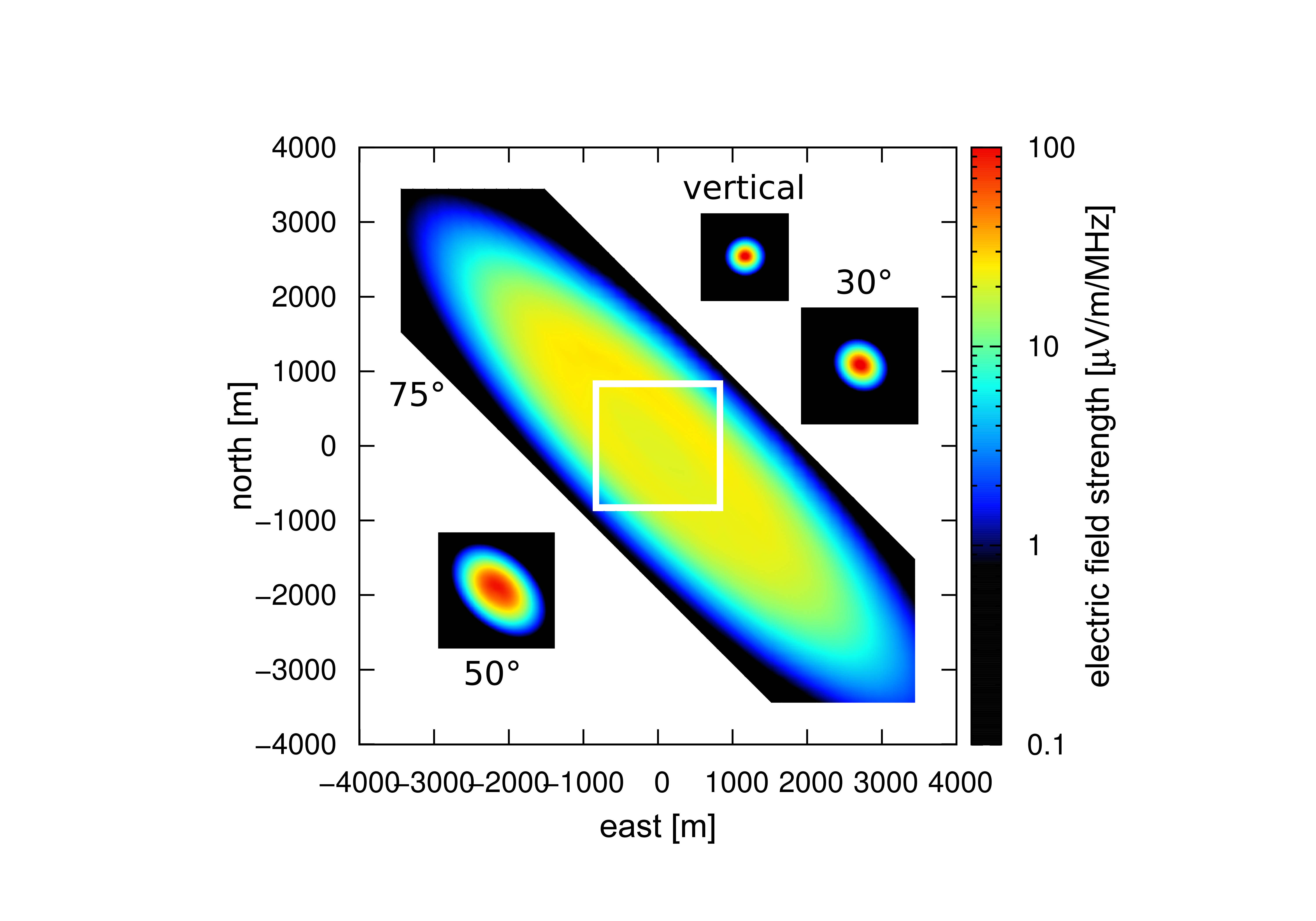}
\caption{Simulated footprints of the radio emission of extensive air 
showers with various zenith angles in the 30-80~MHz frequency band for 
an air shower with an energy of $5 \times 10^{18}$~eV. The detection threshold 
governed by Galactic noise typically corresponds to 
$\approx$~1-2~$\mu$V/m/MHz. The footprint is small for air showers with zenith angles up to $\approx 
60^{\circ}$, but becomes very large for inclined showers with zenith 
angles of $70^{\circ}$ or higher. The white rectangle denotes 
the size of the $50^{\circ}$ inset. The strong increase of the 
area illuminated by inclined air showers is due to the large 
geometrical distance of their emission region from the ground. Adapted 
from \citep{HuegeUHECR2014}.}
\label{fig:inclined}
\end{figure*}

A particularly important effect is the dependence of the radio 
emission on the air shower zenith angle. As the zenith angle 
increases, the traversed atmospheric depth grows as\footnote{This is an approximation for a planar 
atmosphere which is valid up to zenith angles of $\approx 
70^{\circ}$.} $\cos^{-1}(\theta)$. The air shower reaches its maximum 
at a given atmospheric depth, 
thus for more inclined showers this maximum will be at significantly 
larger geometrical distances from the observer than for vertical air 
showers. As a consequence, the forward-beamed radio emission 
illuminates a much larger area, as is illustrated impressively in 
Fig.\ \ref{fig:inclined}. The average electric field amplitude is 
lower (the radiated power is distributed over a larger area), but also 
the LDF is less steep. This makes inclined air showers more favourable 
for detection with a sparse antenna grid \citep{GoussetRavelRoy2004}.

\begin{figure}[!htb]
  \includegraphics[width=0.48\textwidth]{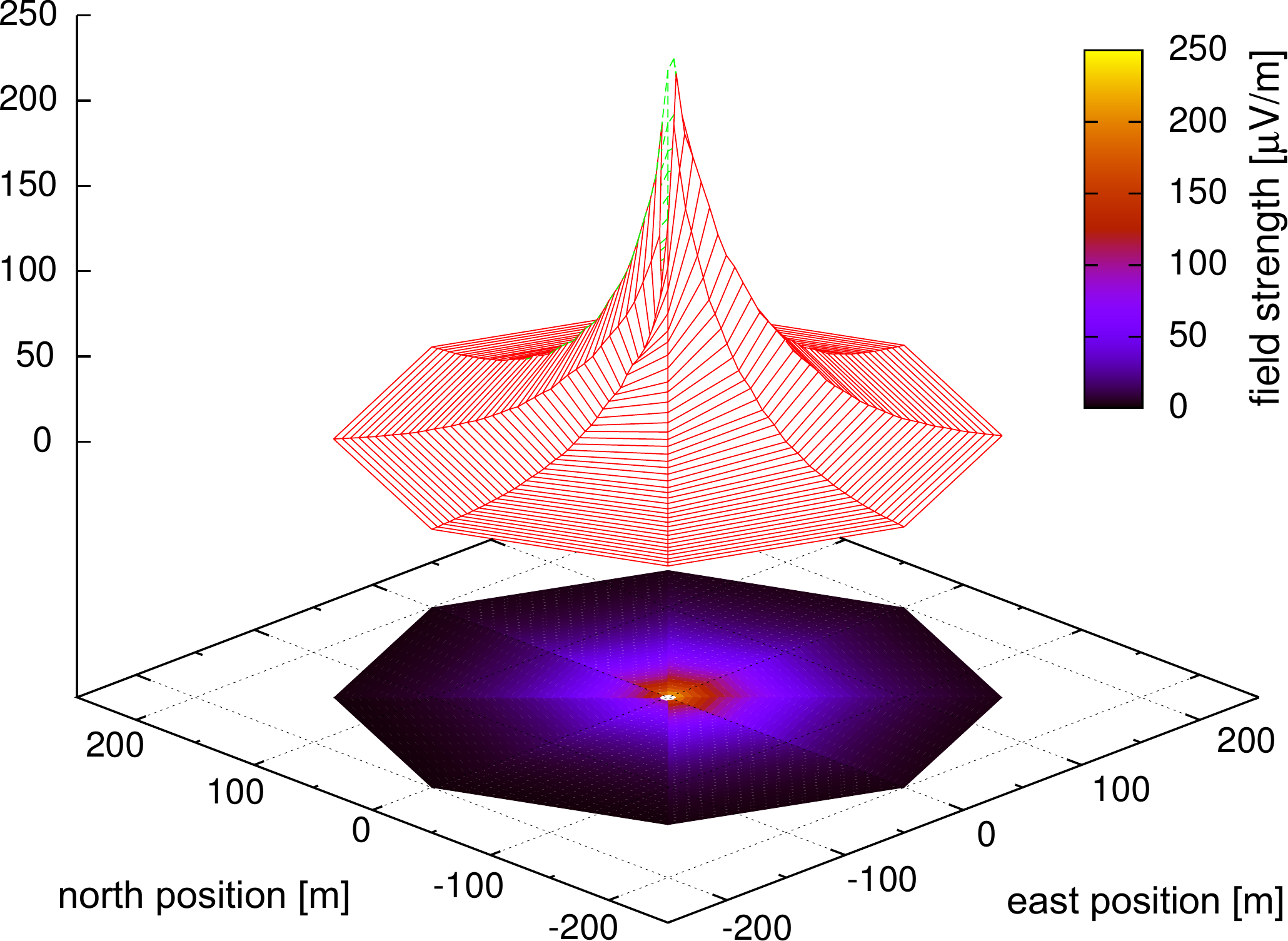}
  \caption{Radio-emission footprint of the 40-80~MHz total field 
  strength of a vertical $10^{17}$~eV air shower induced by a proton 
  primary at the LOPES site. Adapted from \citep{HuegeARENA2012a}. \label{fig:lopesproton}}
\end{figure}

For a fixed zenith angle, another important factor influencing the 
geometrical source distance is the depth of the maximum of an 
individual air shower, $X_{\mathrm{max}}$. This quantity undergoes 
statistical fluctuations, but is one of the most important observables 
to determine the mass of the primary particle. Changes in 
\xmax are also reflected in the geometrical distance between 
radio source and observer, and thus can be exploited to determine 
\xmax from radio measurements. The most obvious way to 
access this information is the LDF, as discussed above and illustrated 
in the comparison between Fig.\ \ref{fig:footprint} for an 
iron-induced air shower (small $X_{\mathrm{max}}$) and Fig.\ \ref{fig:lopesproton} 
for a proton-induced air shower (large $X_{\mathrm{max}}$) (note the different 
scales). For the same geometrical reasons, the geometrical source distance also 
influences the shape of the radio wavefront, which can be determined 
by precise timing measurements, and the pulse shape (or spectral index of the 
frequency spectrum) measured at a given lateral distance. We will discuss the sensitivity of the radio signal to the mass of the 
primary particle in more depth in the chapter on experimental results.

\section{Modern models and simulations of air shower radio emission}

In parallel with the modern experimental efforts, modelling efforts 
for the radio emission from extensive air showers were started. We 
give an overview here of approaches that have been tried out, but will 
focus on those that are still being maintained at the time of writing 
this review.

\subsection{Flawed modern approaches}

A number of efforts that were started in the early 2000s later turned 
out to be flawed. The ``geosynchrotron'' scheme was followed initially 
with a semi-analytic calculation in the frequency domain 
\citep{HuegeFalcke2003a}. The explicit assumption of synchrotron 
radiation of particles on long orbits dominating the emission, 
however, later turned out to be untrue.

Time-domain calculations inspired by the ``geosynchrotron'' idea
were started as well. Approaches by DuVernois et al.\ \citep{DuVernoisIcrc2005} and Suprun et al.\ 
\citep{SuprunGorhamRosner2003} were developed at the same time as the REAS Monte Carlo 
code, which was first based on parameterized air showers (REAS1)
\citep{HuegeFalcke2005a,HuegeFalcke2005b} and later on histogrammed 
particle distributions \citep{LafebreEngelFalcke2009} extracted from CORSIKA 
\citep{HeckKnappCapdevielle1998} simulations (REAS2).  Independently, 
the ReAIRES code \citep{RiviereIcrc2009}, based on the AIRES air shower Monte Carlo code 
\citep{Sciutto1999}, was implemented. A simplified point-like model was 
also formulated \citep{ChauvinRiviereMontanet2010}. The SELFAS1 code 
was developed on the basis of parameterizations of particle 
distributions determined from REAS2 histograms 
\citep{LafebreEngelFalcke2009}.

It turned out later that in all these time-domain approaches the 
discretized implementation of the classical electrodynamics 
calculation was flawed. For details, we refer the reader to 
\citep{HuegeLudwigScholtenARENA2010}. The problem was that these models 
start from the Li\'enard-Wiechert description of the electric field of 
a single moving charged particle \citep{Jackson1975},
\begin{eqnarray} \label{eqn:radiate}
\vec{E}(\vec{x},t) &=& e \left[\frac{\vec{n}-\vec{\beta}}{\gamma^{2}(1-\vec{\beta} \cdot \vec{n})^{3} R^2}\right]_{\mathrm{ret}} \nonumber \\
&+& \frac{e}{c} \left[ \frac{\vec{n} \times \{(\vec{n}-\vec{\beta})\times \dot{\vec{\beta}}\}}{(1-\vec{\beta}\cdot\vec{n})^{3}R}\right]_{\mathrm{ret}},
\end{eqnarray}
with $e$ for the electron charge, $c$ the speed of 
light, $\gamma$ the Lorentz factor of the particle, $\vec{n}$ the unit vector along the line-of-sight 
between particle and observer, $\vec{\beta}$ the particle velocity in 
units of $c$, and $R$ the distance between particle and 
observer. The index ``ret'' specifies that the equations have to be 
evaluated at the appropriate retarded time. Using this expression as 
their building block, the models took into account the emission from acceleration of 
particles in the magnetic field. However, in an air shower, the radio 
emission emanates from an ensemble of $N$ relativistic charged 
particles. Because $N$ varies over the air shower evolution, it 
has a time-dependence $N(t)$. The flawed models calculated the radio 
emission from the ensemble of particles in an air shower as
\begin{equation}
\vec{E}_{\mathrm{tot}}(\vec{x},t) = N(t)\, \vec{E}(\vec{x},t)
\end{equation}
Although this equation seems to take into account the variation of the 
number of relativistic charged particles, radiation associated with 
this variation of the number of charges is neglected. This becomes 
apparent when recalling that the electric field equation 
(\ref{eqn:radiate}) itself is derived from the Li\'enard-Wiechert potentials
\begin{eqnarray}
\Phi(\vec{x},t) & = & \left[ \frac{e}{(1 - \vec{\beta} \cdot \vec{n}) R} \right]_{\mathrm{ret}} \nonumber \\
\vec{A}(\vec{x},t) & = & \left[ \frac{e \vec{\beta}}{(1 - \vec{\beta} \cdot \vec{n})} \right]_{\mathrm{ret}} \label{lwpot}
\end{eqnarray}
via
\begin{equation} \label{eqn:fieldderivation}
\vec{E}(\vec{x},t)=-\nabla \Phi(\vec{x},t) - \frac{\partial}{\partial 
t}\vec{A}(\vec{x},t).
\end{equation}
The fact that the number of charged particles $N(t)$ is changing as a 
function of time thus must be taken into account in the calculation 
of $\vec{A}_{\mathrm{tot}}(\vec{x},t) = N(t)\,\vec{A}(\vec{x},t)$ 
already. The time-derivative applied to calculate $\vec{E}_\mathrm{tot}(\vec{x},t)$ then leads to additional radiation terms appearing because of the time-dependence of $N(t)$.

This problem in the early approaches was only realized around 2009, 
when comparisons were made with the modern macroscopic approaches 
described in the next subsection.

\subsection{Modern macroscopic approaches}

Models for the radio emission from extensive air showers describing the 
radiation physics with macroscopic concepts such as electric currents 
and electric charge rather than individual particles are called ``macroscopic 
approaches''. The advantage of these approaches is that they provide 
direct insight into relevant effects contributing to the radio 
emission from extensive air showers. Also, they are mostly analytic 
and can thus predict a signal with very small computational effort.

The MGMR \citep{ScholtenWernerRusydi2008} approach, a modern representation of 
the original Kahn \& Lerche approach \citep{KahnLerche1966}, 
contributed significantly to our understanding of the radio 
emission from extensive air showers. It describes the transverse 
drift currents that arise in the interplay between the acceleration of 
particles in the geomagnetic field and their deceleration due to 
interactions with atmospheric molecules. These drift currents vary with 
time as the air shower evolves. The time-derivative of the transverse currents then leads to 
the dominating radio emission component. Secondary mechanisms such as a 
radiation from a moving dipole, the contributions from the time-varying charge-excess 
\citep{WernerScholten2008} and the role of the positive ions left behind in the 
atmosphere are also investigated and accounted for. A similar approach was presented in ref.\ \citep{KonstantinovMacroscopic2011}.

Macroscopic approaches have the advantages of speed and 
transparency, and were essential in arriving at today's understanding 
of the radio emission from cosmic ray showers. However, they also have 
important drawbacks.

One problem is that they ``sum up mechanisms'' such as radiation from 
transverse currents, a time-varying charge excess, and other 
effects. Unfortunately, these ``mechanisms'' cannot be always clearly 
separated under realistic conditions \citep{JamesFalckeHuege2012}, and thus there is a risk of 
double-counting contributions. Likewise, relevant effects could be forgotten in the 
description.

Another difficulty is that there are parameters that can be tuned, such as the 
drift velocities in the MGMR approach, which directly scale the 
predicted electric field strengths. A related disadvantage is that 
macroscopic approaches have to make simplifying approximations. It is very difficult to reflect the full complexity 
of the particle distributions in a macroscopic description, in 
particular regarding correlations between particle energy, particle 
position and particle momentum direction. Deviations arising from
non-optimal choices for these free parameters and related 
parameterizations can be sizable \citep{LudwigMGMRvsREAS3Icrc2011}.

Approaches such as EVA \citep{DeVriesARENA2012,WernerDeVriesScholten2012}, which includes the treatment of the 
atmospheric refractive index gradient and couples the macroscopic description 
of the radio emission with a Monte Carlo simulation of the air shower 
cascade, try to mitigate these problems. However, the clarity and speed 
of the analytic calculation is lost to some extent in this approach 
without actually reaching the accurateness of a purely microscopic 
approach.

\subsection{Modern microscopic approaches} 
\label{sec:microscopicmodels}

In microscopic approaches, each single electron and positron in an 
extensive air shower is considered separately. Its radio emission is 
calculated and superposed to arrive at the total radio emission from 
an extensive air shower. Coherence effects are automatically taken 
into account by proper incorporation of the time-delays (phase shifts)
acquired by the emission from individual particles. This means that a combination of a Monte 
Carlo simulation of the electromagnetic cascade in an air shower 
coupled with a formalism for the classical electrodynamics calculation 
of the radio emission from single particles fully determines the 
result.\footnote{In that sense, it is fair to refer to these codes as 
``simulation codes'' rather than ``models'', as the underlying model 
is well-proven classical electrodynamics.} There is no ambiguity in ``mechanisms'' that need to be summed 
up, and there are no free parameters which influence the result. The 
radio signal is basically predicted from first principles, 
uncertainties only arise from the treatment of the air shower cascade 
itself (e.g., due to hadronic interactions). Four independent simulation codes have been developed for the 
microscopic calculation of radio emission from extensive air showers. 

The code of Konstantinov et al.\ \citep{KonstantinovArena2005,Konstantinov2009} was
based on EGSnrc \citep{EGSnrc} and can be considered the earliest that provided a 
self-consistent microscopic simulation of the radio emission from 
extensive air showers. It did, however, not take into account refractive index 
effects and mostly applied to air showers with energies below 
$10^{15}$~eV, i.e., below the typical detection threshold for air shower radio 
signals. The code is no longer actively maintained today.

To provide a self-consistent calculation of the radio emission in the 
REAS line of codes, the ``endpoint formalism'' 
was developed and implemented in REAS3 \citep{LudwigHuege2010}. This 
formalism calculates the radio emission from moving particles
arising from instantaneous acceleration of charges at the 
beginnings and ends of straight track segments 
\citep{JamesFalckeHuege2012}. The follow-up version REAS3.1 included 
the effects of the refractive index gradient in the air. 
Finally, the endpoint formalism was implemented directly in CORSIKA, 
leading to the CoREAS code \citep{HuegeARENA2012a}.

The long-standing ZHS formalism \citep{ZasHalzenStanev1992} for the 
frequency-domain calculation of radio emission from moving charges was 
adapted for application in the time domain \citep{AlvarezMunizRomeroWolfZas2010} and built into the AIRES code to 
arrive at ZHAireS \citep{AlvarezMunizCarvalhoZas2012}. In contrast to 
the endpoint formalism, ZHS describes the radiation as arising from 
the straight track segments themselves (not the acceleration at the 
ends of the tracks). It has been shown that the two approaches are mathematically
equivalent \citep{Konstantinov2009,BelovARENA2012}.
However, there are various advantages and disadvantages between the 
two approaches in the practical implementation. For example, the ZHS 
algorithm builds on the Fraunhofer approximation, i.e., tracks have to 
be sub-divided such that they are small with respect to both the 
wavelength of interest and the distance from radiating particle to observer. 
Such a sub-division is not necessary for the endpoint formalism, which 
leads to a potential performance advantage. On the other hand, the endpoint 
formalism becomes numerically unstable when calculating the emission 
of a particle for an observer near the Cherenkov angle, and a fall-back to a ZHS-style
calculation becomes necessary. A detailed comparison of the computing performance
of the two approaches has not yet been conducted. The agreement on the 
predicted signals is commented on in the next subsection.

Finally, the SELFAS2 code \citep{MarinRevenu2012} was developed using 
an independent formalism for the calculation of electromagnetic 
radiation from track segments. Unlike the other three codes, SELFAS2 
is not based on a Monte Carlo simulation of the underlying air shower. 
Instead, the particle distributions are regenerated from 
parameterizations that were originally derived from histograms made 
for REAS2 \citep{LafebreEngelFalcke2009}, which does neglect some 
potentially important correlations.

\subsection{Agreement between different approaches}

\begin{figure*}[h!t]
  \includegraphics[width=\textwidth]{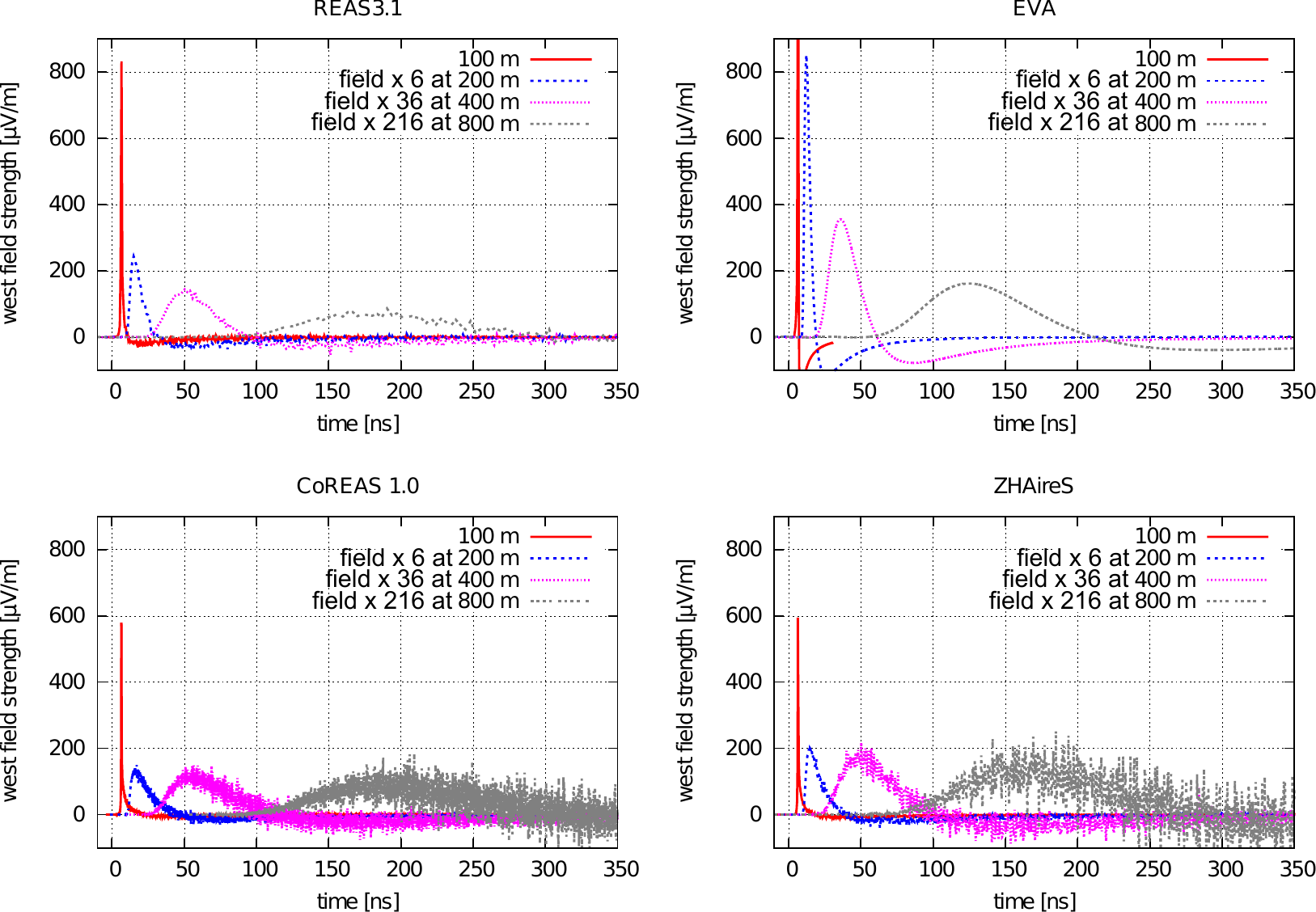}
  \caption{Comparison of the west-component of the electric field 
  predicted for a vertical $10^{17}$~eV proton-induced air shower at 
  the site of the Pierre Auger Observatory by the indicated models.
  The refractive index of the atmosphere was modelled according to 
  atmospheric density. In the Monte Carlo models, particle noise is 
  visible. It results from the finite number of particles in the air 
  shower and is amplified by thinning of the particle cascade to 
  keep computing times reasonable. Adapted from \citep{HuegeARENA2012b}.
  \label{fig:comppulsesnr}}
\end{figure*}

From the point of view of accurateness of the radio emission 
calculation, CoREAS and ZHAireS can be considered the current state of 
the art, and the most directly comparable. Both couple a formalism for the calculation of the radio 
emission directly with a full Monte Carlo simulation of the air 
shower, without any simplifying approximations made in the process. Comparisons of the predictions between
the two models show that they agree within $\approx 20$\% with each other, both 
quantitatively and qualitatively (pulse shape, frequency spectra, shape of the LDF, ...), as is 
illustrated in Fig.\ \ref{fig:comppulsesnr}.

Cross-checks between the ZHS and endpoints formalisms have been made in the context 
of the SLAC T-510 experiment \citep{SLACT510-PRL}, indicating that the formalisms 
produce results agreeing within $\approx 5$\% of each other \citep{AnneIcrc2015}. It is thus likely that 
deviations between CoREAS and ZHAireS are related mostly to the 
underlying air shower simulation between AIRES and CORSIKA, possibly 
related to hadronic interaction models and/or the choice of energy cuts for the simulation of the 
particle cascade. Possibly, a different model for the atmospheric refractive index could also 
explain some of the deviation \citep{ANITAEnergy}. These 
differences need to be studied in further detail.

Semi-analytic models such as EVA as well as microscopic simulations based on histogrammed/parameterized particle 
distributions such as REAS3.11 and SELFA2S 
show qualitative agreement with CoREAS and ZHAireS, but there are 
significant deviations, as is again visible in Fig.\ 
\ref{fig:comppulsesnr}. This indicates that simplifications with 
respect to the full Monte Carlo simulation of the particle cascade 
deteriorate the quality of the prediction.

\subsection{Additional aspects and comments} 
\label{sec:additionalemissionaspects}

At the time of writing, modern microscopic simulation approaches, in 
particular CoREAS and ZHAireS, can explain all experimentally observed 
features of the radio emission (for details see section \ref{sec:expresults}).
There thus does not seem to be any pressing 
need to improve the codes at this time. A few ideas have been voiced 
for effects that could be relevant and could be investigated in the 
future. These include the effect of scattering of the radio emission 
by the plasma of the air shower disk, which propagates in front of the 
radio emission. Another factor that could be investigated is the 
influence of water vapor on the propagation of the radio signals. 
For detectors at high altitudes, the reflection 
of radio signals off a reflective surface have been treated within 
ZHAireS \citep{ZHAireSReflection}. In the same analysis it was ruled 
out that ray bending plays a significant role.

A practical limitation arising in full microscopic Monte Carlo 
simulations is that of particle thinning. To keep computation time for 
high-energy air showers reasonable, several low-energy particles are 
approximated by single particles with a higher weight. This introduces 
artificial coherence and thus leads to artifacts in the resulting 
predicted radio signals. As long as these artifacts are at a level below the 
typical level of Galactic noise, they do not pose a problem.
When high-energy particles are simulated, however, these 
artifacts can reach amplitudes above the noise 
floor. The only currently available way to deal with this is to thin the simulation less, at the cost 
of computation time. Similarly, it is necessary to thin less if 
high-frequency emission is to be predicted. Parallelized simulations, 
especially on GPUs, could help mitigate problems of computing
resources for high-quality simulations in the future. Unthinned 
simulations have so far only been presented for individual air showers 
\citep{HuegeIcrc2013CoREAS}, showing some interesting features such as 
small-scale ripples on the radio emission footprints.

\begin{figure*}[h!t]
  \includegraphics[width=0.48\textwidth]{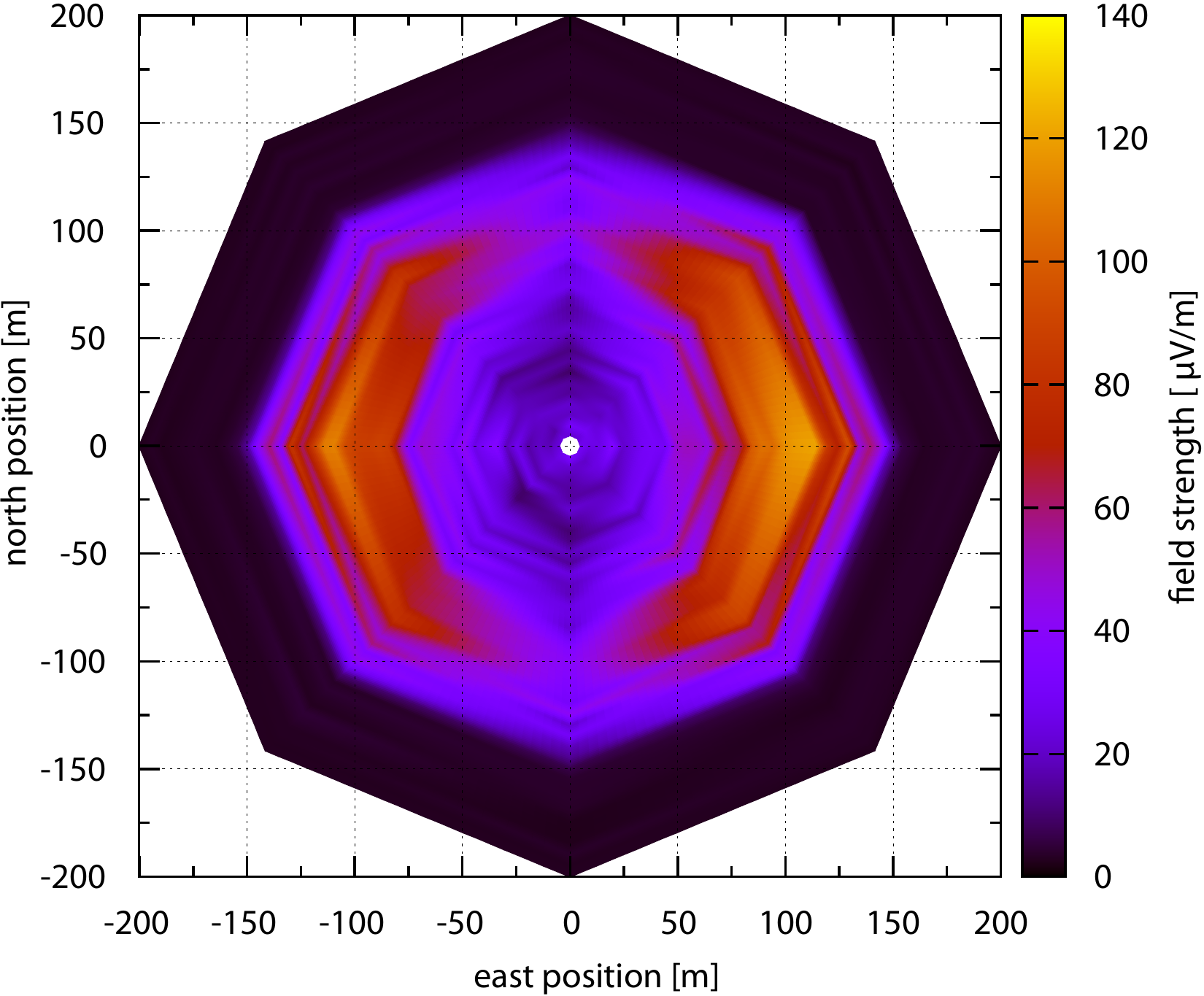} \hspace{0.5 cm}
  \includegraphics[width=0.48\textwidth]{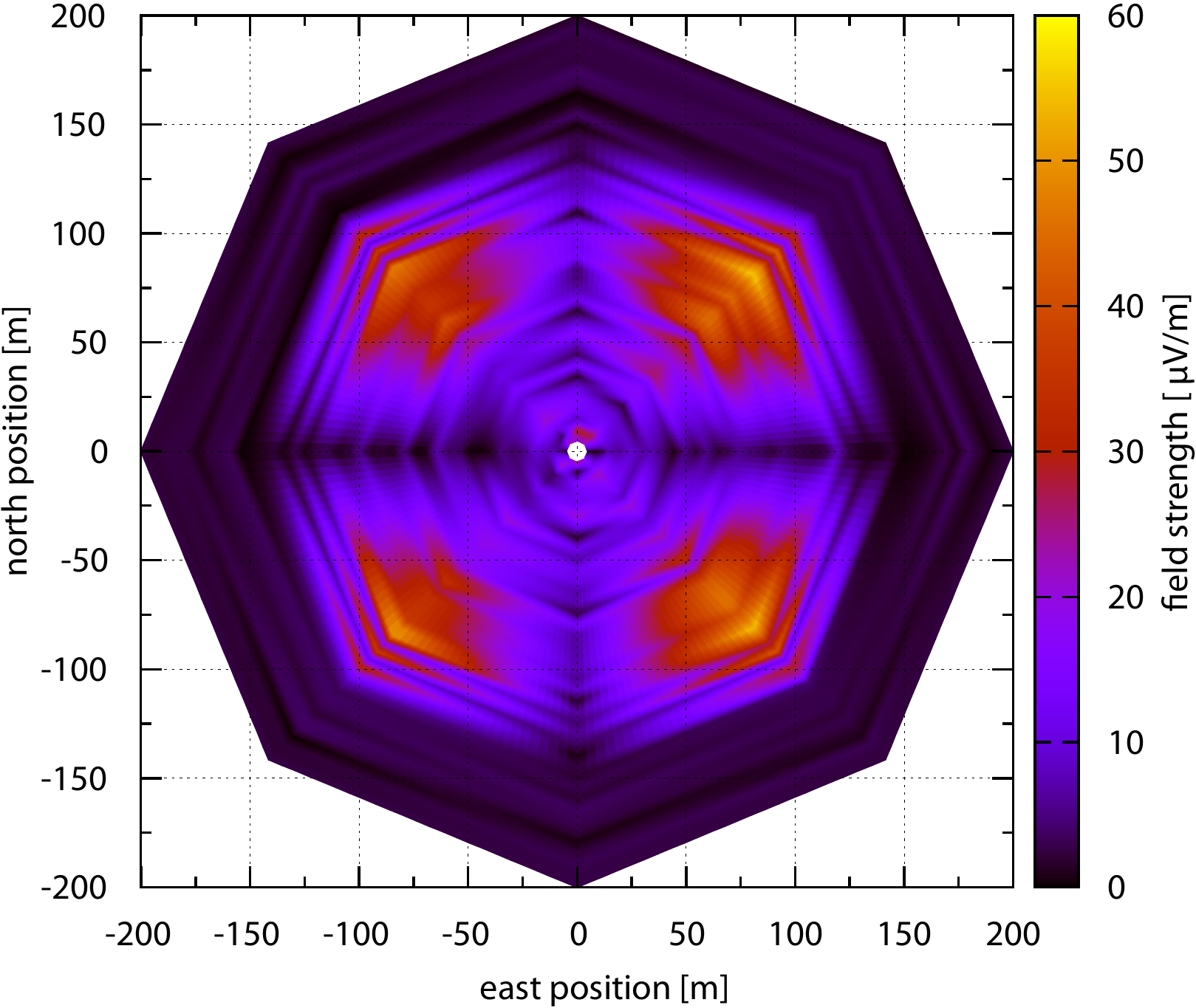}
  \caption{Radio-emission footprints of the total field strength (left) and 
  north-south component of the electric field (right) for a 
  vertical $10^{17}$~eV air shower at the LOPES 
  site in the frequency range from 3.4 to 4.2~GHz. Adapted from 
  \citep{HuegeARENA2012a}.\label{fig:cromefreqs}} 
\end{figure*}

Although the codes discussed here have been developed for the 
application at MHz frequencies, they also predict the radio signal at 
GHz frequencies. In fact, the predictions show that the
characteristics of the radio emission at frequencies beyond $\sim 2$~GHz 
change: For a vertical air shower, CoREAS simulations predict 
the absence of radio signals along the north-south axis from the 
shower core (the Cherenkov ring is ``broken''), and the north-south polarisation component of the signal 
shows a ``clover-leaf'' pattern as is indeed expected for synchrotron 
radiation, see Fig.\ \ref{fig:cromefreqs}. One possible explanation is that the originally proposed 
``geosynchrotron radiation'' could indeed be relevant at high 
frequencies \citep{HuegeARENA2012a}.

Predictions of the current codes at low frequencies (well below a MHz) 
should be treated with some caution. At these frequencies the positive 
ions left behind in the air shower can play an important role, and 
they are neglected in most of the codes.


\section{Experiments for radio detection of cosmic rays}

\begin{figure*}
\centering
\includegraphics[width=1.0\textwidth]{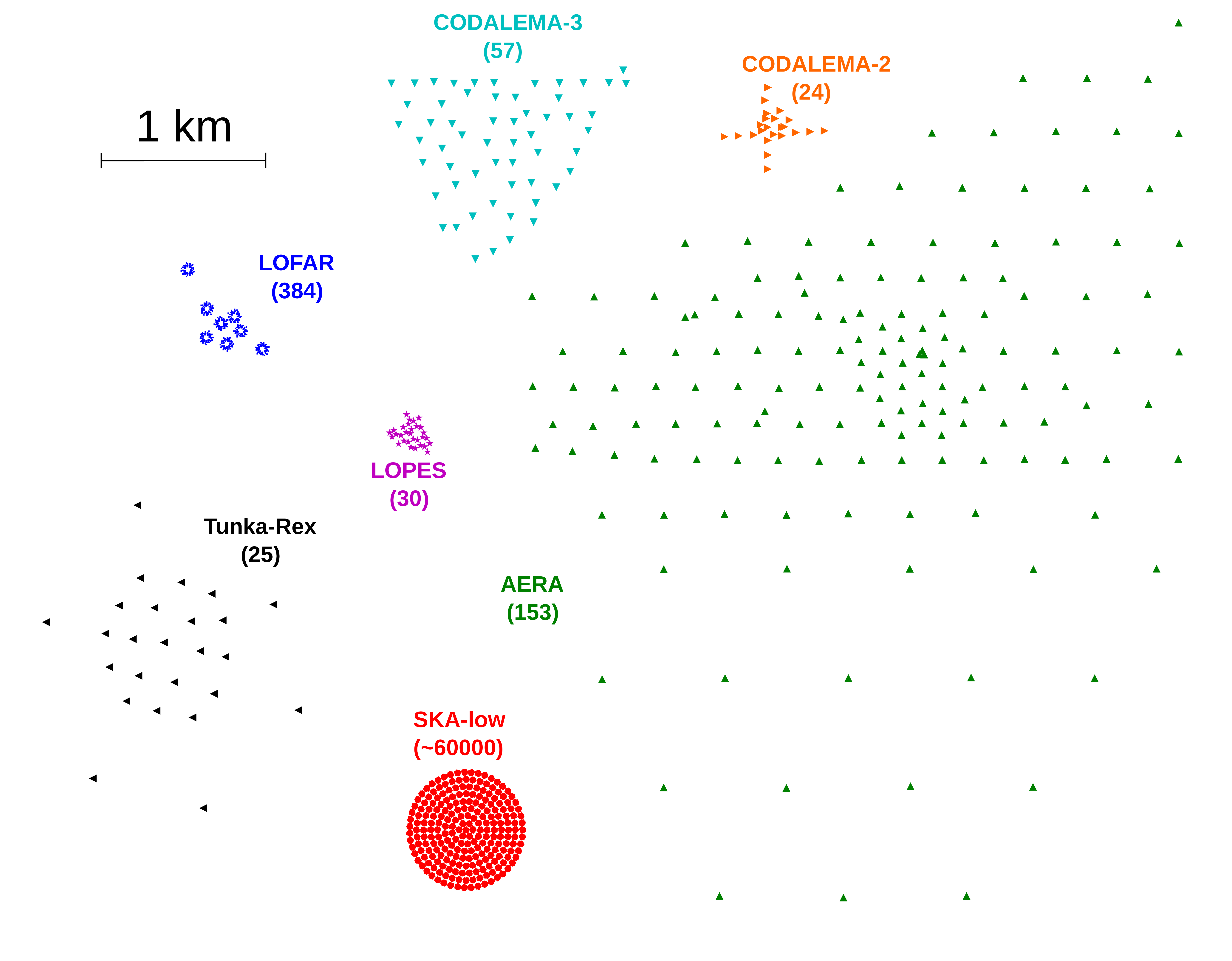}
\caption{Overview of digital radio detection experiments plotted on the 
same scale. Each symbol represents one radio detector (typically with 
dual-polarised antenna), except for the SKA where individual detectors 
are not discernible due to their very high density. The number in 
brackets denotes the total number of antennas.\label{fig:experiments}}
\end{figure*}

In this section, we give a concise overview of the experiments that have been performed
in the past decade, compared to scale in Fig.\ \ref{fig:experiments}. It is our goal to
shortly discuss the various approaches and highlight the
differences, but not go into any technical details. Results gathered 
by the various experiments will be discussed in section 
\ref{sec:expresults} to allow a better discussion of the physics, detached from the 
specific experiments.

\subsection{First generation digital experiments}

\begin{figure*}[!htb]
  \vspace{2mm}
  \centering
  \includegraphics[width=0.45\textwidth]{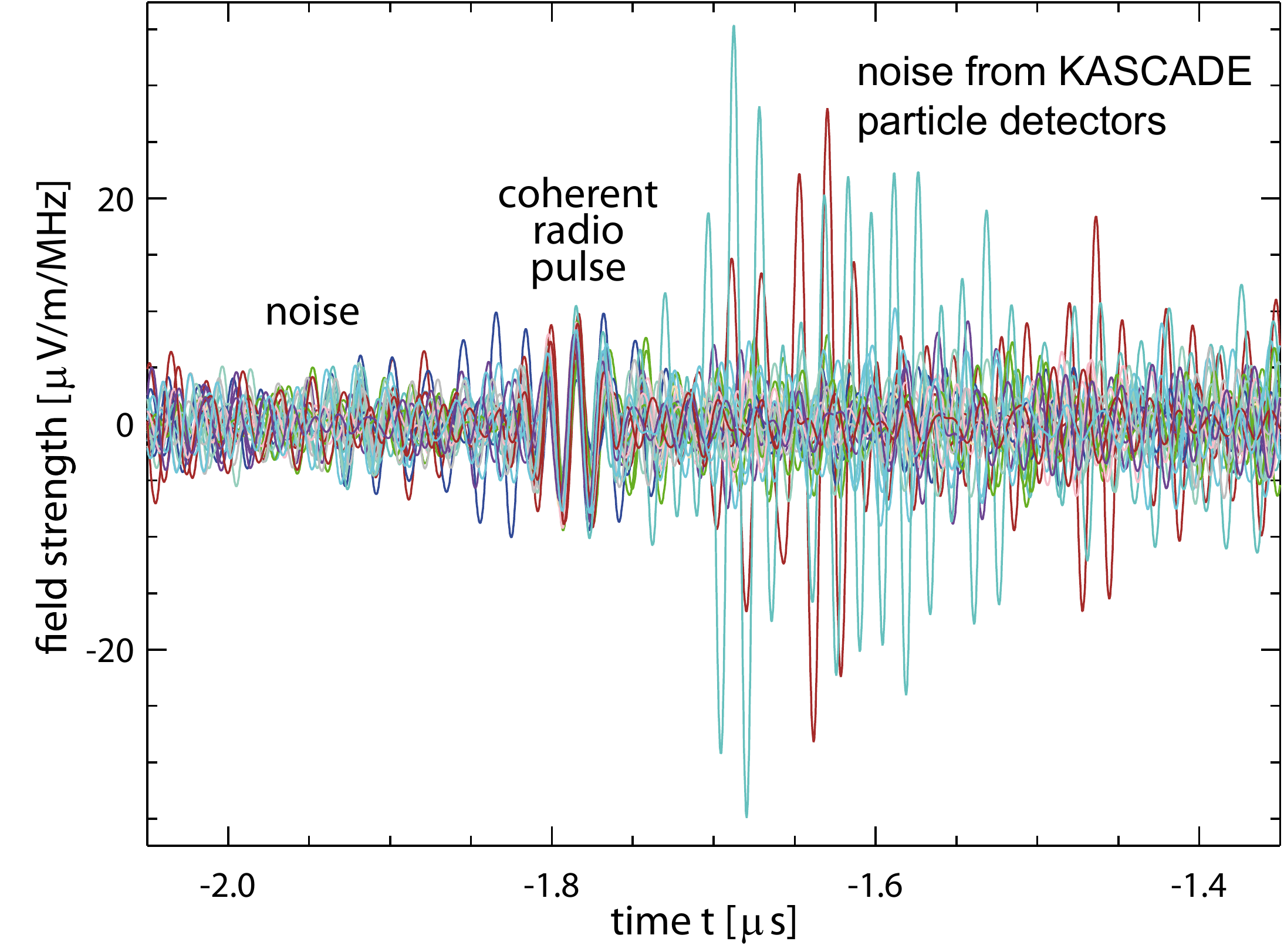}
  \includegraphics[width=0.45\textwidth]{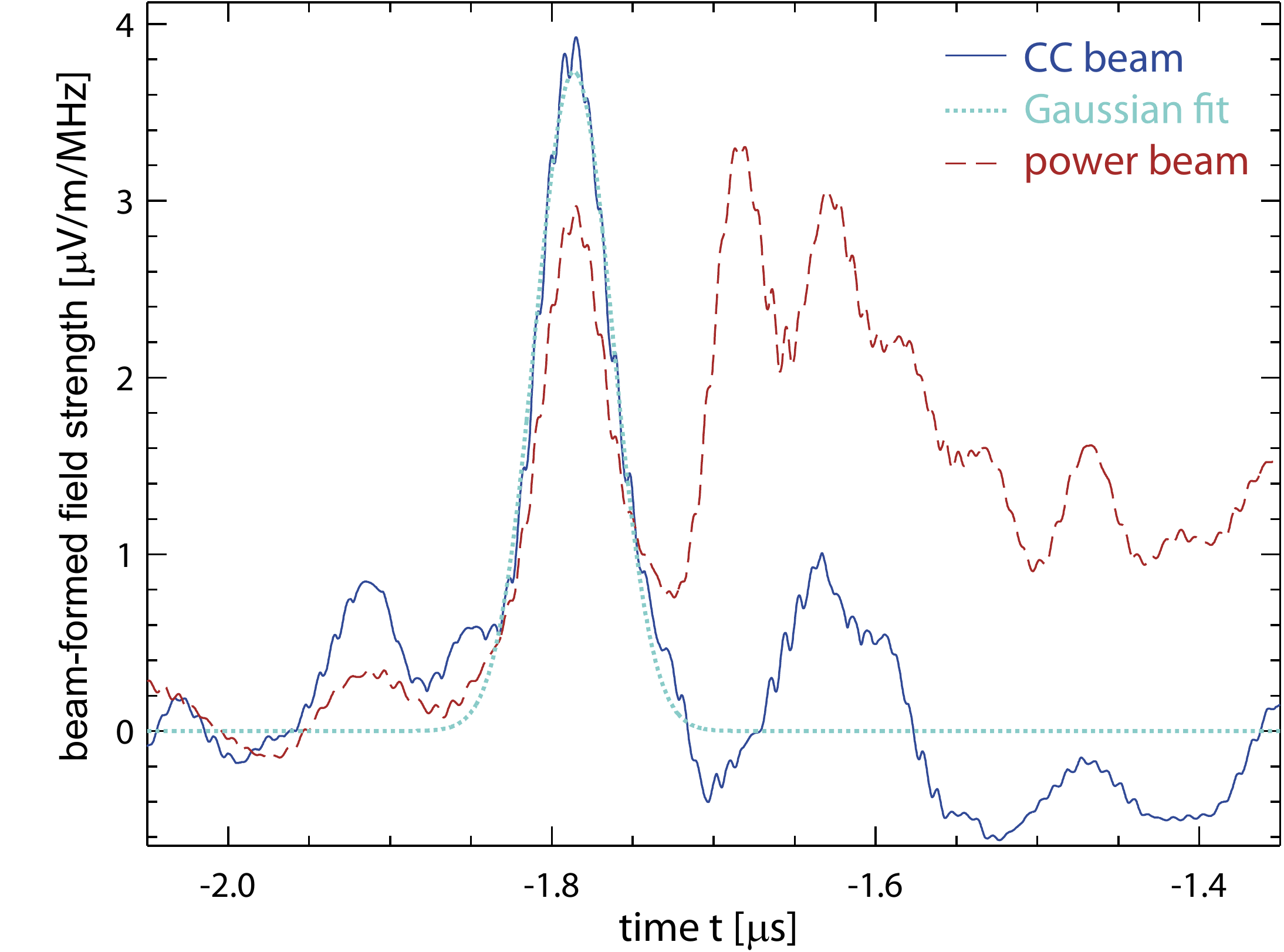}
  \caption{Left: Radio signals measured in various LOPES antennas 
  during the arrival of an extensive air shower. The most prominent 
  pulses originate from the high-voltage feeds of the KASCADE particle 
  detectors. The radio pulse from the extensive air shower is smaller 
  in comparison. It is only discernible from the noise because the signal 
  is coherent in all antennas. Right: Cross-correlation beam of the 
  signal in the LOPES radio antennas. The radio pulse from the 
  extensive air shower correlates strongly between antennas and can 
  thus be clearly identified in the presence of much stronger 
  incoherent pulses from the particle detectors. Adapted from 
  \citep{SchroederLOPESCoREAS2013}.}
  \label{fig:interferometry}
 \end{figure*}

The first generation of experiments for digital detection of radio emission from cosmic
ray air showers comprised the CODALEMA \citep{ArdouinBelletoileCharrier2005} and LOPES \citep{FalckeNature2005} experiments. Both of them began
in 2003, approximately at the same time. In a sense, the approaches were complementary to
each other. In the case of CODALEMA, an existing astronomical radio array (the decametric
array in Nan\c{c}ay, France) was equipped with a special readout system and a small array
of particle detectors for triggering purposes. Later, dedicated antennas of various types were set up. The advantage
of the CODALEMA approach was that the site was very well-suited for radio observations,
as it was located in a sparsely populated area underlying special regulations for radio
frequency interference. On the downside, the particle detector array was fairly simple
and not very optimized or well-studied. In contrast, the LOPES experiment followed the strategy
of complementing an existing precision particle detector experiment, KASCADE-Grande \citep{ApelArteagaBadea2010}, with
radio antennas and readout electronics that were prototypes developed 
for the Low Frequency Array (LOFAR). The disadvantage of this approach was that the site was almost the worst
imaginable for radio detection activities, as the environment at former Forschungszentrum
Karlsruhe can best be qualified as an industrial one including the regular operation of
heavy machinery, welding, an on-site particle accelerator and a lot of high-frequency-emitting
equipment such as computers, etc. In the end, this meant that LOPES data analysis had to
exploit sophisticated interferometric analysis techniques (see Fig.\ 
\ref{fig:interferometry}), without which the radio signals
from cosmic ray showers would never have been identified among the strong noise. This required
supreme timing resolution of $\approx$~1 ns, which was achieved in particular with the development
of a ``beacon timing calibration'' approach 
\citep{SchroederAschBaehren2010}. In this scheme, a transmitter emits 
sine-waves with defined frequencies within the measurement band of the 
experiment. The relative phasing of these sine waves can then be used 
to correct for clock-drifts from event to event. A more advanced 
version of this approach is also used today within the Auger
Engineering Radio Array (AERA) \citep{SchulzIcrc2015}. A 
cross-check of the AERA beacon timing correction using pulses emitted 
by commercial air planes at known positions has recently confirmed that 
the beacon technique, applied in a distributed detector on the scale 
several km$^{2}$, indeed yields a timing precision of 2~ns or 
better \citep{AERAAirplane}. 

\begin{figure}[!htb]
\centering
\includegraphics[width=0.48\textwidth,clip=true,trim=1.8cm 2cm 11cm 9cm]{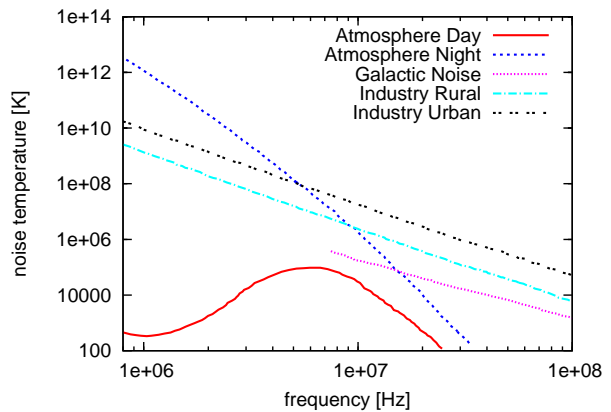}
\caption{Minimum world-wide background noise temperatures according to CCIR report 670 \citep{CCIR670}.\label{fig:noise}}
\end{figure}
 
Both CODALEMA and LOPES focused on the frequency range between low-frequency atmospheric noise dominating
over Galactic noise at frequencies below $\approx 30$~MHz, cf.\ Fig.\ 
\ref{fig:noise}, and the FM band above $\approx 85$~MHz.
Due to the second-Nyquist sampling scheme of 
LOPES\footnote{Conventionally, an analog signal has to be sampled at 
least with double the frequency of the highest-frequency 
component in the signal to avoid aliasing effects in digitization. A measurement in the 
second Nyquist zone can be performed with the same frequency as that 
of the highest-frequency signal component when ensuring that 
no signal components below the sampling frequency are present.}, its frequency band was strictly limited to
40-80~MHz, whereas CODALEMA recorded also lower and higher frequencies. Another important technical difference was that LOPES used deep ring buffers in which the raw waveform
data was continuously sampled digitally, while CODALEMA used an ``analogue-memory'' type of approach
ever only buffering short snapshots of data.

CODALEMA underwent several development stages, from 11 circularly 
polarised antennas of the decametric array in Nan\c{c}ay (CODALEMA-1, \citep{ArdouinBelletoileCharrier2006}) to 
24 cabled linearly polarised fat dipole antennas (CODALEMA-2, 
\citep{CodalemaGeoMag}) with single polarisation only
(either north-south or east-west) to 57 autonomous stations using 
dual-polarised bow-tie antennas (``butterfly'', later also 
deployed in AERA) which are self-triggered on radio signals 
(CODALEMA-3) \citep{CODALEMA3}. The CODALEMA-3 setup covers an area of 
roughly 1~km$^2$ and measures over a broad frequency band from 20 to 
200~MHz. In its context, new concepts continue to be explored 
\citep{DallierIcrc2015}. This includes the search for low frequency emission in the 
EXTASIS setup (see section \ref{sec:lowfrequency}) as well as a 
compact array performing real-time interferometric analysis to improve 
the efficiency and purity of self-triggering on radio signals from 
extensive air showers.

The LOPES experiment also evolved strongly over the course of the decade that it was taking data.
First, there were 10 linearly polarised inverted V dipole antennas 
measuring in the 40-80~MHz band with east-west polarisation
only (LOPES-10, \citep{FalckeNature2005}), followed by an enlarged array with 30 antennas 
(LOPES-30, \citep{ApelArteagaAsch2010}). Afterwards, a setup mixing antennas with either north-south, either east-west and some dual-polarised antennas
(LOPES-30pol, \citep{IsarIcrc2007}) followed. Finally, LOPES tested the concept of 3D measurements using tripole
antennas (LOPES-3D, \citep{ApelArteagaBaehren2012a}). However, this was severely limited by the ambient noise, especially in its
latest stage when a refinery was going online only dozens of meters away from the array. Another
activity was the LOPES-STAR \citep{LOPESSTAR} setup for the tests of self-triggered 
radio detection.

Although both CODALEMA and LOPES had prototype character and were not intended at precision
measurements, they yielded very important results, which we will highlight in section
\ref{sec:expresults}.

\subsection{Second generation digital experiments}

Based on the experience gathered with CODALEMA and LOPES, a second generation of
experiments was designed and deployed. These consisted in particular of 
the Auger Engineering Radio Array (AERA) \citep{SchulzIcrc2015},
the cosmic ray detection capabilities of the Low Frequency
Array (LOFAR) \citep{LOFARCosmicRays} and the Tunka Radio Extension 
Tunka-Rex \citep{TunkaRexInstrument}.

Activities to set up the Auger Engineering Radio Array started in 2007, first with
small-scale prototype experiments \citep{RevenuRAuger2012}, and then later with various deployment phases of
the actual array. The science goals of AERA are to do the necessary engineering for a
larger-scale application of the radio detection technique, then determine the capabilities
and limitations of the detection method at energies beyond 10$^{18}$~eV and finally
exploit radio detection to contribute to actual cosmic ray research in the region of
transition from Galactic to extragalactic sources. The technological challenges that
had to be overcome were in particular the design of a rugged, autonomous, wirelessly
communicating radio station that can be deployed on areas as large as 20 km$^{2}$ in the
harsh environment of the Argentinian pampa. At the same time, one of the goals was to
investigate the possibility of self-triggering on the radio detectors. In a first phase
in 2011, 24 radio detection stations using dual-polarisation logarithmic periodic
dipole antennas based on a design initially developed for LOPES-star were deployed on a
triangular grid of 144 m. The measurement frequency band was 30 to 80 MHz. Two types of
readout electronics were used, one of which providing the possibility to buffer data for
up to 7 seconds. This provides enough time to wait for a trigger from the Auger particle
detector array in addition to triggering on the radio signals themselves. In a second
stage in 2013, an additional 100 radio detection stations were deployed on grids of
250 and 375~m distance. The antennas used here were of the ``butterfly'' type, originally
developed in the context of CODALEMA. Again, two different kinds of electronics were used,
one with a deep buffer and one incorporating small scintillators to provide a local trigger
used in conjunction with the radio self-trigger. In a third phase in spring 2015, AERA
was extended by an additional 25 radio detection stations with deep buffering. Those
antennas are spaced on a grid of up to 750 m, so that in total an area of roughly
17~km$^{2}$ is covered. One of the main advantages of AERA is its co-location
with the very sophisticated particle detection and fluorescence 
detection instruments of the Pierre Auger Observatory. The
latter in particular will allow a direct cross-check of the sensitivity of radio detection
to mass-sensitive parameters such as the depth of shower maximum, which is directly
accessible with fluorescence detectors.

While AERA comprises a ``sparse array'' covering a large area with a homogeneous array
of radio antennas, LOFAR can be characterised as a ``dense array''. LOFAR is a general-purpose
radio astronomy instrument for which cosmic ray detection is only one mode of observation.
To facilitate cosmic ray detection, transient buffer boards have been installed which act
as a ring buffer for the continuously sampled radio signals of individual LOFAR antenna
elements. Upon a trigger from a dedicated particle detector array, LORA \citep{LORANIM}, the buffers are
frozen and the buffered data are read out for analysis. The scheme is similar to the one
originally used at LOPES, yet on a much larger scale. In the dense 
core, roughly 300 antennas sensitive to the frequency band from 10 to 
90~MHz are distributed over an area of $\sim 0.1$~km$^{2}$. Further 
antennas are located outside the core. Independent sets of 
high-band antennas sensitive to the frequency range of 110 to 240~MHz 
are co-located with the low-band antennas, but these are only usable when dedicated 
high-frequency observations have been scheduled for astronomical 
targets, thus statistics are lower than in the low-band mode. The 
antenna spacing in LOFAR has been optimized for the needs of 
interferometric radio-astronomical observations with long integration times. As a consequence, the antennas are distributed
in stations that consist of very dense rings of antennas with large distances between the
stations (cf.\ Fig.\ \ref{fig:experiments}). If a cosmic ray shower does fall within a favorable location 
near the core, however, several hundreds of antennas are illuminated by the radio signal, giving an extremely detailed measurement
of individual air shower radio footprints. LOFAR is thus the most powerful tool to date to
test the details of the radio emission physics and compare them with 
model predictions. Also, the wealth of information measured for individual air showers can be used to determine
characteristics of the underlying cosmic ray shower with very high precision, as is described
in section \ref{sec:expresults}.

The Tunka-Rex experiment focuses on the aspect of determining the capabilities of a radio
detection array built with a dedicatedly economic approach. It comprises an extension of
the Tunka-133 optical Cherenkov light detector array with currently 44 radio antennas measuring in the
30-80 MHz band. The grid size is $\sim 200$~m. The antennas are short aperiodic loaded loop
antennas (SALLA antennas) \citep{KroemerSALLA2009,AERAAntennaPaper2012}, which were 
originally developed within LOPES for one of the prototype systems
at the Pierre Auger Observatory. These antennas can be built very 
cheaply (less than 500~USD including analog electronics). Also, 
antennas are cabled and integrated with the pre-existing infrastructure of the Tunka array. As the Tunka Cherenkov detectors provide information on the depth of shower maximum of the measured
air showers, Tunka-Rex can also directly evaluate the mass sensitivity of radio measurements.
In an upgrade campaign in September 2014, 19 antennas were deployed which
are triggered by scintillators rather than optical Cherenkov detectors. This allows duty
cycles of nearly 100\% rather than the $\lesssim$~10\% achieved with triggers from
optical Cherenkov detectors.

\subsection{Air shower measurements at higher frequencies} 
\label{sec:highfreqmeasurements}

Although the main focus of radio detection of cosmic rays has been at 
frequencies below 100~MHz where the coherence of the emission is 
maximized, several experiments have also been performed to search for 
radio emission at higher frequencies, up to several GHz. The main 
motivation for GHz measurements was given by particle accelerator 
experiments in which microwave emission at GHz frequencies had been found after 
shooting a particle beam into air in an anechoic 
chamber \citep{GorhamMBR}. The presumed source of the measured radio signal was so-called 
``molecular bremsstrahlung'' from low-energy electrons in the particle 
cascade. As the low-energy particles are non-relativistic, the emission 
should be isotropic. Thus, the air shower development should be 
observable ``from the side'' in an approach analogous to imaging 
fluorescence telescopes --- however, with the tremendous advantage that receivers 
can be bought cheaply off-the-shelf and the possibility to observe 
with 100\% duty cycle.

Several projects have been started to search for this ``molecular 
bremsstrahlung''. The CROME \citep{CROMEPRL} experiment within the 
framework of the KASCADE-Grande array consisted of various  
antennas covering frequencies up to 12~GHz. Its main focus was on the 
C-band from 3.4 to 4.2 GHz which was measured with 3 x 3 receivers measuring radio emission from 
near the zenith after receiving a trigger from the KASCADE-Grande 
particle detector array. Other experiments searching for microwave 
emission from cosmic ray air showers have been developed and later 
deployed within the framework of the Pierre Auger Observatory, namely 
MIDAS \citep{MIDAS} as well as EASIER and AMBER \citep{GHzAtAuger}.
As will be discussed in more detail in section \ref{sec:highfreqresults},
none of these efforts were able to confirm the original measurement
reported in ref.\ \citep{GorhamMBR}.

Another experiment measuring radio emission at higher frequencies is 
the ANITA balloon-borne radio detector which was flown in so far three 
flights over the antarctic ice. Its original purpose was to search for 
Askaryan radio emission arising from neutrino-induced particle showers 
in the antarctic ice in the 200 to 1200~MHz band. Somewhat 
unexpectedly, however, radio pulses from cosmic ray air showers, 
understood today as arising from time-compression of geomagnetic and 
charge-excess emission, have been detected instead \citep{HooverNamGorham2010}.

\subsection{Laboratory measurements}

To verify the approaches developed for simulations of radio 
emission from extensive air showers, the SLAC T-510 experiment 
\citep{SLACT510-PRL} was conceived and carried out. Its goal was to reproduce a particle shower in the lab, 
including a tunable magnetic field that mimics the geomagnetic field,
and measure the arising radio emission with well-known antennas of the 
ANITA experiment.

The SLAC T-510 experiment builds on the experience gathered with previous accelerator 
experiments that successfully measured Askaryan radiation from 
particle showers in dense media \citep{SaltzbergGorhamWalz2001,AskaryanSLAC2}, which is directly relevant for radio 
detection of showers in ice or the lunar regolith. As the exact 
configuration of the particle shower can be controlled, such a 
laboratory experiment allows a precise cross-check with the simulation 
codes.

The activities for radio detection of air showers in the GHz range were
complemented by the AMY \citep{AMY} and MAYBE \citep{MAYBE} projects, the goal of which was to
verify the ``molecular bremsstrahlung'' measurement with independent
experiments using particle accelerators. With the Telescope Array 
Electron Light Source \citep{ELSMBRIcrc2015} microwave radiation was 
also searched for in artificially generated air showers.

\subsection{Related activities and projects}

A number of related activities have been performed, which we will not 
discuss in detail, but want to mention here shortly. The RASTA 
\citep{RASTAArena2010} project 
intended to complement the IceCube neutrino detector with a cosmic ray 
radio detector, in particular for veto purposes. While the project 
could not be realized, it yielded the result that the rate of 
transient noise is very low in Antarctica. This is a result to be kept 
in mind as it is an important prerequisite for successful 
self-triggered radio detector setups. The Yakutsk array has set up a 
handful of modern radio antennas and measured radio signals from air 
showers \citep{PetrovYakutsk2013}, but the results are only very sparsely documented in the 
literature. The TREND project \citep{Trend2011} was initiated with the long-term aim to detect 
radio signals from air showers induced by tau particles arising from neutrino 
interactions in the Earth or mountain ridges \citep{BrusovaIcrc2007}. 
TREND itself reported the successful self-triggering on radio emission 
from air showers. Based on the experience gathered with TREND, plans 
have recently emerged to build a very large detector for air showers 
initiated by taus from neutrino interactions, called GRAND \citep{GRANDIcrc2015}.
Finally, a number of projects have continued to investigate the oldest proposed technique for the 
detection of air showers with radio waves: radar detection of the ionization trails 
left behind by the passage of air showers, originally proposed 1940 by 
Blackett and Lovell \citep{BlackettLovell1940}. No successful detection 
has been reported to date. Recent results from the TARA experiment 
\citep{TARAIcrc2015} put very stringent upper limits on the radar 
cross-section of extensive air shower ionization trails, far below 
levels exploitable for practical detection. This pessimistic view is also confirmed by recent theoretical 
calculations \citep{StasielakIcrc2015}.


\section{Analysis aspects}\label{sec:analysis}

Here, we discuss some important aspects related to analysis of radio 
detector data. The goal is to explicitly state some pitfalls and 
subtleties that can make interpretation and comparison of results 
difficult and should hence be kept in mind.

\subsection{Considerations on signal and signal-to-noise definitions}

An important aspect to keep in mind is that there is a variety of 
signal definitions, noise definitions and consequently signal-to-noise 
definitions used throughout the field. In fact, often signal and noise 
are defined in different ways, leading to somewhat arbitrary 
``signal-to-noise'' ratios. One thus has to be 
very careful when interpreting and comparing results. It is important 
to realize that radio signals, unlike other measurements such as 
the energy deposited in particle detectors, possess phase 
information and can thus not only sum up but also interfere 
destructively. While this might seem a trivial statement, it has 
important non-trivial consequences.

One important question is how to define ``the radio signal'' in the 
first place. Options include

\begin{itemize}
\item voltages measured at antenna foot-points or some other point in 
the electronics chain
\item electric field vectors (or components thereof) at the location of the antenna 
\item power quantities derived from peak amplitudes of electric fields 
or voltages
\item energy quantities derived from time-integrals of power quantities
\end{itemize}

Voltages measured at antenna foot-points or later in the signal chain have not been deconvolved for 
antenna responses and thus are not comparable between different 
experiments. However, experiments that only measure with one 
antenna polarisation per location (as was the case for a long time for 
LOPES and CODALEMA) have to rely on analysis of voltage traces, as a 
reconstruction of the electric field vector is not 
possible without the measurement of a second polarisation component. 
(Possibly, a detailed model for the signal polarisation could be used 
to replace the missing information of a second polarisation component, 
but this has not been demonstrated in practice.)

Going from voltages to the electric field vector (V/m) in air at the 
location of the antenna requires measurements in at least two 
polarisations. As electromagnetic waves in air are transverse waves, 
the third piece of information needed to reconstruct the 
three-dimensional electric field vector is then provided by the 
arrival direction of the electromagnetic wave, which is accessible 
from the arrival time distribution of the pulses in the array of 
antennas. Ideally, the electric field vector reconstructed from the 
measurements has been deconvolved from all experimental effects 
(antenna characteristics, electronic gain and dispersion) except for 
the limited frequency window measured by the experiment. Therefore, 
electric field vectors are much more suitable than voltages for comparison 
between different experiments. They can also be directly compared with 
emission simulations.

From the squared electric field vector, the Poynting flux in units of power per area can be 
calculated. Power quantities are easier to handle because they can 
only sum up and not interfere destructively. An immediate consequence 
is that one can subtract noise from a measurement that contains signal 
and noise to estimate the pure signal. However, phase information is 
lost when analysing only signal powers.

In comparison with power quantities, the influence of noise on voltages and electric field vectors is much 
more difficult to describe, as depending on the (random) phase of the
noise, signal and noise can add constructively or destructively. The fact that 
signals are usually identified when they exceed some signal-to-noise 
threshold leads to a bias at low signal-to-noise levels: pulses 
amplified by constructive interference of signal and noise are 
selected, while pulses diminished by destructive interference of 
signal and noise are deselected. A careful treatment of the influence 
of noise at low signal-to-noise ratios is thus required to estimate 
the uncertainties on reconstructed signals correctly 
\citep{LOPESNoiseTreatment}.

Often, analyses refer to the maximum amplitude (in voltage or electric 
field strength) of a radio pulse only. Maximum amplitude was in particular the quantity of 
choice for the CODALEMA and LOPES experiments. As the spatial 
extension of these experiments was rather small, the 
width of the pulses measured in different antennas was fairly 
constant, usually dominated by the impulse response (due to bandwidth 
limitation) of the filters used in the experiment. When data from 
larger experiments are analyzed, this assumption is no longer 
valid and in addition to the maximum amplitude, the width 
of the varying pulses should be taken into account. One way to do this 
is to integrate over the time of the pulses (determined by some 
criterion such as FWHM), which in case of the Poynting flux leads to 
quantities of energy deposited per area. Another integration over 
area, which needs an interpolation/extrapolation of the radiation pattern illuminating 
the ground between the sampled locations, can then yield the energy 
deposited in the radio signal on the ground \citep{AERAEnergyPRD}. This quantity 
has the benefit of an intuitive physical interpretation, and it should 
be largely insensitive on the distance between source and antennas, 
provided that the measurement allows the determination of the complete 
radiation pattern.

In summary, readers should pay particular attention when signal, noise and 
signal-to-noise quantities are referred to in the literature, as they can have very different 
underlying definitions.

\subsection{Determination of pulse arrival times} 
\label{sec:timedefinitions}

Special care has to be taken when determining arrival times of radio 
pulses from extensive air showers. Different choices can be made for 
the time of arrival: The time of the maximum amplitude of a pulse. The 
middle of some fit to the radio pulse. The rise time of the 
signal to a certain fraction of its maximum amplitude. These choices 
can all be made, but they are influenced differently by instrumental 
effects (dispersion broadening pulses) and by the intrinsic pulse 
characteristics changing as a function of observer position.

It was in particular realized that wavefront analyses trying to 
determine cosmic ray characteristics from the arrival time 
distribution of radio signals at individual antennas are strongly 
influenced by these choices \citep{LOPESIcrc2015}. It is therefore imperative to clearly 
define the way in which arrival times are determined when publishing 
results.

\subsection{Time domain versus frequency domain data}

Particular confusion can arise in comparisons of data in the 
frequency domain. While time-domain quantities such as the 
instantaneous electric field as a function of time are 
well-determined, there are ambiguities in the definition of 
frequency-domain quantities. One particular example are spectral 
amplitudes with units of $\mu$V/m/MHz. First, an ambiguity arises from the freedom in normalizing 
Fourier transformations where a factor of $1/(2 \pi)$ can be introduced in 
the forward transformation only, the backward transformation only or 
symmetrically in each of the two directions with a factor of $1/\sqrt{2 \pi}$. The 
last option seems to be the most natural definition as it ensures equality 
of the integral over the squared entries in both time-domain and frequency-domain 
data. However, it is far from safe to assume that everybody 
consistently uses this convention. Additionally, there is an ambiguity 
of a factor of two involved because one can describe the frequency 
spectrum with frequencies from $-\infty$ to $\infty$ or just from $0$ 
to $\infty$ as the values for negative frequencies are just the 
complex conjugate of those at positive frequencies for real time 
series data. To make things even worse, in many occasions authors 
use quantities that also have units of spectral amplitudes such as 
$\mu$V/m/MHz, but have not been determined from frequency-domain 
quantities but rather represent a maximum amplitude ($\mu$V/m) 
normalized by the effective bandwidth of an experiment (MHz). The 
resulting quantity is of course related to spectral amplitudes, but 
it implies a flat spectrum and a particular distribution of phases for 
the spectral components (the pulse shape is mostly dominated by the 
phases). The exact relation of bandwidth-normalized amplitudes to spectral
amplitudes thus depends on all of the aspects described above.

In summary, frequency-domain data should be treated with great care. 
If possible, comparison of time-domain quantities is strongly 
preferred.

\subsection{Interferometry versus single pulse analysis} 
\label{sec:interferometry}

Typically, today's analysis approaches are based on the detection of radio pulses in individual radio detector 
stations. Once the pulse is identified, its characteristics can be 
determined. From the relative arrival times at different radio 
detectors, the arrival direction can be deduced. This approach is 
analogous to the analysis of particle detector data. It does, however, 
not exploit the full information content of the radio signal, as it 
does not take into account the phase information.

The full information content can be exploited when an interferometric 
analysis technique is employed. For ground-based arrays, this has so 
far only been applied by the LOPES experiment \citep{HuegeArena2010a},
and as mentioned before (cf.\ Fig.\ \ref{fig:interferometry}), 
interferometry has been a key element in making radio measurements in the noisy environment of the LOPES 
experiment feasible. The ANITA experiment has also 
been using interferometric techniques \citep{RomeroWolf201572} in 
analysing their data, and for the fourth flight a real-time interferometric  
trigger is being developed. In the LOPES interferometric approach, the 
time traces $s(t)$ of the radio signals measured with $N$ different antennas are correlated with each other, 
multiplying the data from each pair of antennas time-bin by time-bin 
and then averaging while keeping the sign of the term under the 
square-root (positive means correlated, negative means anti-correlated):

\begin{equation}
cc(t) = \pm\sqrt{\left|\frac{1}{(N-1)N/2}
                 \sum_{i=1}^{N-1}\sum_{j>i}^{N}
                 s_i(t)s_j(t)\right|
                }
\end{equation}

\begin{figure}[!htb]
\centering
\includegraphics[width=0.45\textwidth]{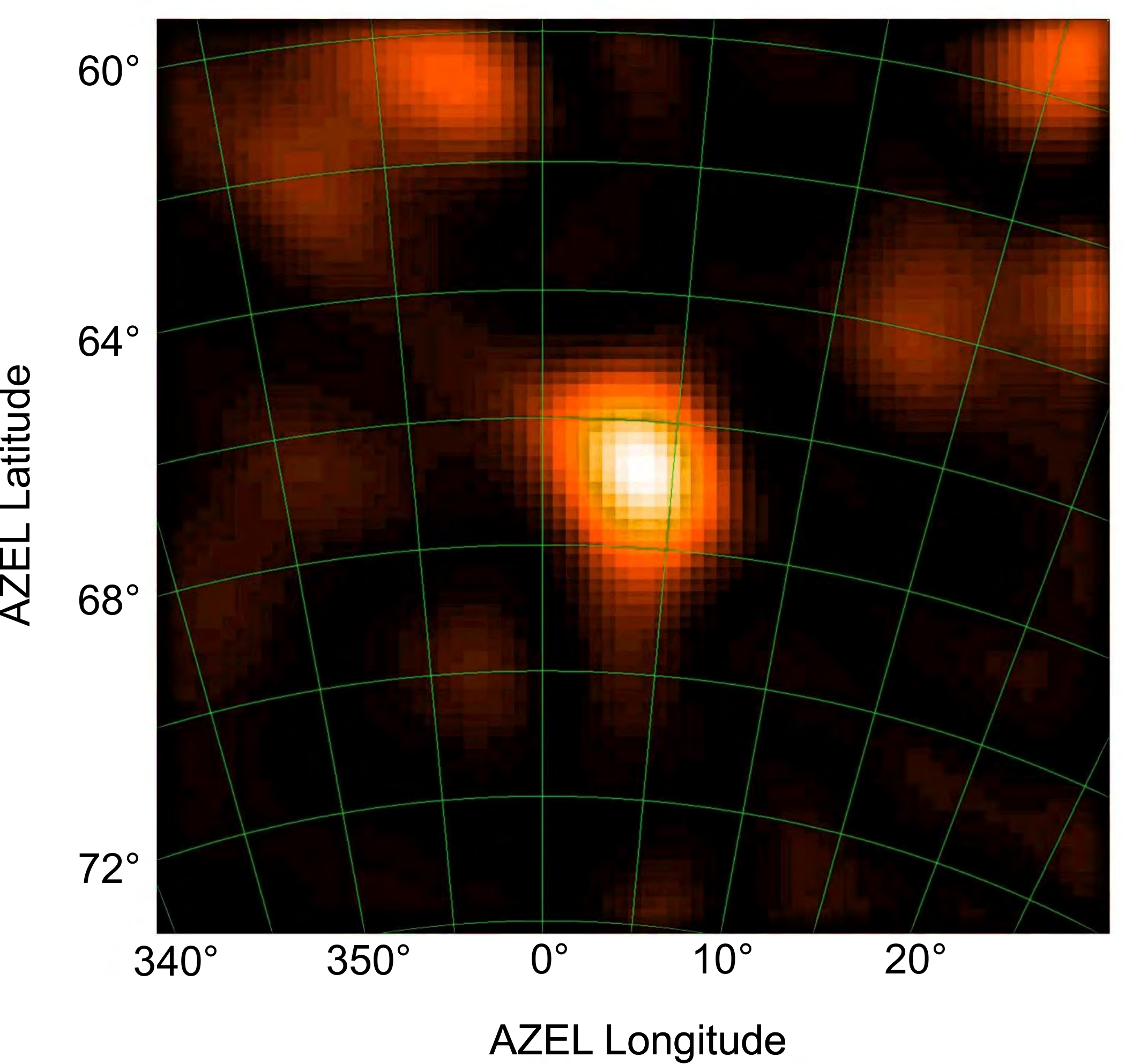}
\caption{Interferometric sky-map of the radio emission from an 
extensive air shower measured with LOPES. The main source is clearly 
identifiable and its direction is consistent with the direction reconstructed by the 
KASCADE particle detectors. Secondary emission regions are an artifact 
of the antenna array layout (``grating lobes''). Adapted from 
\cite{FalckeNature2005}.\label{fig:skymap}}
\end{figure}

This approach allows the calculation of a sky map (cf.\ Fig.\
\ref{fig:skymap}) that allows the 
identification of the radio source. There has been some discussion on 
whether the signal-to-noise ratio of bandwidth-limited pulses ideally achievable
with this analysis approach scales as $\sqrt{N}$ (see, e.g., \citep{RomeroWolf201572}) or as 
$\sqrt{(N-1)N/2}\approx \sqrt{1/2}N$, i.e., linearly with the number 
of antennas for a large number of antennas \citep{HuegeFalcke2003a}.
Another caveat for application to ground-based arrays is that this analysis approach is based on a 
classical far-field assumption, pretending that the radio emission 
observed at all locations is identical (except arriving with a 
different time-delay). Only for the calculation of the delays, a 
wavefront model different than a plane wave (source at infinity) has been 
used \citep{LOPESWavefront2014}. We know, however, that the radio 
pulses at different axis distances vary not only in amplitude but also in width. For LOPES, this effect 
was fairly small due to the limited size of the array. For larger 
arrays, however, it will play a central role, and the far-field 
approach will have limited use (except probably for inclined air 
showers, see section \ref{sec:inclinedshowers}).

Advanced analysis techniques for ``near-field interferometry'' have yet to be developed, but there 
is high potential. Ideally, one could not only make a two-dimensional 
sky map of the signal distribution, but go to a three-dimensional 
tomography of the signal distribution and thus the radiating 
electromagnetic component of the air shower.

\subsection{Polarisation characteristics}

In radio astronomy, signal polarisation typically refers to signals 
received (and usually integrated) over significant time-scales, in 
particular over many oscillation periods of the frequency components contained in 
the signal. The electric field vector of polarised signals performs a defined motion within the plane perpendicular 
to the transverse electromagnetic wave, a line for pure linear 
polarisation, a circle for pure circular polarisation (cf.\ Fig.\ 
\ref{fig:lopespolarisation}). In contrast, unpolarised 
radiation would follow a random path in this plane. But this raises interesting question: Can an 
impulsive radio signal (present only for one oscillation of the 
contained frequency components) be unpolarised? And what is the polarisation of 
the maximum amplitude of a detected radio pulse? In fact, it is not 
possible to define the polarisation of a radio signal at a certain 
point in time (e.g., the maximum), as polarisation is related to the 
\emph{time-evolution} of the electric field, the path that the 
electric field vector follows in the plane perpendicular to the 
propagation direction. Statements about polarisation thus always have 
to be made for the evolution of impulsive signals over a certain 
time-scale. This can be an explicit-time scale defined in an analysis, 
or it can be an implicit time scale introduced for example by the 
bandwidth-limitation of an experimental measurement (which broadens 
the pulses to a minimum width) or an enveloping procedure such as a 
Hilbert-Envelope. If the time scale over which the signal evolution is 
observed is short enough, the signal will always have a high degree of 
polarisation. Therefore, it can be questioned whether a quantification 
of the degree of polarisation as presented in ref.\ \citep{LOFARChargeExcess2014} is actually meaningful 
for impulsive emission. Characterizing the fraction of circular versus 
linear polarisation, however, can extract useful information.

In astronomy, it is usual to express polarisation using Stokes 
parameters (for an excellent review of polarisation and various 
ways to represent it see the review by Radhakrishnan 
\citep{Radhakrishnan}). Adoption of this approach is also possible for 
impulsive radio emission, but has the disadvantage of being somewhat 
obscure.


\section{Results of digital radio detection}\label{sec:expresults}

In this chapter, we review important experimental results that have 
been achieved over the past decade. We will not report these along the 
lines of specific experiments, but rather structure this section with 
respect to the relevance for the understanding of the radio emission 
physics and how it can be used for cosmic ray research. Where 
appropriate, we directly compare the experimental results with the predictions from the emission 
modelling and simulations.

\subsection{Self-triggering vs.\ external triggering, noise background 
and detection threshold}

Encouraged by the successes of the externally triggered 
first-generation digital radio experiments for air shower detection, 
the community strove to develop the radio detection technique towards a 
full-fledged approach which could be used completely independently of 
other detectors, ideally in large-scale radio detection arrays that 
could potentially be built at much lower cost than comparable particle detector arrays. One 
necessary prerequisite for this is the ability to self-trigger 
on cosmic ray air showers from radio signals alone.

In particular in the context of AERA, the goal for self-triggered 
detection was followed with significant efforts. Original site-surveys 
had indicated that the \emph{continuous} background noise in the Pampa 
Amarilla is very low, dominated by Galactic noise. When the first 
radio detector prototype stations had been deployed, it became 
evident, however, that \emph{transient} noise (short bursts of radio 
emission) was prominent. This is not a contradiction, as 
transient RFI contributes little power to the continuous noise,
i.e., a survey with a spectrum analyzer or any other approach 
integrating over time-scales larger than a few hundreds of nanoseconds will 
not reveal its presence. The sources at the site of AERA were investigated and seemed 
to include a mixture of faulty power-lines a few kilometres away, 
transformers in a nearby village and other, unidentifiable origins 
\citep{SchmidtThesis}. 

Transient noise, however, makes it very hard to develop a self-trigger 
on radio pulses which is both efficient and pure. Data rates that can 
be handled are limited by several factors. For detectors communicating 
wirelessly such as those in AERA, bandwidth poses significant 
limitations on the number of individual-detector-level triggers that can be 
communicated to a central data acquisition for coincidence search. 
In the presence of frequent noise pulses, this means that the 
trigger threshold has to be raised to lower the trigger rates. In turn, this 
means that the detection threshold rises to high cosmic-ray energies. 
Even if one can handle the high rates, another problem appears: the 
trigger becomes strongly contaminated by noise pulses, and cosmic ray 
pulses only constitute a vanishing fraction of the data set. This 
could in principle be accepted if one could reliably identify cosmic 
ray signals in the huge data set of noise pulses. The only reliable 
way to do so, however, so far has been to check for coincident 
detection with particle detectors. This approach has been 
demonstrated to work in principle at a prototype detector of AERA 
\citep{RevenuRAuger2012}. However, it is not entirely convincing for two reasons: First,
this is not the independent radio detection that was the 
original goal. Second, the efficiency reached with this approach is 
much worse than can be reached with a direct external trigger by particle 
detectors.

Significant efforts were made to overcome the problems involved in 
self-triggering. Much effort has been made to reduce the trigger rate on the level of individual 
detectors. These included on-the-fly cleaning of the measurement 
spectrum of narrow-band transmitters as well as approaches to 
identify pulses and compare them to the expectation of cosmic ray 
signals \citep{SchmidtThesis}. However, the rate reduction achievable with these approaches 
is not sufficient. Many transient signals of anthropic origin 
constitute bandwidth-limited pulses, just like cosmic ray pulses, and cannot be 
suppressed without additional information. The rise-time of pulses has 
been investigated as an additional means to differentiate cosmic ray pulses 
from RFI, as is currently being investigated with CODALEMA-3 
\citep{MachadoIcrc2013}. It has not yet been shown, however, that this will 
bring the rates of false triggers down far enough. The situation is better when 
information from several detectors is combined in the coincidence 
search. All signals that repeatedly arrive from similar directions 
over the course of a few minutes are clearly of anthropic and can be rejected.
This can lower the false trigger rate significantly. An 
interferometric real-time trigger as currently investigated in the compact 
array of CODALEMA-3 \citep{DallierIcrc2015} could lower the detection threshold 
and suppress anthropic sources from the horizon. Again, however, the 
practical use of this approach still needs to be demonstrated. 
Additional characteristics of air-shower radio signals such as the lateral 
amplitude distribution, the signal polarisation and the signal 
wavefront could be used in self-triggering to identify 
cosmic rays reliably. However, this would require a very sophisticated 
online (real-time) analysis of large amounts of radio data. If 
a (simple) particle detector, which has virtually no false positive 
coincidences, is available to generate the trigger, this seems to be 
the much easier and more promising approach, yielding both a much 
more efficient and purer trigger. Therefore, this author is convinced
that the true power of the radio detection 
technique lies in combining it with other detection methods so that 
systematic uncertainties can be controlled better.

Using an external trigger, radio detection has been demonstrated to be able to 
reliably detect radio pulses as of cosmic ray energies $\gtrsim 10^{17}$~eV, 
where the radio pulses become clearly visible above the 
Galactic noise background in the signals recorded by individual radio 
detectors. This threshold, of course, depends on the strength of the 
local geomagnetic field, the air shower arrival direction (geomagnetic 
angle) and also the altitude of the detector. There is potential to 
lower this threshold significantly using interferometric analysis 
techniques which do not rely on the identification of pulses within 
the data of individual detectors (cf.\ section 
\ref{sec:interferometry}).

Radio self-triggering might be feasible in environments where the rate 
of transient RFI is very low. One site that could be suitable is in 
Antarctica, as indicated by RASTA \citep{RASTAArena2010} measurements. 
The successful self-triggering of cosmic ray events at high frequencies with ANITA 
\citep{HooverNamGorham2010} over Antarctica lends further support to 
this. Also the TREND project in rural China reported successful identification of cosmic ray pulses 
on the basis of radio data \citep{Trend2011}. Nevertheless, combining 
radio detectors with particle detectors should be considered where 
possible, as the combination ensures a robust, efficient and pure 
trigger.

\subsection{Absolute amplitude calibration}

One particularly important, and particularly challenging, 
experimental aspect is an accurate absolute calibration of the radio 
measurements. In the early days of radio detection, comparisons 
between cosmic ray radio measurements made with different experiments 
revealed apparent discrepancies by orders of magnitude in amplitude 
\citep{AtrashkevichVedeneevAllan1978}. The authors speculated at the 
time that these problems were related to the amplitude calibrations of 
the different experiments.

In the modern experiments, tremendous efforts have been made to get 
the calibration under control. Two approaches, and combinations 
thereof, have been followed. First, the single components 
comprising a radio detector can be characterized individually to 
estimate the overall response. This includes antennas with their 
frequency-dependent directivity patterns, cables, filters and 
amplifiers. Dispersion plays a significant role in many of these 
elements and needs to be taken into account properly. Care 
also has to be taken in adequately modelling the impedance matching of 
individual components. While cables, filters and amplifiers can easily 
be characterized individually in the lab using 
vector network analyzers, antenna characteristics are usually modelled 
with simulation codes such as NEC-2 \citep{NEC2}, which are then cross-checked 
with measurements in the field. These cross-checks are difficult for 
antennas in the frequency range of 30-80~MHz, as the antennas are 
large and far-field measurements require significant distances from 
the antennas (the wavelength at 30~MHz corresponds to 10~m). They are, 
however, very important, as previous experience has shown that 
simulations do not always accurately predict the antenna response in 
the field (probably due to environmental affects not properly 
accounted for in the simulations). The second approach in 
characterizing a radio detector consists of end-to-end measurements in the field using 
an external reference source placed appropriately in the field-of-view 
of the antenna, or relying on the universal calibration source 
available to radio detectors, the Galactic noise.

The LOPES experiment has been calibrated early-on using a commercial external 
reference source that emits a frequency-comb with a defined power at 
steps of 1~MHz \citep{NehlsHakenjosArts2007}. In involved calibration 
campaigns this source was placed $\sim 10$~m above LOPES antennas for 
an end-to-end calibration of the analogue chain of the experiment. Recently, a 
re-calibration of the reference source revealed that amplitudes published by LOPES before 2015 
were on average a factor of 2.6 too high \citep{LOPESrecalibration}
--- calibration data for \emph{free-field} conditions had been used while air shower 
measurements actually correspond to \emph{free-space} conditions. The 
absolute scale of the amplitudes in LOPES has a systematic 
uncertainty of 16\%, as specified by the manufacturer 
of the calibration source. The calibration source has since been provided also to the Tunka-Rex and 
LOFAR experiments, which means that these three experiments share the 
same amplitude scale and can thus be directly compared, unaffected by 
the 16\% systematic scale uncertainty. In a recent publication, a 
cross-check of this calibration scale has also been performed with the 
Galactic noise, yielding agreement within the systematic uncertainties 
\citep{LOFARcalibration}.

The most accurate absolute calibration quoted so far is the one of the 
Logarithmic Periodic Dipole Antennas (LPDA) of AERA, which has been 
quantified at 14\% \citep{AERAAntennaPaper2012}. This value has been 
determined with a combination of measurements of the analog signal 
chain, simulations of the antenna and cross-checks of the antenna with 
a reference transmitter on a balloon. The LPDAs have the advantage 
that they are fairly insensitive to ground conditions 
(wet/dry/snowy/...). The butterfly antennas deployed in later phases of AERA 
use the ground as a reflector for improved sensitivity at high zenith 
angles. This comes at the cost of increased sensitivity to ground 
conditions, which yet need to be quantified. The SALLA antennas used 
within Tunka-Rex \citep{TunkaRexInstrument} are mostly insensitive to 
the ground and should thus yield a well-calibrated measurement without 
the need for monitoring environmental conditions closely 
\citep{TunkaRexStationIcrc2013}.

\subsection{Validation of emission models and simulations} 
\label{sec:simcomparison}

To efficiently exploit the information encoded in the measured radio 
signals from cosmic ray air showers, a detailed understanding of the 
underlying radio emission physics is imperative. In section 
\ref{sec:emissionparadigm} we have discussed the current paradigm of 
the emission physics. Here, we present experimental results that 
validate the radio emission models and simulations.

Radio emission from air showers can be interpreted as a superposition of geomagnetic and charge-excess 
radiation. The geomagnetic effect was already known to dominate the emission in 
the historical experiments, and was also immediately confirmed by both 
the LOPES \citep{FalckeNature2005} and CODALEMA 
\citep{ArdouinBelletoileCharrier2006,CodalemaGeoMag} results. The 
charge-excess emission, however, took much longer 
to confirm explicitly.

Already in the 1970s, there were some results pointing to emission in 
addition to the geomagnetic effects. One of them 
\citep{PrescottHoughPidcock1971} demonstrated that the radio emission 
does not vanish completely for air showers propagating parallel to the 
geomagnetic field, for which the geomagnetic emission should vanish. It was presumed that this 
could be related to charge excess (Askaryan) emission. However, no 
additional characteristics of the signal were known. Other results at 
the time seemed to be compatible with pure geomagnetic emission.

\begin{figure}[!htb]
\centering
\includegraphics[width=0.45\textwidth]{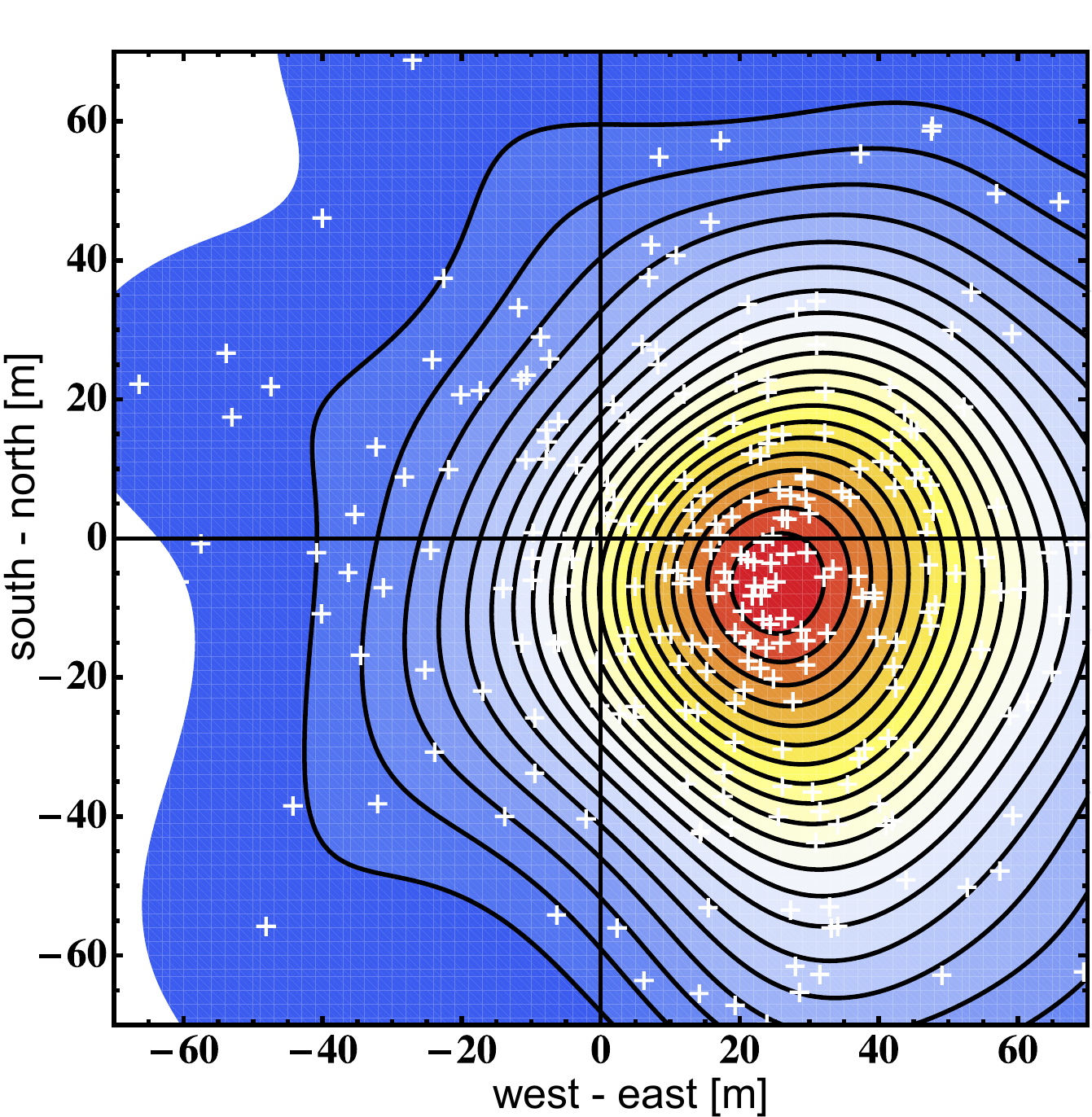}
\caption{CODALEMA measurements revealed a systematic offset 
  to the east in the core positions reconstructed from the lateral 
  distributions of air shower radio signals (white crosses and contours
  derived from the distribution of these) with respect to the cores measured with 
  particle detectors (origin of the diagram). This was an 
  indication of the asymmetry in the radio emission footprint 
  introduced by the charge-excess mechanism. The observed offset to the south 
  is not explained by the asymmetry in the radio signal. Adapted from 
  \citep{CODALEMACoreShift}.\label{fig:CODALEMACoreShift}}
\end{figure}

A first modern result showing the presence of a secondary mechanism in 
addition to the geomagnetic radiation was shown at the ICRC 2011 by 
the CODALEMA experiment \citep{CODALEMACoreShift,Belletoile201550} (see Fig.\ 
\ref{fig:CODALEMACoreShift}). An analysis on 
CODALEMA data showed that the core position determined from a 
one-dimensional lateral distribution fit to the radio data was 
systematically offset to the east from the position determined 
with an analysis of the particle detector data. This is expected for a 
secondary contribution on top of the geomagnetic component in case that this 
secondary contribution is linearly polarised with electric field 
vectors oriented radially with respect to the shower axis, as is 
expected for charge excess emission. A comparison with SELFAS2 
simulations confirmed that the charge excess emission can explain the 
observed core shift.

\begin{figure}[!htb]
  \centering
  \includegraphics[width=0.38\textwidth]{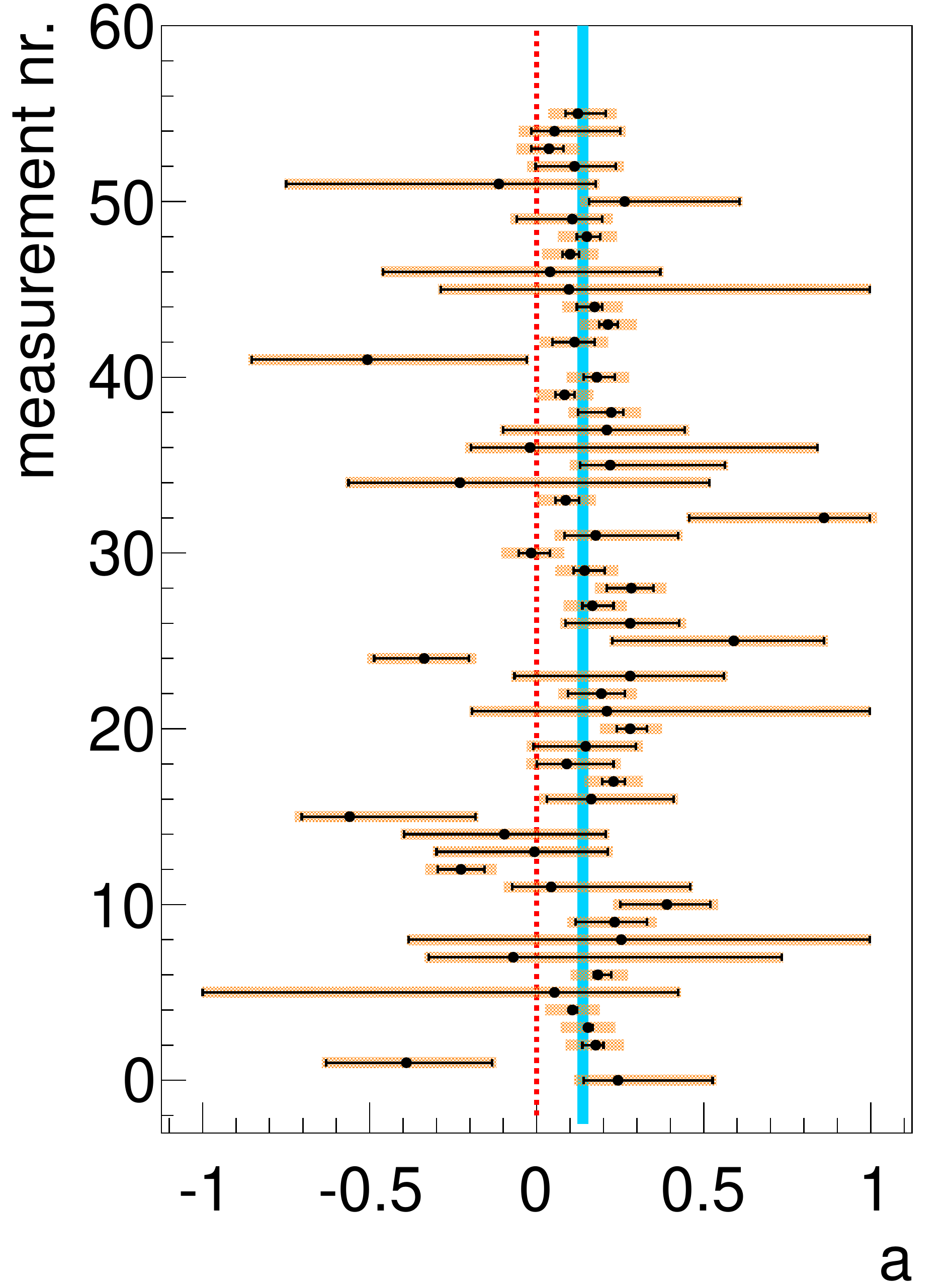}
  \caption{AERA measured a radially polarised emission contribution 
  $a$ with an average strength of 14\% relative to that of geomagnetic emission. Positive values 
  denote the radial polarisation expected for charge-excess emission, 
  negative values denote orientation of the radial 
  electric field component opposite to that expected for charge-excess 
  emission. Adapted from \citep{AERAPolarization2014}.\label{fig:AERAchargeexcess}}
\end{figure}

The first result demonstrating directly the presence of a  
contribution with the polarisation characteristics expected for charge 
excess radiation was given by the Auger Engineering 
Radio Array \citep{AERAPolarization2014} (see Fig.\ 
\ref{fig:AERAchargeexcess}). The analysis showed that the 
orientations of the electric field vectors measured in individual 
antennas depend on the relative locations of the given antennas to the 
shower axis, and that they behave as expected for the superposition of 
geomagnetic and charge excess radiation with their corresponding 
polarisation characteristics. Furthermore, the relative strength of 
this contribution could be quantified, yielding an average value of 14\%. There were already 
indications that the scatter in the relative strength for different 
measurements is larger than expected from statistical fluctuations for 
a constant strength.

\begin{figure}[!htb]
  \includegraphics[width=0.5\textwidth]{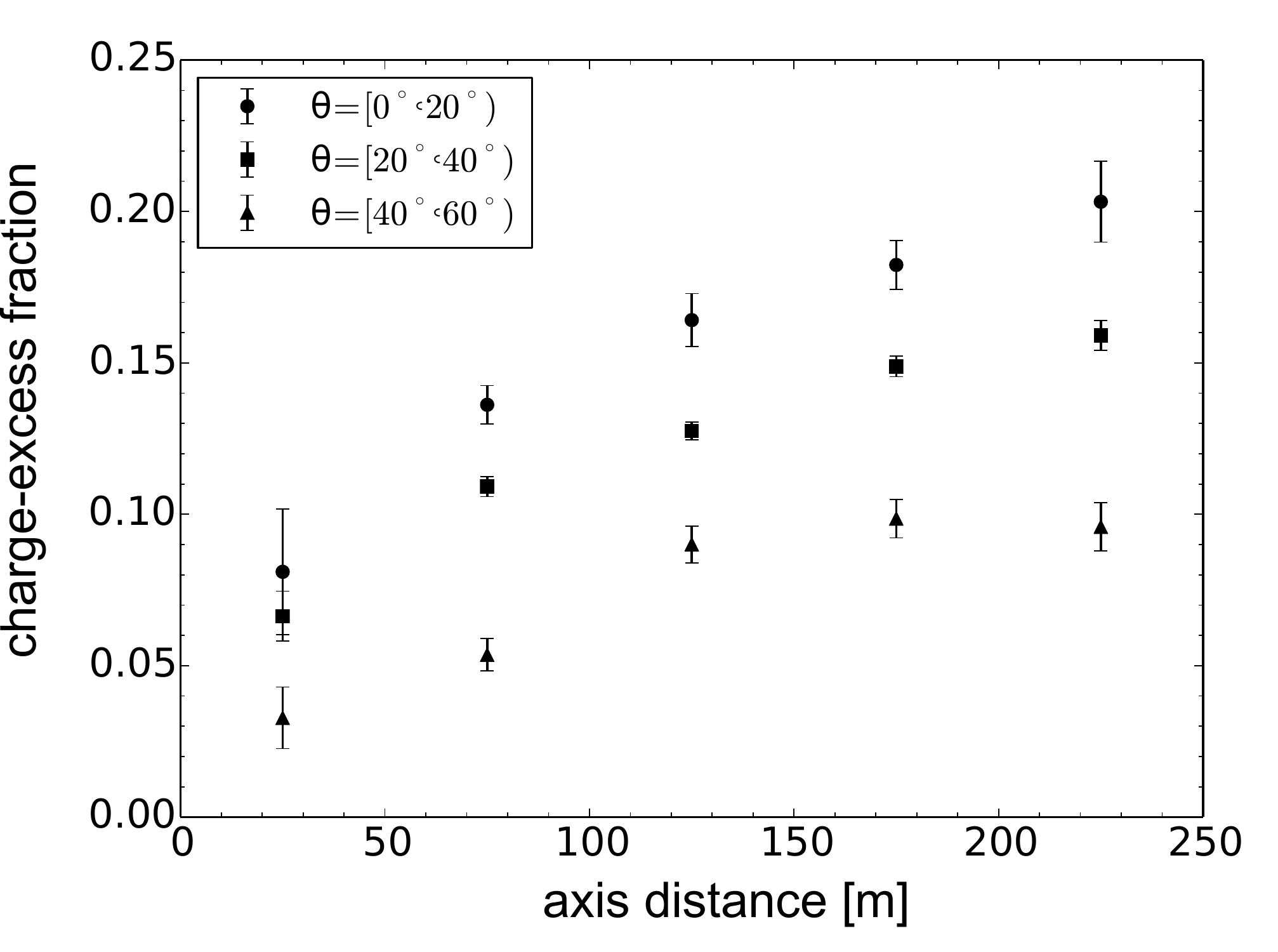}
  \caption{Analyses of LOFAR data revealed that the relative 
  charge-excess contribution depends on the air shower zenith angle 
  and observer lateral distance from the shower axis. Adapted from 
  \citep{LOFARChargeExcess2014}.\label{fig:lofarchargeexcess}}
\end{figure}

In fact, LOFAR later demonstrated \citep{LOFARChargeExcess2014} that the relative strength of the 
charge excess contribution is not a constant but depends on the 
lateral distance from the shower axis and the shower zenith angle, as 
is shown in Fig.\ \ref{fig:lofarchargeexcess}. A dependence on 
lateral distance had previously been predicted by simulations \citep{deVriesCherenkovXmax}.
Of course, the relative strength also depends on the strength of the 
local magnetic field, the observing frequency window also 
plays a significant role, and the altitude of observation can have an 
effect, too. 
  
Measurements thus have clearly confirmed the predictions of 
macroscopic models and microscopic simulation codes that the emission 
can be described with a superposition of a dominant geomagnetic and a 
sub-dominant charge-excess contribution, which leads to a 
characteristic asymmetry in the lateral distribution with peculiar 
polarisation features. The presence of a rising part of the lateral 
distribution function arising from the Cherenkov-like compression 
effects has also been observed early on in LOPES data, which 
found a small but significant fraction of events with a ``rising 
lateral distribution function'' \citep{ApelArteagaAsch2010}, before 
being shown much more explicitly by higher-frequency experiments (see 
below).

Given that the \emph{qualitative} description of the radio emission 
features observed in data agrees with our current paradigm for the emission physics,
the question arises, however, how good the \emph{quantitative} 
agreement with today's state-of-the-art simulations is. We focus on predictions
of the microscopic simulation codes CoREAS and ZHAireS for such a 
comparison, as these provide the most precise description of the emission physics in 
an extensive air shower.

\begin{figure}[!htb]
    \centering
    \includegraphics[width=0.45\textwidth]{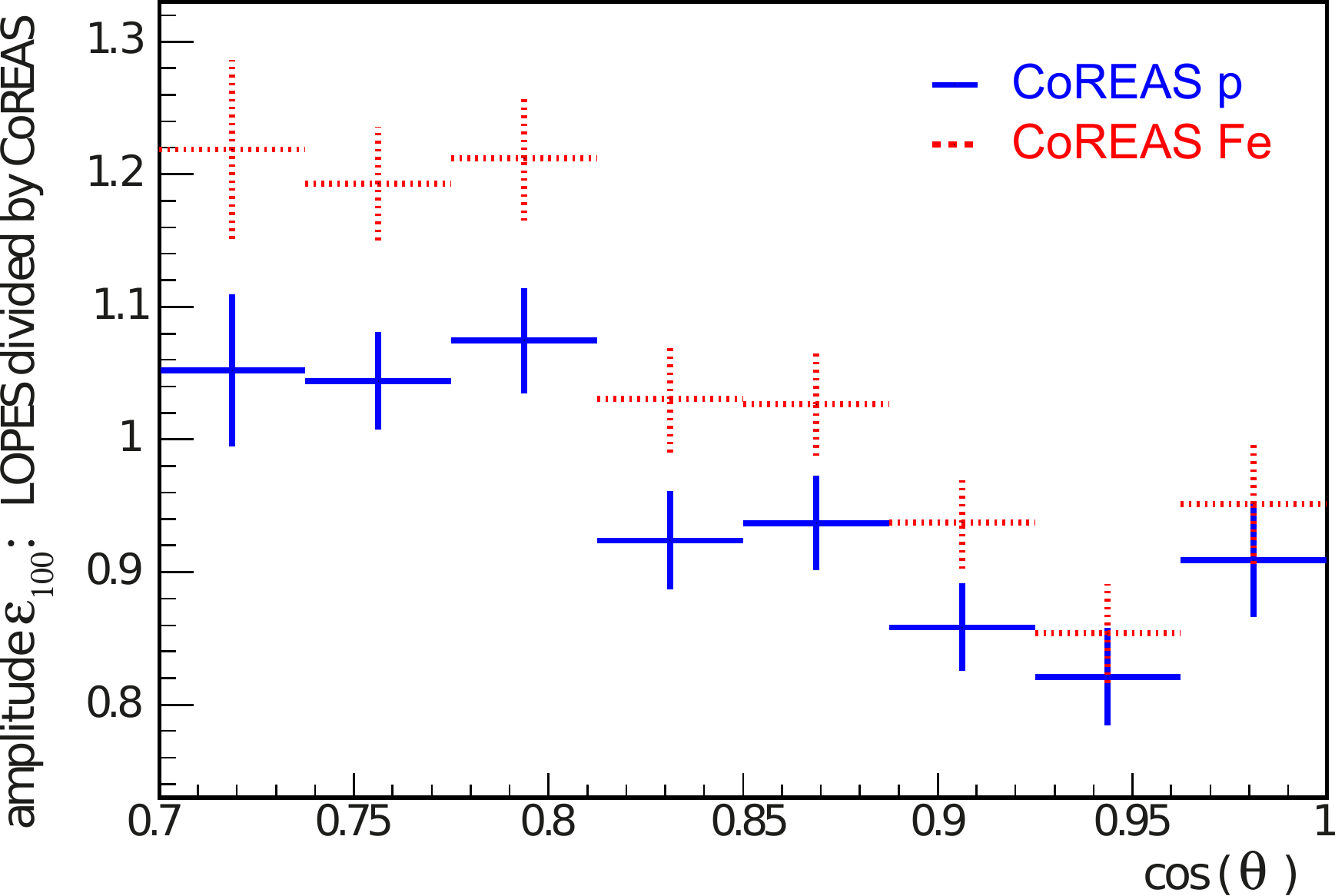}
    \caption{Offset factor for the comparison of the amplitude at 100~m lateral distance between LOPES data and 
    CoREAS simulations as a function of air shower zenith angle 
    $\theta$. A complete simulation of the LOPES detector has been performed for 
    the comparison. Adapted from \citep{LOPESIcrc2015}.}
    \label{fig:simszenithangle}
\end{figure}

The LOPES experiment was the first to publish a quantitative, 
high-statistics comparison between its measurements and CoREAS simulations. The 
data were in good agreement in all characteristics except the 
absolute amplitudes, which were approximately a factor of two higher 
in the LOPES data than predicted by CoREAS simulations \citep{SchroederLOPESCoREAS2013}. With the 
aforementioned re-calibration of the absolute amplitude scale of 
LOPES, the comparison has recently been repeated 
\citep{HuegeLOPESIcrc2015,LOPESrecalibration}. For a set of 
$\sim 500$ cosmic ray events, simulations for proton- and iron-induced 
showers were performed with CoREAS. The measured and simulated radio 
signals were fitted with a simple 1-dimensional exponential lateral 
distribution function
\begin{equation}
\epsilon(d) = \epsilon_{100}\,\exp[-\eta(d - 100\,\mathrm{m})]
\end{equation}
\begin{figure*}[t]
    \centering
    \includegraphics[width=0.48\textwidth]{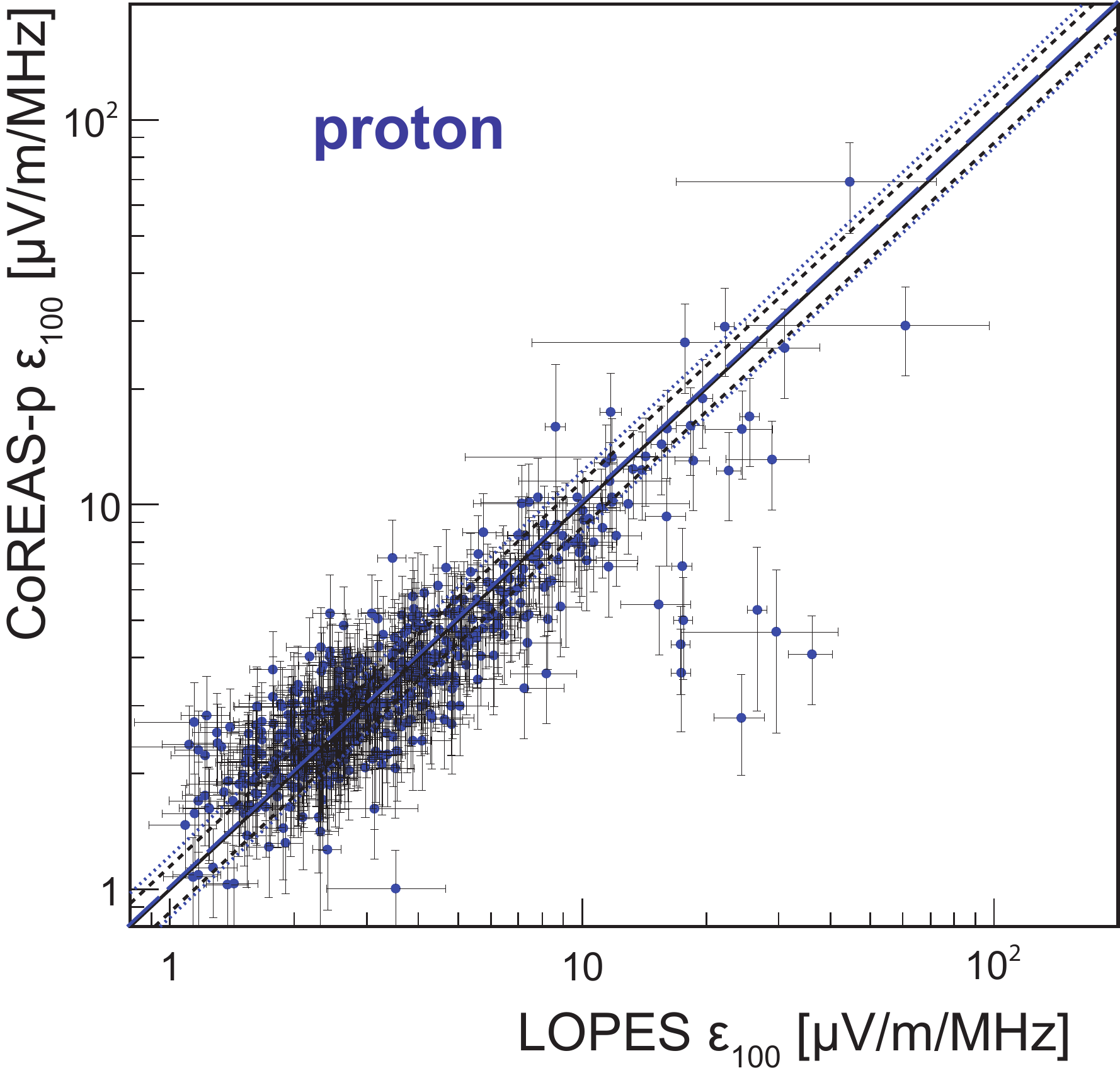}
    \hspace{3mm}
    \includegraphics[width=0.48\textwidth]{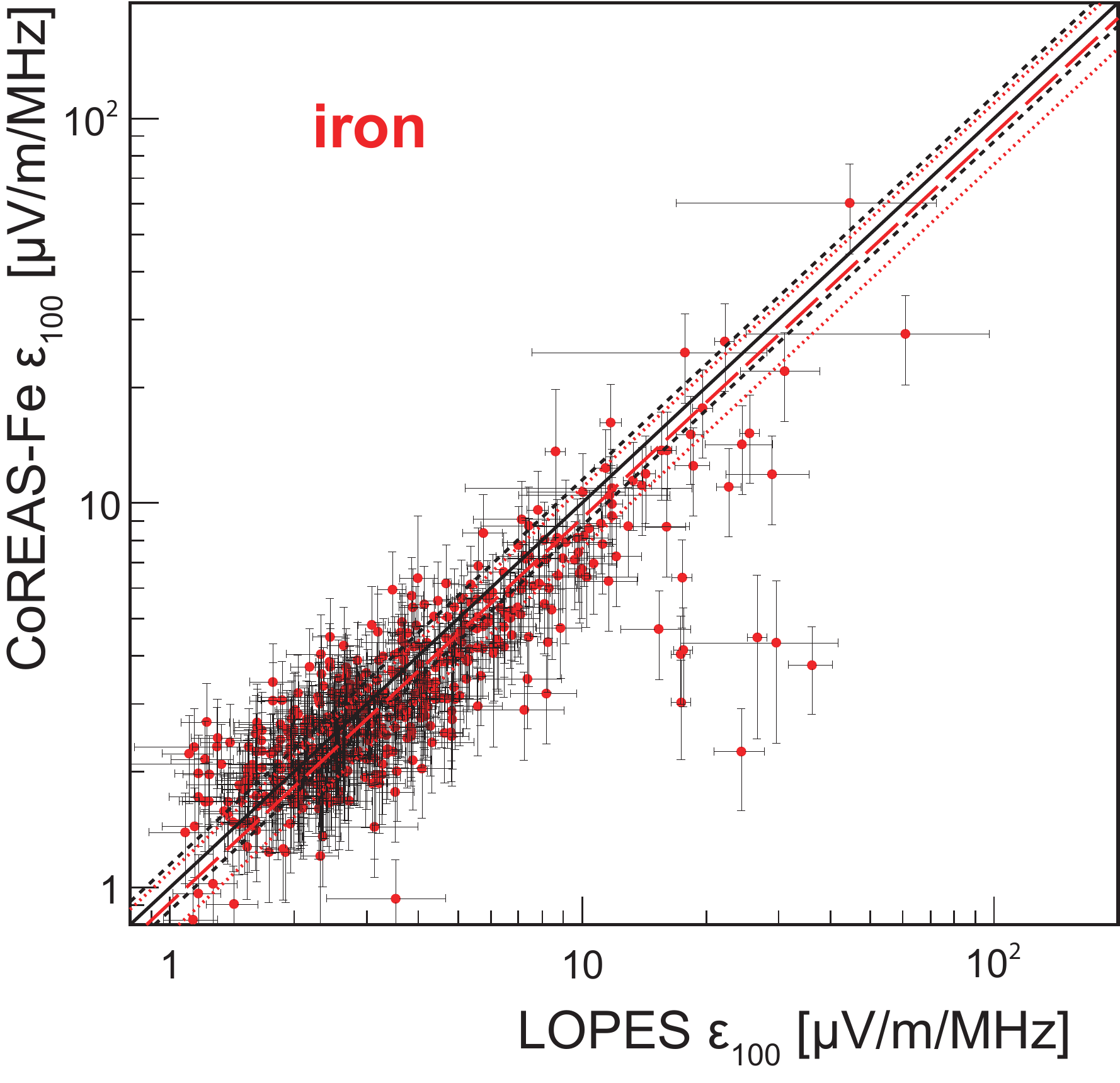}
    \caption{Event-by-event comparison of the amplitude at a lateral 
    distance of 100~m ($\epsilon_{100}$) derived from 
    LOPES measurements and from CoREAS simulations for simulations of 
    proton-induced showers (left) and iron-induced showers (right). 
    The black lines indicate the 1:1 expectation (solid) and the 
    16\% systematic scale uncertainty of the LOPES amplitude 
    calibration (dashed). The colored lines mark the actual 
    correlation between simulations and data (long-dashed) and the 
    systematic uncertainty of the simulated amplitudes arising from the 
    20\% systematic uncertainty of the energy 
    reconstruction of KASCADE-Grande (dotted). The few outlier events 
    in the lower-right parts of the diagrams 
    are not understood (they were not recorded during thunderstorm 
    conditions), but constitute less than 2\% of the data. Adapted from 
    \citep{HuegeLOPESIcrc2015}.}
    \label{fig:epscomparison}
\end{figure*}%
with two free parameters: the amplitude at a lateral distance of 100~m 
and a slope parameter characterising the steepness of the lateral 
distribution. The result obtained after the LOPES re-calibration is shown in 
Fig.\ \ref{fig:epscomparison}. LOPES data and simulations are in 
very good agreement. The mean offset is only 2\% for proton 
simulations and 9\% for iron simulations, well within the systematic 
amplitude scale uncertainty of the LOPES experiment of 16\% (at 
68\% confidence level). The observed scatter is in good agreement with
the one expected from the measurement uncertainties \citep{HuegeLOPESIcrc2015}.
Also the slope parameters of the measured and simulated events
were confirmed to be in good agreement. The only hint at a discrepancy 
visible in LOPES data is a slight deviation in the scaling of the mean 
$\epsilon_{100}$ values between simulations and data with zenith 
angle, shown in Fig.\ \ref{fig:simszenithangle}. The deviations are at 
the level of the uncertainty in the LOPES antenna directivity pattern, 
and should thus not be over-interpreted. Nevertheless, it is important 
to check this effect with additional measurements of other experiments.

\begin{figure*}[!htb]
  \vspace{2mm}
  \centering
  \includegraphics[width=0.9\textwidth]{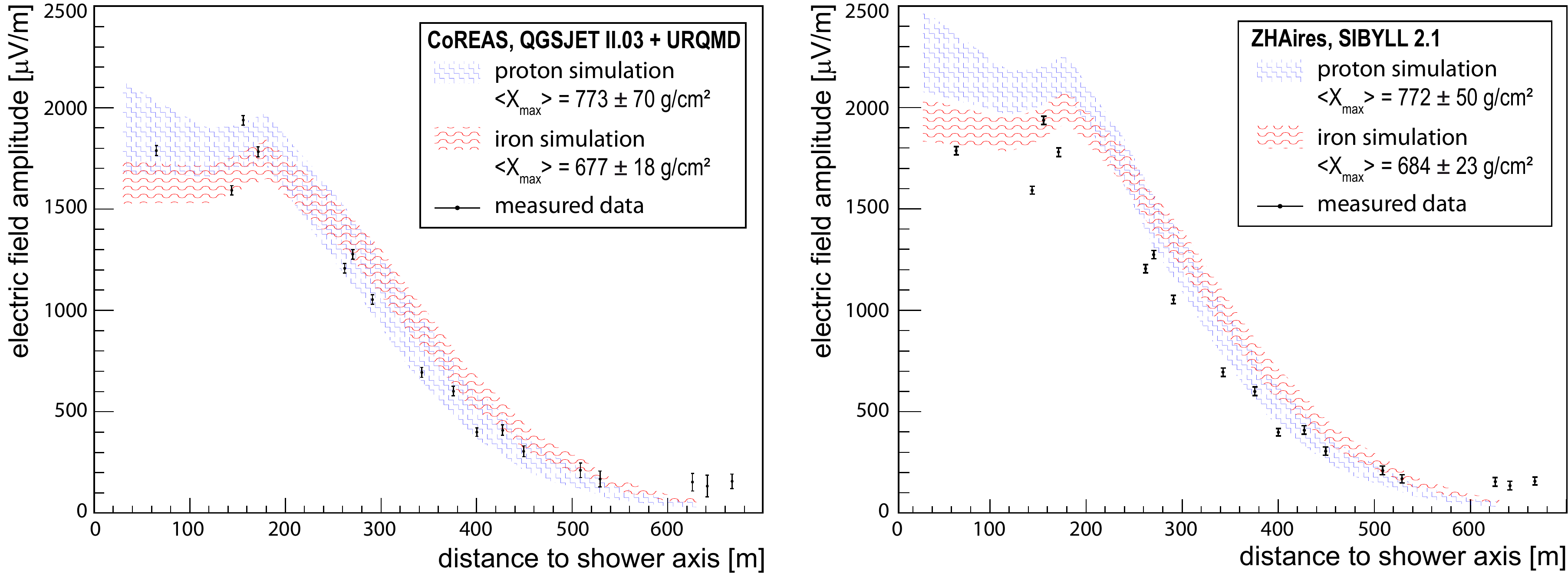}
  \caption{Predictions of cosmic ray radio signals as a function of 
  lateral distance from the shower axis with CoREAS and ZHAireS 
  compared to data for a particular air shower event recorded with 
  AERA. Different hadronic interaction models (as indicated) have been used in the 
  two codes, which can explain at least part of the differences in the 
  simulation predictions. Adapted from \citep{SchroederAERAIcrc2013}.}
  \label{fig:aeracomparison}
 \end{figure*}

An example for a quantitative comparison between simulations and data
with AERA is given in Fig.\ \ref{fig:aeracomparison} which shows 
direct comparison of an individual cosmic ray radio event with 
dedicated CoREAS and ZHAireS simulations. As simulation input, the 
arrival direction, core location, and particle energy as reconstructed 
from Auger particle detector measurements have been varied 50 times 
within their uncertainties, taking into account the covariances 
between these parameters appropriately. Then, CoREAS and ZHAireS have 
been run for these 50 parameter sets for both proton and iron 
primaries. The resulting predicted electric field traces have been fed 
through a complete detector simulation of the Auger Engineering Radio 
Array \citep{AbreuAgliettaAhn2011} and then reconstructed in the same way as have the measured 
radio data, finally reading off the maximum of the total electric field. The comparison shows that there is a good agreement 
between the measured and simulated signals, within the uncertainties 
given by the variation of the input parameters. The 
absolute scale of the emission is well-reproduced within $\approx 
20$\%. (The previously mentioned deviations between the absolute amplitudes predicted by 
CoREAS and ZHAireS are, however, apparent once more.) The error bars on the data points only denote the statistical 
uncertainty, and not the systematic uncertainty on the absolute 
calibration of the experiment. Nevertheless, also for AERA, there seems 
to be good agreement between data and state-of-the-art simulations.

\begin{figure}[!htb]
\centering
\includegraphics[width=0.49\textwidth]{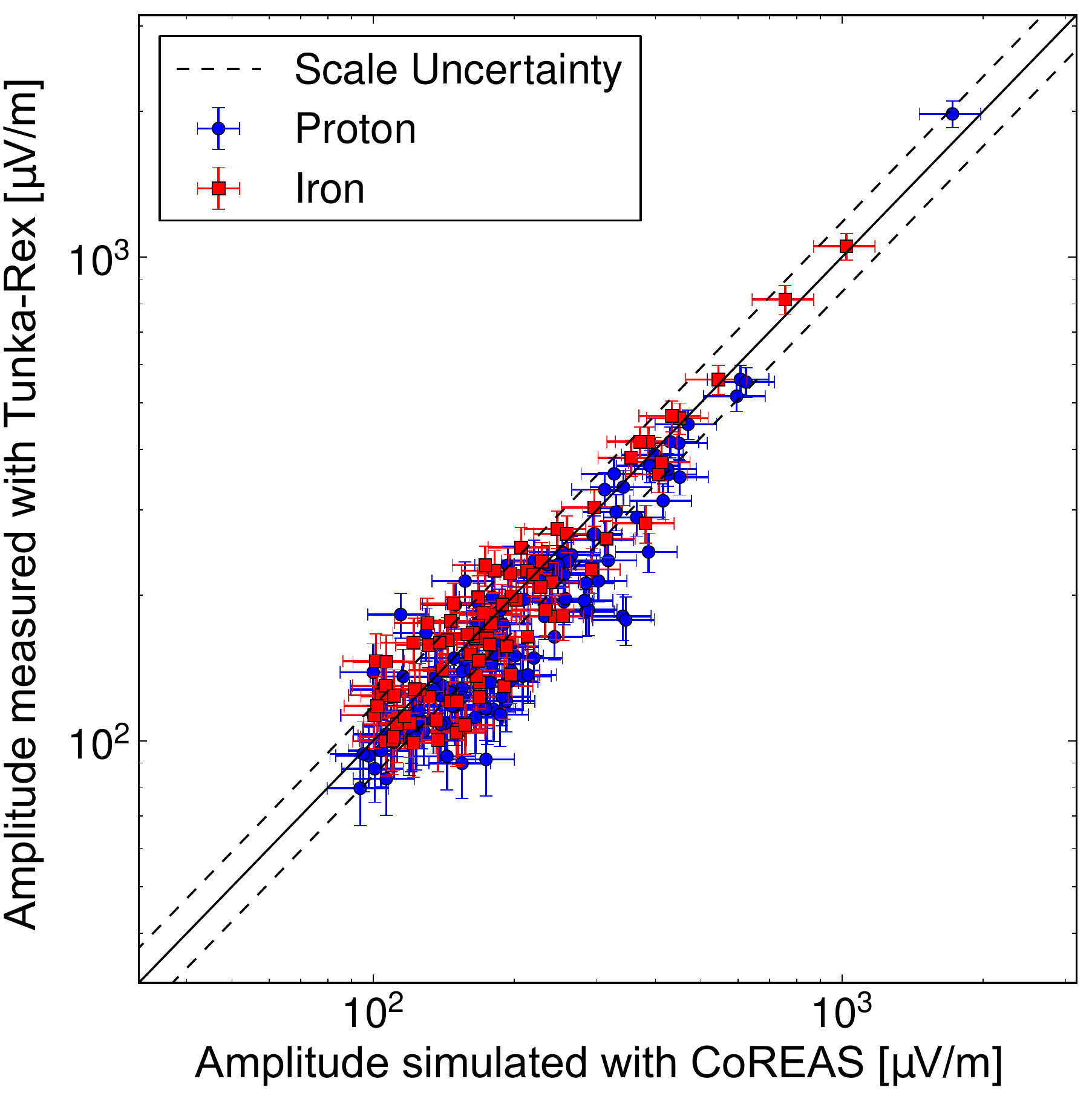}
\caption{Comparison of electric field amplitudes measured in 
individual Tunka-Rex antennas for several events with 
corresponding CoREAS simulations. Adapted from 
\citep{TunkaRexInstrument}.\label{fig:tunkarexvscoreas}}
\end{figure}

A third quantitative comparison of simulated and measured signal 
amplitudes has been published recently by Tunka-Rex \citep{TunkaRexInstrument} and is shown in 
Fig.\ \ref{fig:tunkarexvscoreas}. Here, the signals measured in 
individual antenna stations are compared with the corresponding CoREAS 
simulation. Again, the quantitative agreement is very convincing. 

Three different experiments have thus confirmed that the absolute 
amplitudes predicted with state-of-the-art simulation codes (in 
particular CoREAS) are in agreement with the measurements. This is a very important achievement, 
first because this implicitly indicates that the absolute calibrations of 
the experiments are in agreement, unlike the orders-of-magnitude 
discrepancies in the 1970s. (Please note that Tunka-Rex and LOPES share the same 
absolute calibration scale, still the agreement illustrates that the 
two independent data analyses yield consistent amplitudes.) Second, 
and more importantly, this also means that simulations based on 
first-principle calculations, without any free parameters that can be 
tuned, are indeed able to correctly predict the radio emission 
amplitudes. This has important consequences for using radio detection 
as a technique to determine the energy scale of cosmic ray detectors, 
cf.\ section \ref{sec:energyscale}.

\begin{figure}[!htb]
  \centering
  \includegraphics[width=0.45\textwidth]{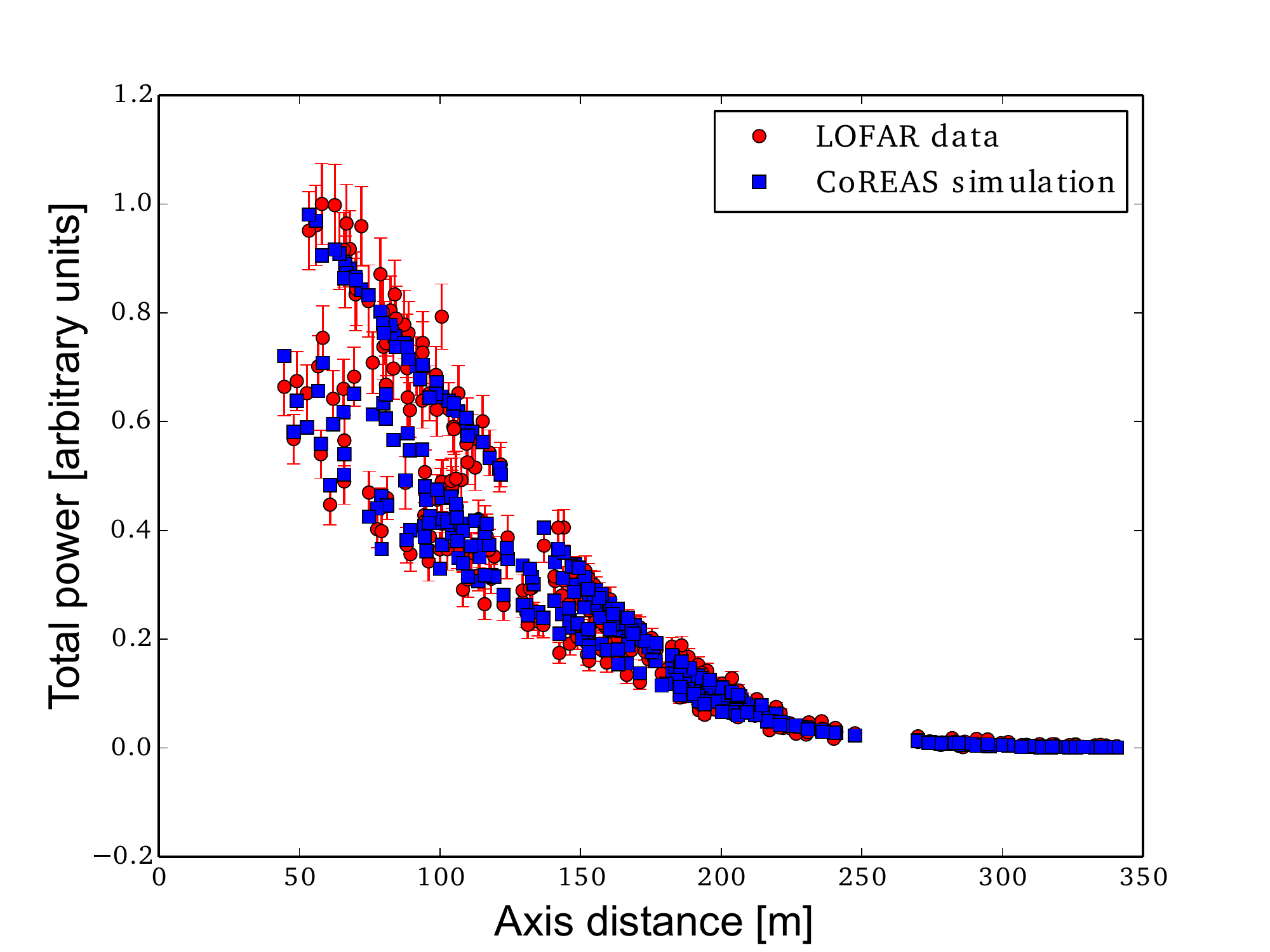}
  \caption{Comparison of the lateral distribution of the radio signal 
  measured by LOFAR and simulated with CoREAS. The data do not lie on a line as there are significant 
  asymmetries in the two-dimensional lateral distribution function 
  which is here projected to a one-dimensional representation. Adapted 
  from \citep{LOFARXmaxMethod2014}.
  \label{fig:lofarcomparison}}
 \end{figure}

Another question is how well the simulations describe the complex 
asymmetric lateral distribution of the radio emission, shaped by the 
superposition of the different emission mechanisms as well as the 
Cherenkov-like compression in the refractive index gradient of the 
atmosphere. This has been probed in a very powerful way with the 
detailed per-event measurements performed with LOFAR. Here, not the maximum amplitude of the electric 
field is used as the quantity of comparison, but rather the 
time-integral over the power in the pulses measured at individual 
antennas. A number of simulations with various input energies has been run and the core position 
has been varied to find the best possible agreement between simulated 
and measured radio signals, a procedure that profits from the high 
number of antennas available per shower in LOFAR. Examples are shown in Fig.\ 
\ref{fig:lofarcomparison}, the left and middle panels of Fig.\ 
\ref{fig:xmaxlofar} (both 30 to 80~MHz) and Fig.\ \ref{fig:lofarhba} 
(110 to 190~MHz).  The agreement is impressive, and it is on a very similar level of quality 
for almost all events measured with LOFAR. While one has to keep in 
mind that this comparison does not test the absolute scale of the 
emission (the LOFAR calibration scale has been established recently, 
quantitative comparisons have yet to be published), it becomes clear that the radio 
emission simulations can correctly reproduce the measured data with 
impressive detail.

\begin{figure}[!htb]
  \centering
  \includegraphics[width=0.35\textwidth]{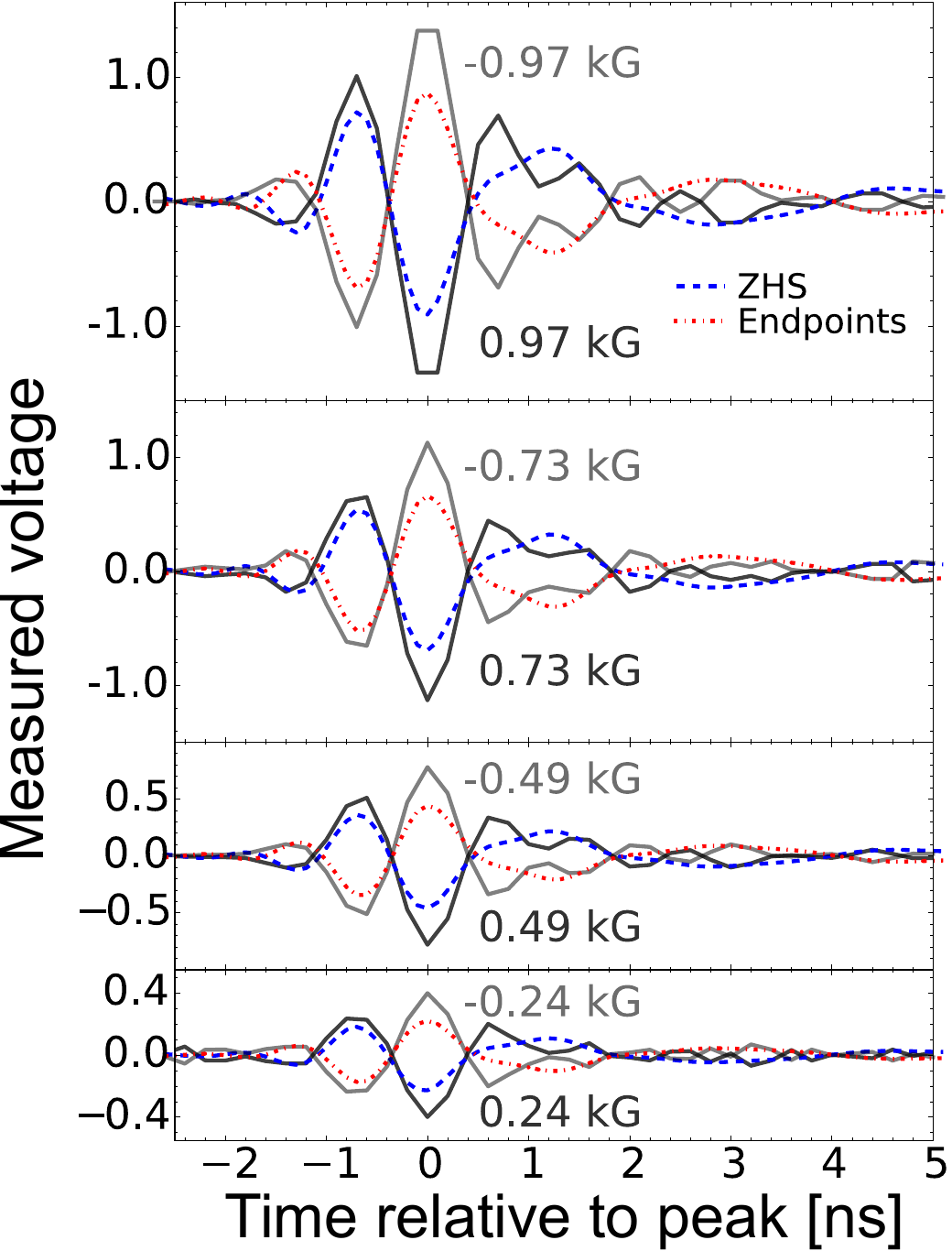}
  \caption{Pulses arising from magnetically induced radio emission (Voltage at oscilloscope) measured
  in the SLAC T-510 experiment for various applied magnetic field configurations, in comparison with 
  predictions from microscopic simulations of the radio emission emitted by
  the electromangetic particle shower. The simulations slightly underpredict the measured radiation strength, but the deviation is within systematic uncertainties of the measurement. Adapted from \citep{SLACT510-PRL}.
  \label{fig:t-510result}}
 \end{figure}

The evidence for the agreement between simulations and 
measured data discussed so far has been provided by air shower 
experiments. Important complementary information, unaffected by systematic 
uncertainties in hadronic interactions and the mass-composition of the 
primary cosmic rays, can be gathered with laboratory experiments. In 
particular, the SLAC T-510 experiment has recently demonstrated on the 
basis of a well-defined electromagnetic particle shower in a well-known 
target encompassed by a strong magnetic field that microscopic simulations describe all aspects of the 
measured radio signals both qualitatively and quantitatively \citep{SLACT510-PRL}.
This includes the signal polarisation, confirming the superposition of magnetic and 
charge-excess emission, the linear scaling of the magnetic emission component
with the strength of the magnetic field, the presence of a Cherenkov cone, and even 
the absolute strength of the emission, within the systematic 
uncertainties of the measurement of $\approx 40$\% in absolute 
amplitude. An example result is is shown in Fig.\ \ref{fig:t-510result}.
The systematic uncertainty of SLAC T-510 is currently dominated by 
uncertainties in the reflection of radio emission at the bottom of the 
target, and can likely be reduced with further measurements.

In summary, it can be stated that the radio emission 
physics has been understood well within the systematic uncertainties 
of the experimental data available today. This quantitative 
understanding of the radio emission is a major achievement of the modern 
studies, and can be considered a true breakthrough. Today's 
state-of-the-art radio emission simulations can thus be used with confidence for the 
development of analysis strategies and reconstruction algorithms, as 
well as an independent cross-check of the energy scale of cosmic ray 
detectors.

\subsection{Detection of inclined air showers}

Very inclined extensive air showers were detected early on with 
the LOPES experiment \citep{PetrovicApelAsch2006}, up to zenith angles 
of 77${^\circ}$ and later 82$^{\circ}$. Already in this early analysis, it could be shown that the 
detection efficiency for inclined air showers is higher than for 
near-vertical showers. Unfortunately, LOPES was too small to determine the extent of the 
illuminated area and verify if indeed it becomes as large as 
expected from the source-distance effects described in section 
\ref{sec:distanceeffects}. Recent results from AERA 
\citep{SchulzIcrc2015}, however, show very convincingly that the radio 
emission footprint becomes very large, with radio pulses at axis 
distances of 1000~m clearly observed. This confirms the predictions by 
event simulations and illustrates the potential for the measurement of 
highly inclined extensive air showers with the radio technique.

\subsection{Direction reconstruction and radio wavefront} \label{sec:wavefront}

Reconstruction of the arrival direction of a cosmic ray air shower 
from radio measurements is usually performed on the basis of the arrival 
times of radio pulses in the individual detector stations (but 
note the caveat described in section \ref{sec:timedefinitions}). 
Alternatively, interferometric techniques can be used to find the sky 
position from which the measured radio signal exhibits the strongest 
correlation between antennas (cf.\ section \ref{sec:interferometry}). In both approaches, an assumption has to 
be made on the shape of the radio emission wavefront.

\begin{figure}[!htb]
  \vspace{2mm}
  \centering
  \includegraphics[width=0.49\textwidth,clip=true,trim=0cm 0cm 2.5cm 15.5cm]{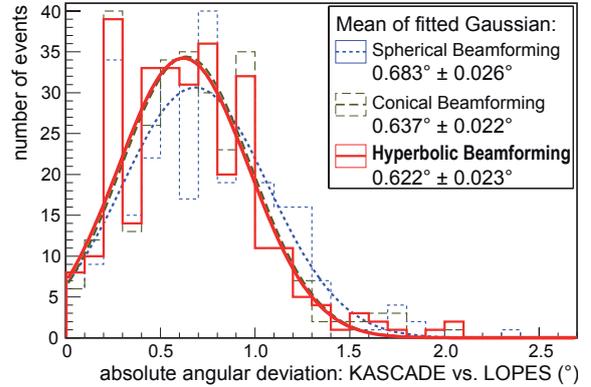}
  \caption{Combined LOPES-KASCADE direction resolution achieved with 
  interferometric reconstruction of the radio source on the basis of 
  various models for the radio wavefront. Adapted from \citep{LOPESWavefront2014}.\label{fig:direction}}
 \end{figure}

The simplest approach is to use a plane wave front, as is expected for 
a source at infinity. This gives a robust reconstruction on 
scales of $\approx 1$ to 2$^{\circ}$. As the radio source, however, is not at infinity, the wavefront is not 
planar. This was seen very early in the analyses of LOPES data, which 
showed that a spherical wavefront works much better in the 
interferometric reconstruction, in the sense that both the fraction of 
reconstructable events and the achievable direction resolution improves.
A spherical wavefront corresponds to a (static) point source at a
finite distance, which obviously is not an 
adequate description for the case of air showers, either. With LOPES data it 
could be shown that a conical wavefront, as can be expected for a 
source extended along a line, provides a better 
reconstruction \citep{SchroederARENA2012}. Finally, it was realized that a hyperbolic 
wavefront, which constitutes a mixture between a spherical (near the 
shower axis) and conical (far from the shower axis) wavefront, can 
describe the data best, both in CoREAS simulations and in a statistical 
analysis of LOPES data \citep{LOPESWavefront2014}. The achieved 
direction resolution using the interferometric reconstruction on the 
basis of the various wavefront models is shown in Fig.\ \ref{fig:direction}.

\begin{figure}[!htb]
  \centering
  \includegraphics[width=0.48\textwidth,clip=true,trim=0cm 1cm 9cm 20cm]{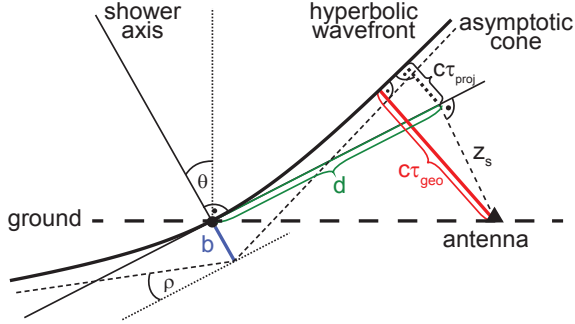}
  \caption{Geometrical delays $\tau_\mathrm{geo}(d,z_s)$ as a function 
  of the antenna position in shower coordinates $(d,z_s)$ for a 
  hyperbolic wavefront. $\theta$ denotes the zenith angle of the air 
  shower, and $\tau_\mathrm{proj}(d)$ is the geometric delay after 
  projecting the antenna position to the shower plane. Adapted from \citep{LOPESWavefront2014}.}
   \label{fig:hyperbola}
\end{figure}

The hyperbolic wavefront is characterized by two parameters, the 
opening angle of the asymptotic cone $\rho$ and an offset at the 
shower axis $b$. Using the geometrical quantities defined in Fig.\ 
\ref{fig:hyperbola} and $c$ as the speed of light, the hyperbolic 
wavefront is described by \citep{LOPESWavefront2014}
\begin{equation}
c \, \tau_\mathrm{geo}(d,z_s) = \\ \sqrt{(d\sin\rho)^2+(c\cdot b)^2} + 
z_s\cos\rho + c\cdot b.
\label{eqn:hyperbola}
\end{equation}

\begin{figure}[!htb]
  \includegraphics[width=0.5\textwidth]{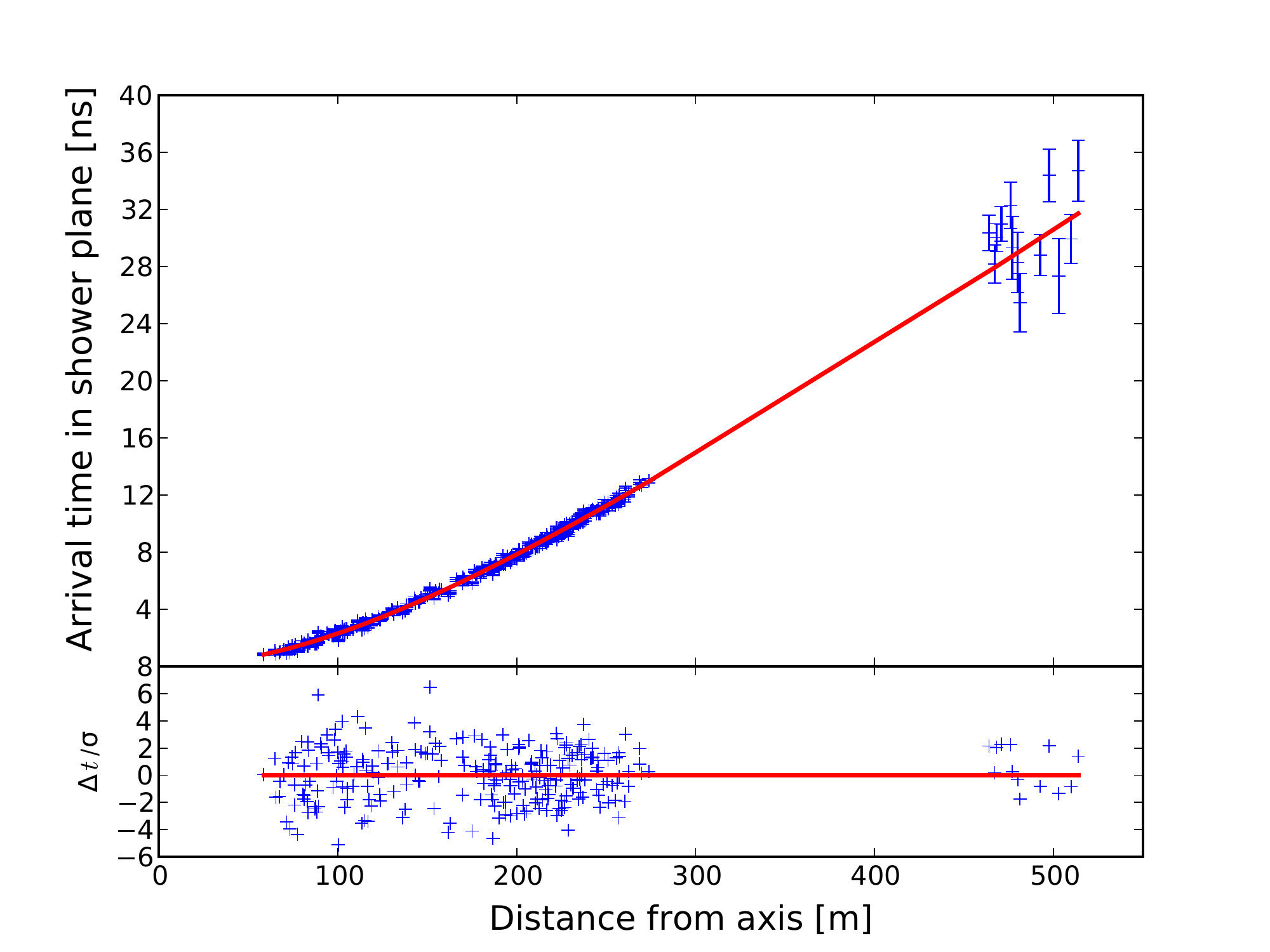}
  \caption{hyperbolic wavefront measured in an individual air shower 
  event measured by LOFAR. Adapted from \citep{LOFARWavefront2014}.\label{fig:wavefront}}
\end{figure}

The result that the radio wavefront has hyperbolic shape was 
confirmed by LOFAR \citep{LOPESWavefront2014}, which could measure the shape of the 
wavefront with very high precision in individual measured events with 
signals detected in hundreds of antennas (Fig.\ \ref{fig:wavefront}). 
Using the hyperbolic wavefront, the precision of the reconstructed 
direction of LOFAR events becomes as small as 0.1$^{\circ}$ (versus 
1$^{\circ}$ when using a planar wavefront). However, 
a possible bias cannot be ruled out with the current data. Another 
feature that remains yet to be investigated experimentally is a slight 
azimuthal asymmetry in the wavefront predicted from CoREAS simulations 
\citep{SchroederARENA2014}.

\begin{figure*}
  \centering
  \includegraphics[width=0.33\textwidth,clip=true,trim=2cm 0cm 2cm 0cm]{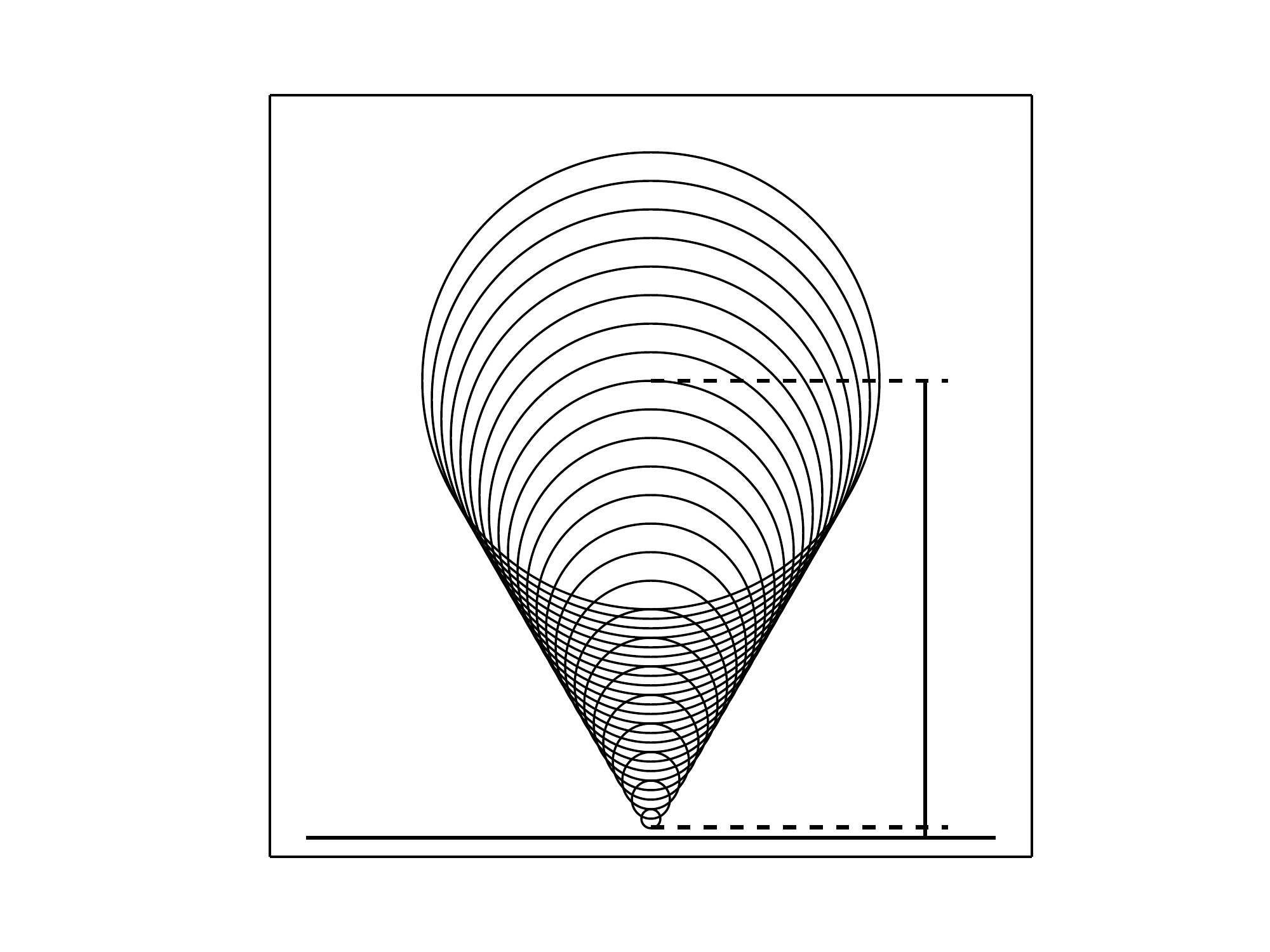}
  \includegraphics[width=0.33\textwidth,clip=true,trim=2cm 0cm 2cm 0cm]{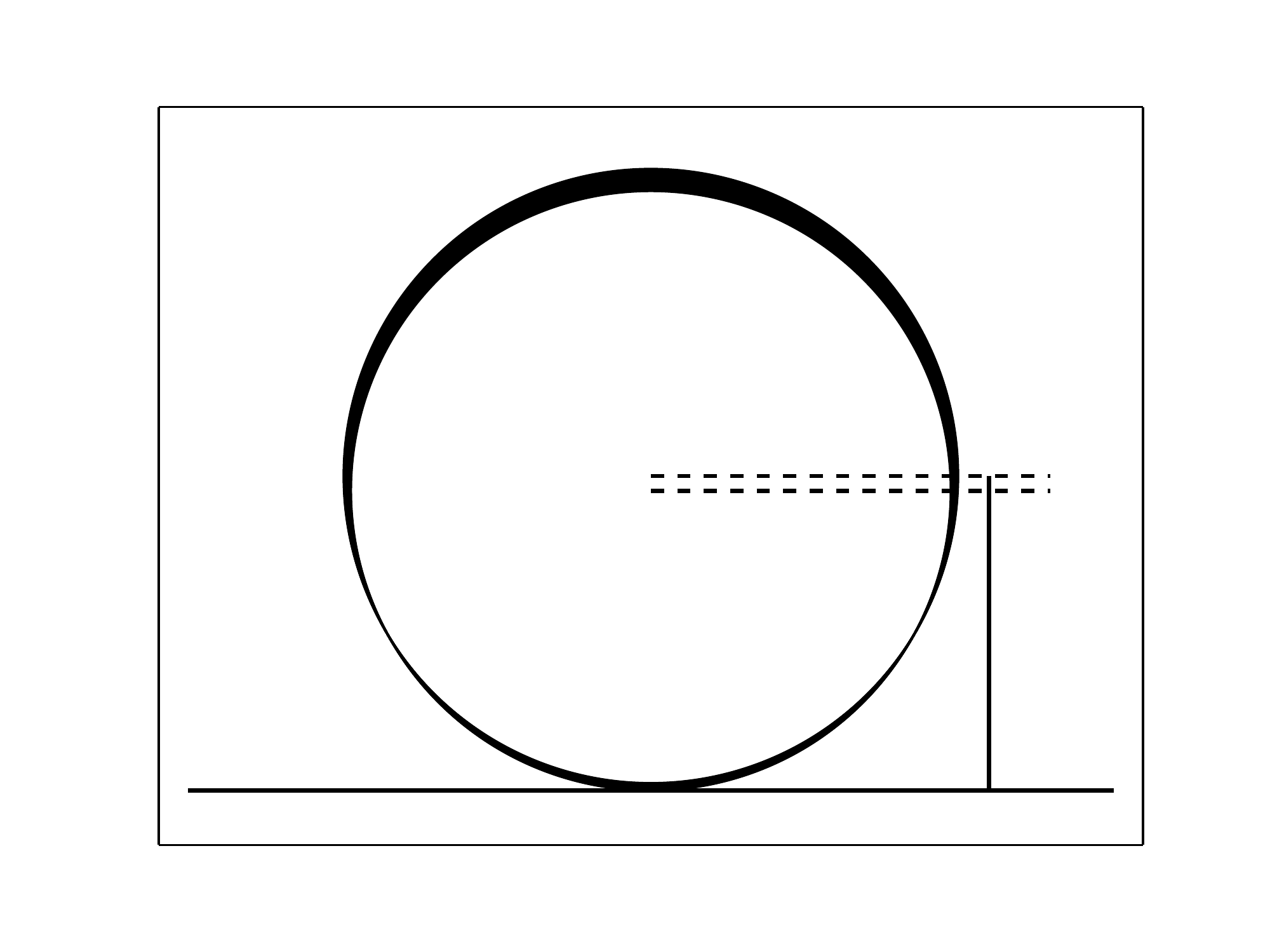}
  \includegraphics[width=0.33\textwidth,clip=true,trim=2cm 0cm 2cm 0cm]{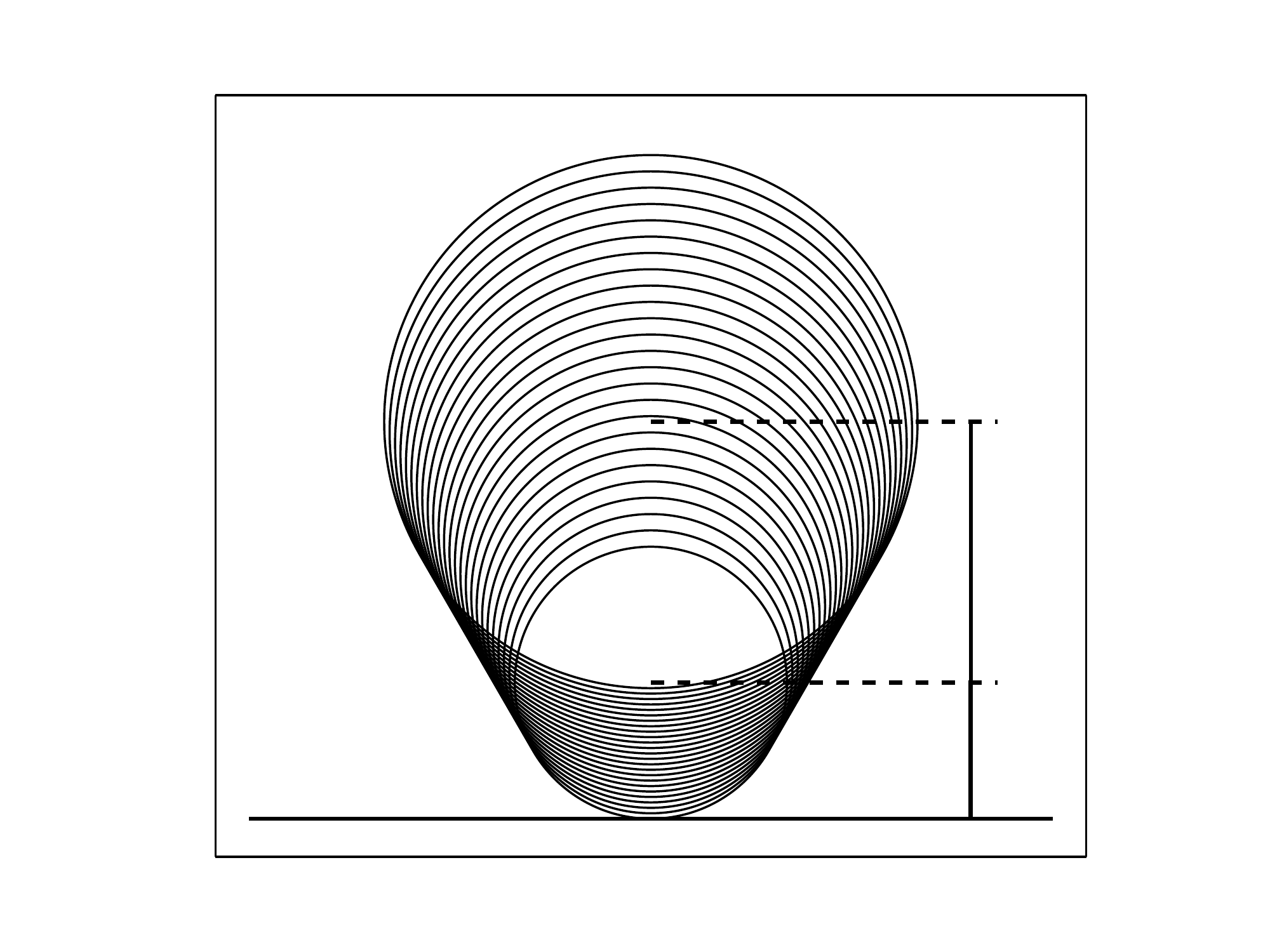}
  \caption{Empirical model for the expected radio wavefront shape. 
  Depending on the length of the emission region in relation to the 
  closest distance between emission region and observer, the wavefront 
  can have a dominantly conical shape (left), a dominantly spherical 
  shape (middle) or the intermediate hyperbolic shape (right). Adapted 
  from \citep{LOFARWavefront2014}.}\label{fig:wavefrontmodel}
\end{figure*}

The LOFAR authors have also provided a simple geometrical model 
explaining why the radio wavefront, depending on the length and 
distance of the air shower cascade, generally has hyperbolic shape 
but can also be nearly spherical or conical for individual air showers.
The governing factor is the length of the emission region in relation 
to the closest distance between emission 
region and observer, as is illustrated in Fig.\ 
\ref{fig:wavefrontmodel}. 

The sensitivity of the wavefront parameters on the distance between 
radio source and observer can be exploited to measure the depth of shower 
maximum of extensive air showers. We will review this approach in section 
\ref{sec:xmaxreconstruction}.

\subsection{Lateral distribution function and core reconstruction} 
\label{sec:LDFs}

In addition to the arrival direction, the location of the shower core 
on the ground needs to be known for quantitative analyses of the 
measurements. There are in principle three ways to determine this core 
position: from the amplitude distribution of the radio signal, from 
the wavefront timing information (cf.\ ref.\ \citep{LafebreFalckeHoerandel2010}),
or from the signal polarisation. Naturally, a combination of these approaches
would be the most powerful, but has not been used in any analyses to 
date. In fact, the only approach that has been studied in some detail 
is the one based on the amplitude distribution, as it is closely 
linked with the important question of how to correctly describe the 
lateral distribution of the radio signal.

Due to the asymmetry of the radio emission footprint (cf.\ Fig.\ 
\ref{fig:footprint}), any approach assuming a rotationally symmetric lateral distribution 
function will lead to a biased result, as was demonstrated in the 
CODALEMA analysis that revealed the charge-excess contribution 
\citep{CODALEMACoreShift}. This added complexity might seem 
like a disadvantage at first. If, however, the functional form of 
the complex lateral distribution function is understood well enough, 
the complexity of the lateral distribution function is in fact an 
advantage: there is a wealth of information encoded in the radio footprint, 
which can be exploited for an accurate reconstruction of the core 
position and even the depth of shower maximum, as we will discuss below.

Several approaches have been made at describing the radio emission 
LDF. The first and easiest approximation is that of rotationally 
symmetric one-dimensional exponential LDF. This has been used for many 
analyses, in the historical experiments and also in CODALEMA and LOPES, cf.\ section 
\ref{sec:simcomparison}. The advantage of an exponential LDF is that it 
only needs two free parameters, namely an absolute scale (e.g.\ the 
amplitude at the axis or at a fixed distance such as 100~m) and a 
scale radius or, equivalently, slope parameter. The one-dimensional exponential LDF 
can neither describe the asymmetries nor the Cherenkov bump observed 
in the radio footprint. For some limited analyses of small-scale 
arrays, it can however be useful because it is robust and, if applied 
consistently to simulations and measured data, allows direct 
comparison of the two. Another approach has been to use 
rotationally symmetric Gaussian LDFs \citep{ApelArteagaBaehren2014}. While adding one free parameter, 
these can accommodate the rising part of the LDF near the shower axis, 
introduced by the Cherenkov-like compression effects. The disadvantage 
of the implied rotational symmetry is that biases occur, the 
asymmetries in the radio LDF only average out if the radio emission 
footprint is sampled sufficiently homogeneously, which is not 
necessarily the case. For high-quality studies, it is therefore 
necessary to take the asymmetry in the radio LDF into account.

One approach is to correct for the charge-excess effects, thereby 
removing the asymmetry and allowing the use of a rotationally 
symmetric LDF. This approach was followed in some analyses of AERA 
\citep{GlaserARENA2012}. A more evolved version of this approach has 
also recently been published by members of the Tunka-Rex 
collaboration \citep{KostuninLDF}, it is the approach that will be 
used for the analysis of the Tunka-Rex data. Both approaches rely on 
information on the relative strength of the charge excess to perform 
the correction. The former approach was based on an estimate from 
polarisation measurements, while the latter approach relies on the predictions 
from CoREAS simulations. As the agreement between data and 
measurements is excellent (see above), however, this seems well-justified.

Another possibility is to use a non-rotationally symmetric 
two-dimensional LDF. This approach has been developed in the frame of LOFAR. 
First, a simulation study based on CoREAS simulations was used to develop a two-dimensional 
parameterization which is based on the empirical result that the 
two-dimensional LDF (in this case of time-integrated power pulses) can 
be described well as the superposition of a large positive and a 
smaller negative Gaussian with some offset and scaling parameters 
\citep{NellesLDF}.\footnote{This 
two-dimensional parameterization in fact contains the essence 
of hundreds of CoREAS simulations and can thus be used as a powerful tool 
whenever the received power as a function of observer position needs to 
be estimated quickly, e.g., in array simulation studies.} One important aspect of this approach is the choice of
a well-suited coordinate system for the shower plane (the plane perpendicular to the shower axis) 
using one unit vector along the $\vec{v} \times \vec{B}$ direction (the 
direction of the Lorentz force and thus of the polarisation 
arising from the dominant geomagnetic emission) and one unit vector 
along the $\vec{v} \times \vec{v} \times \vec{B}$ direction which does not contain any emission 
contribution from the geomagnetic effect. Many of the parameters in 
the parameterization exhibit strong correlations and can thus be expressed as functions of 
each other, reducing the number of free parameters. The version used to 
fit LOFAR events \citep{NellesApplicationLDF} has been reduced to 
six free parameters, two for the core position, two for the arrival 
direction and two related to the energy and depth of shower maximum, which can be 
well-constrained in measurements of LOFAR events with hundreds of data points. Recently, the same empirical two-dimensional function with 
parameters adapted to the appropriate altitude and local geomagnetic 
field of the site of the Auger Engineering Radio Array has also been 
successfully employed in AERA analyses \citep{AERAEnergyPRD}. If the 
core position has been established with another approach (such as the 
Auger surface detector information), the two-dimensional LDF can be applied to AERA events 
with only three signal detections. If at least five signal detections 
are available, the core position can be estimated from radio data using the 
two-dimensional LDF.

A related work, motivated by the peculiar polarisation 
characteristics of the geomagnetic and charge excess contributions, was 
presented in \citep{AlvarezMunizLDFScheme}. It does not describe the lateral distribution 
function of the radio emission {\it per se}, but can interpolate the 
asymmetric radio-emission footprint from simulations performed only 
along two major axes of the radio footprint (such that information on 
both emission contributions is obtained). This approach 
could in principle be developed further to give a more physically 
motivated two-dimensional lateral distribution function than the 
empirical two-dimensional LDF based on two summed Gaussians.

\subsection{Energy reconstruction} \label{sec:energyreconstruction}

One of the primary interests in cosmic ray measurements is to 
precisely and accurately determine the energy of cosmic ray particles. It has by 
now been successfully demonstrated that radio measurements can give a 
very direct and precise access to this energy.

The main reason for this is the coherent nature of the radio emission. 
The amplitude of the radiated radio pulses is proportional to the 
number of electrons and positrons in the cascade, which in turn is 
proportional to the energy of the primary particle. As essentially 
only electrons and positrons contribute to the radio signal (all other 
particles have a much lower charge/mass ratio, their radio emission is 
thus very strongly suppressed), radio detection directly probes the electromagnetic 
component of air showers. This is the best-understood air shower component, and 
it also harbors the vast majority of the energy of the cascade, more than 80\% 
up to almost 100\% depending on the primary energy, primary mass and the hadronic interaction model 
adopted to interpret the data \citep{ThePierreAuger:2013eja}. Furthermore, the radio emission from all 
along the shower evolution is integrated when it arrives at the 
ground, as the radio emission undergoes no relevant absorption or scattering in 
the atmosphere. In other words, radio detection provides a calorimetric 
measurement of the energy in the electromagnetic cascade of an air 
shower.

\begin{figure}[!htb]
\includegraphics[width=0.5\textwidth]{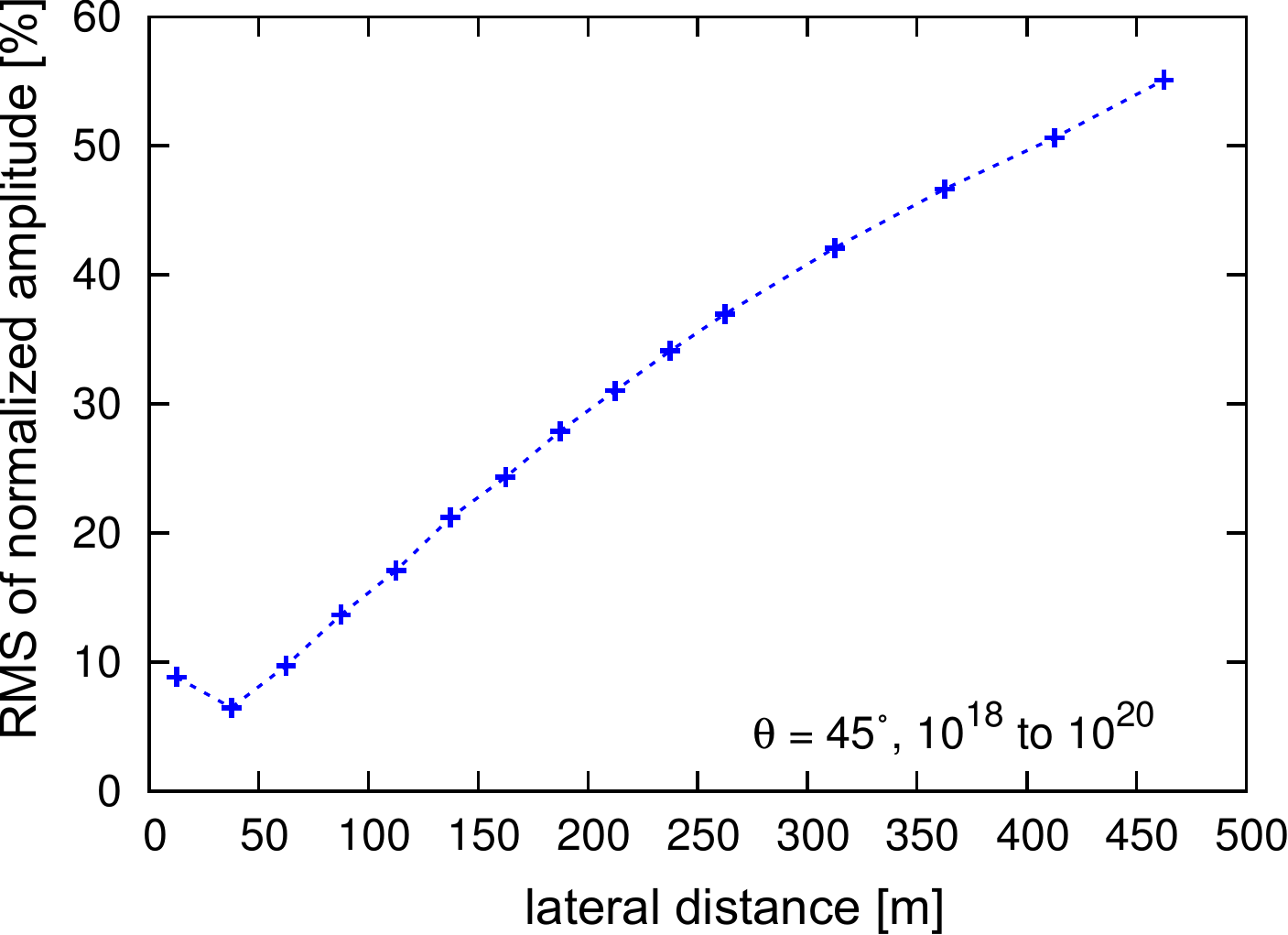}
\caption{There is a lateral distance at which the energy-normalized
radio amplitude is influenced little by shower-to-shower fluctuations. 
Measurements at this distance can thus be used for a precise energy estimation. Adapted from 
\citep{HuegeUlrichEngel2008}.\label{fig:energyestimator}}
\end{figure}

The important question is how precisely the energy 
can be determined and how strongly the determination suffers from intrinsic 
shower-to-shower fluctuations, which --- unlike instrumental 
uncertainties, are not addressable. According to a simulation 
study, the intrinsic energy resolution of air shower radio 
measurements was expected to be very good, with intrinsic resolutions below 
10\% \citep{HuegeUlrichEngel2008}, illustrated in Fig.\ 
\ref{fig:energyestimator}. (This study 
was based on the flawed REAS2 approach, but as these effects are 
purely geometry, the main results are independent of the emission 
model. Consequently they have later been confirmed also by other simulation
approaches \citep{Konstantinov2009,deVriesEnergyXmax2010}.) By now,
several experiments have published analyses regarding the reconstruction
of the primary particle energy from radio measurements. We shortly
review the different approaches here and state the achieved resolutions.

\begin{figure*}[htb]
  \includegraphics[width=0.47\textwidth]{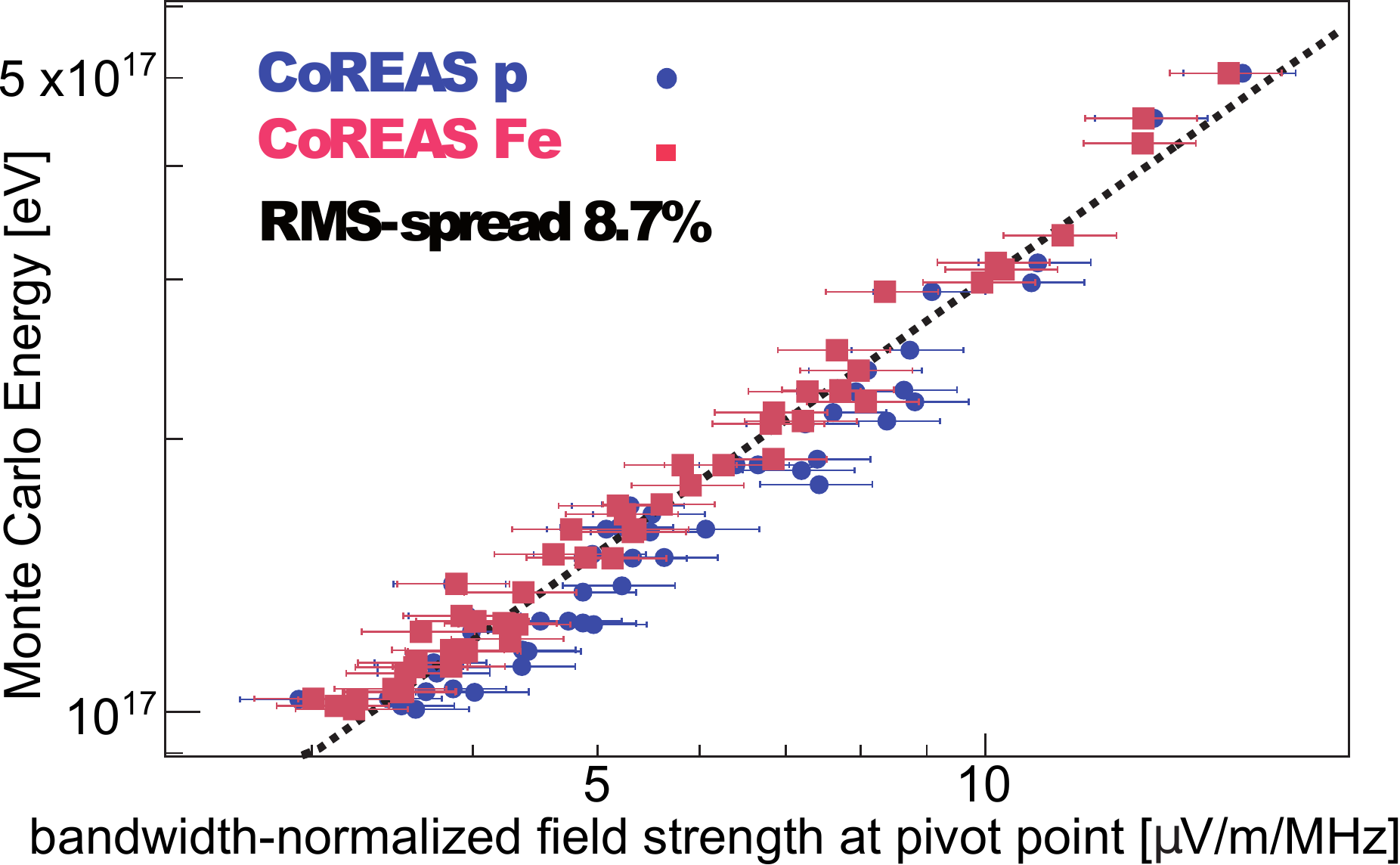}
  \hspace{3mm}
  \includegraphics[width=0.51\textwidth]{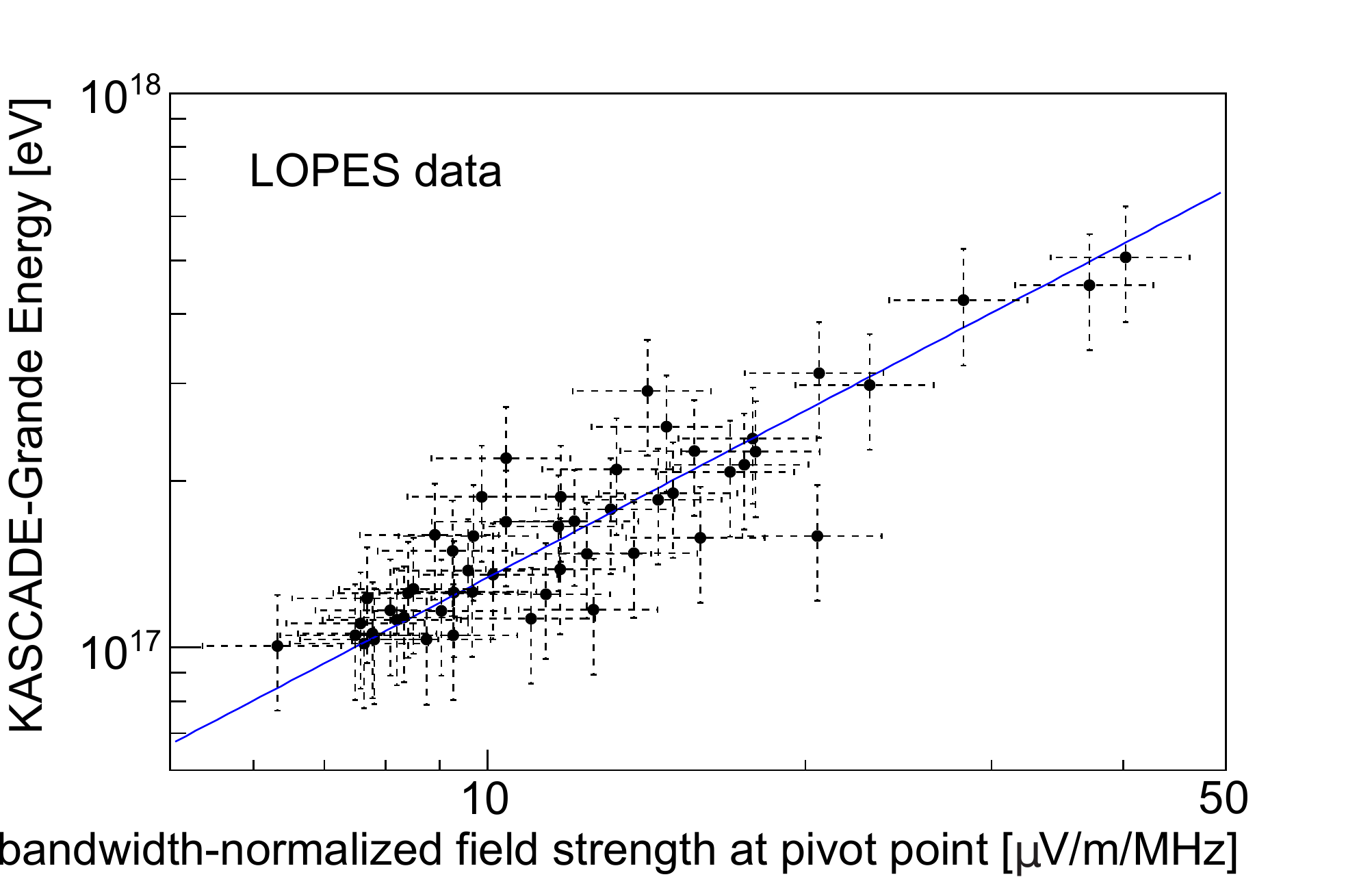}
  \caption{Correlation between the radio amplitude at the 
  pivot-point of the radio LDF and the Monte Carlo true 
  energy as predicted with CoREAS simulations of LOPES events (left). 
  The predicted linear correlation is also observed in data measured 
  with LOPES as compared with the energy reconstructed by 
  KASCADE-Grande (right). Amplitudes have been normalized appropriately for the angle
  to the geomagnetic field. The amplitudes reported for LOPES data 
  still have to be scaled down by a factor of 2.6 due to the revised 
  calibration of the experiment. Adapted from 
  \citep{ApelArteagaBaehren2014}.}\label{fig:lopesenergy}
\end{figure*}

The first quantitative analysis on the reconstruction of the cosmic 
ray energy from radio data was published by the LOPES 
experiment \citep{ApelArteagaBaehren2014}. This analysis exploits the 
result of the above-mentioned simulation study \citep{HuegeUlrichEngel2008}: A characteristic lateral distance from 
the shower axis exists at which the influence of shower-to-shower-fluctuations on 
the radio amplitude is minimized. This is a geometrical effect 
directly related to the forward-beaming of the radio emission. In 
ref.\ \citep{ApelArteagaBaehren2014}, an updated simulation study with CoREAS, 
tailored to the specific situation of the LOPES experiment and based 
on a rotationally symmetric Gaussian LDF, confirmed 
the expectation for the presence of such a characteristic distance 
($\emph{pivot-point}$) in the LDF. The intrinsic resolution of an energy measurement with a radio array the size and 
density of LOPES was predicted to be better than 10\% (Fig.\ 
\ref{fig:lopesenergy}, left). Part of the scatter in this distribution 
is due to the systematic difference of energy in the electromagnetic 
cascade for proton- and iron-induced air showers. If the energy 
in the electromagnetic cascade of the shower (rather than the total 
energy of the primary particle) needs to be determined, then intrinsic 
fluctuations are expected to be even as low as 5\%. Hence, radio detection
should be a very precise technique for energy determination with small intrinsic fluctuations.
The predicted correlation between the radio amplitude at the 
pivot-point, normalized for the scaling of geomagnetic emission 
with geomagnetic angle, and the primary particle energy determined by KASCADE-Grande has 
then been verified in LOPES data (Fig.\ \ref{fig:lopesenergy}, right). There is a clear correlation, and 
the combined uncertainty on the energy determined with LOPES and 
KASCADE-Grande is $\approx 20$-25\%. As the energy resolution of 
KASCADE-Grande alone is $\approx 20$\%, the intrinsic resolution of the 
radio-based energy determination is probably indeed much smaller than 
20\%, even for a non-ideal prototype experiment such as LOPES and an 
analysis procedure which does not take the asymmetries of the 
radio footprint into account explicitly.

\begin{figure}[!htb]
\centering
\includegraphics[width=0.48\textwidth]{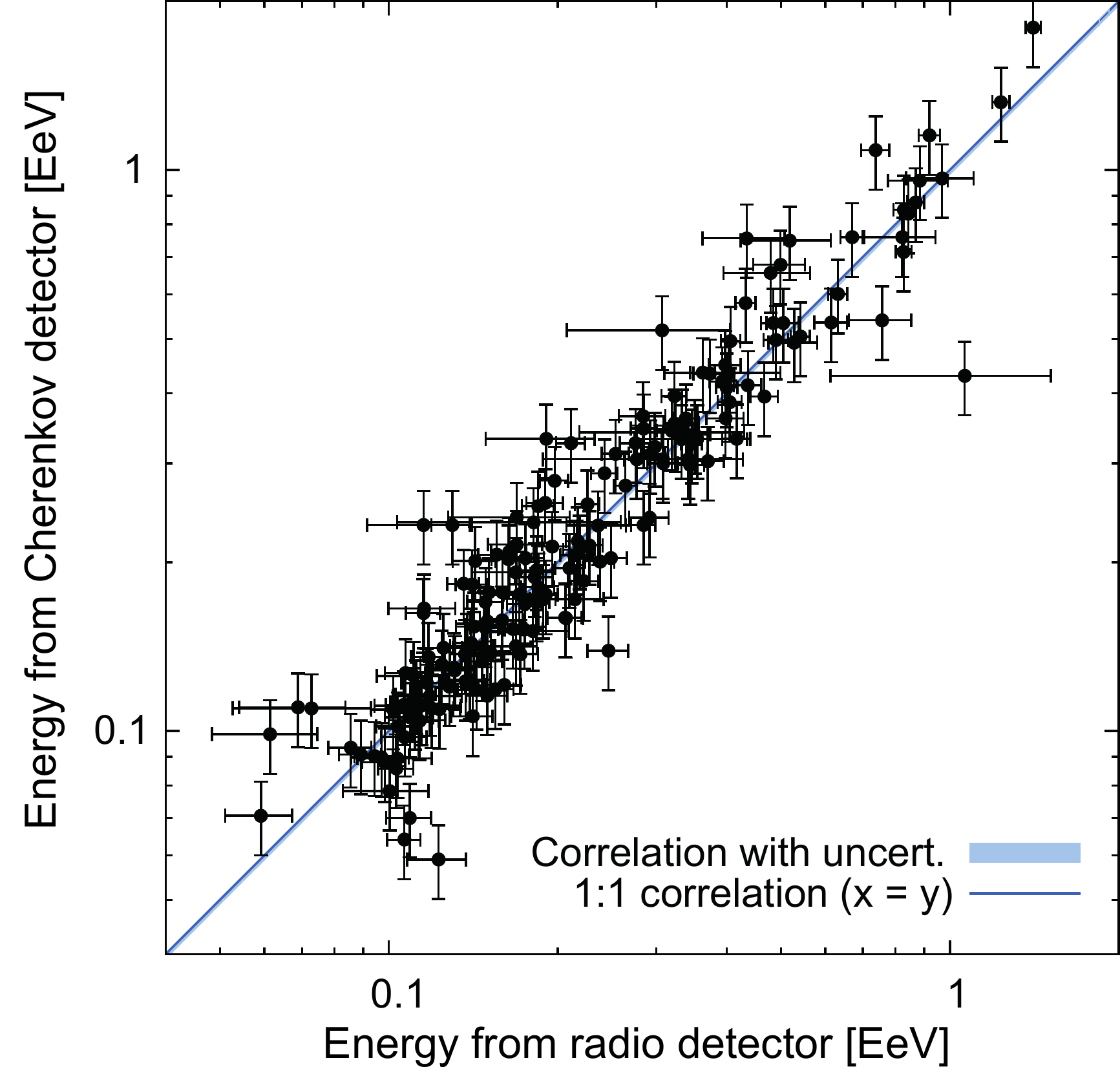}
\caption{Correlation of the radio energy estimator of Tunka-Rex, 
determined from a Gaussian LDF after correction of charge-excess 
asymmetries, in comparison with the energy reconstructed with the 
Tunka-133 optical Cherenkov detectors. Adapted from 
\citep{TunkaRexCrossCalibration}.\label{fig:tunkarexenergy}}
\end{figure}

A conceptually similar approach as in LOPES, exploiting the minimum 
intrinsic fluctuations in the radio amplitude of the LDF, but
this time using an approach that correctly compensates for the effects of the charge-excess 
contribution \citep{KostuninLDF}, has been applied to Tunka-Rex data. 
The main result is depicted in Fig.\ \ref{fig:tunkarexenergy}. Again, 
a very good correlation of the energy determined from the radio signal 
with the energy determined from another detector, in this case an 
optical Cherenkov detector, is observed. The combined resolution of 
the two energy estimators amounts to 20\%, which as in the case of LOPES
is only slightly larger than the energy resolution of the reference detector
alone (amounting to 15\%).

LOFAR has applied the two-dimensional lateral distribution function 
\citep{NellesLDF} discussed in section \ref{sec:LDFs} to measured 
events. One of the LDF fit parameters, $A_{+}$, the amplitude of the 
dominant Gaussian in their fit to the measured signal powers, is
expected to correlate with the energy of the 
primary cosmic ray. Indeed, a good correlation between this LDF fit parameter 
and the energy reconstructed with the LOFAR particle detector array LORA 
\citep{Thoudam2015} is observed. However, this result so far lacks an absolute 
calibration of the radio amplitude scale.

\begin{figure}[!htb]
\centering
\includegraphics[width=0.48\textwidth]{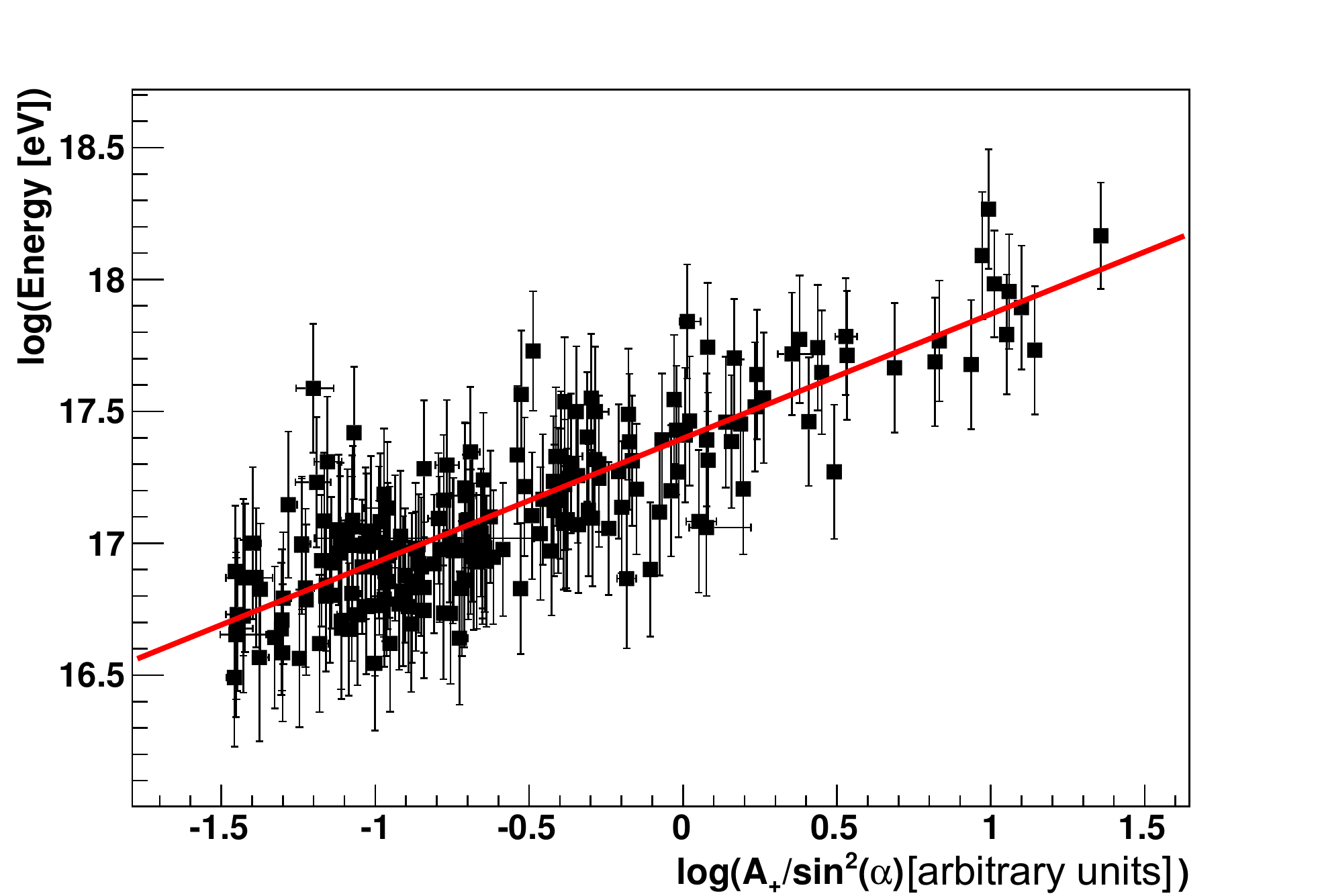}
\caption{Correlation of the normalized $A_{+}$ parameter in the two-dimensional 
LDF \citep{NellesLDF} applied to LOFAR data in comparison with the 
cosmic ray energy reconstructed with LORA. Adapted from 
\citep{NellesApplicationLDF}.}
\end{figure}

\begin{figure}[!htb]
\centering
\includegraphics[width=0.5\textwidth]{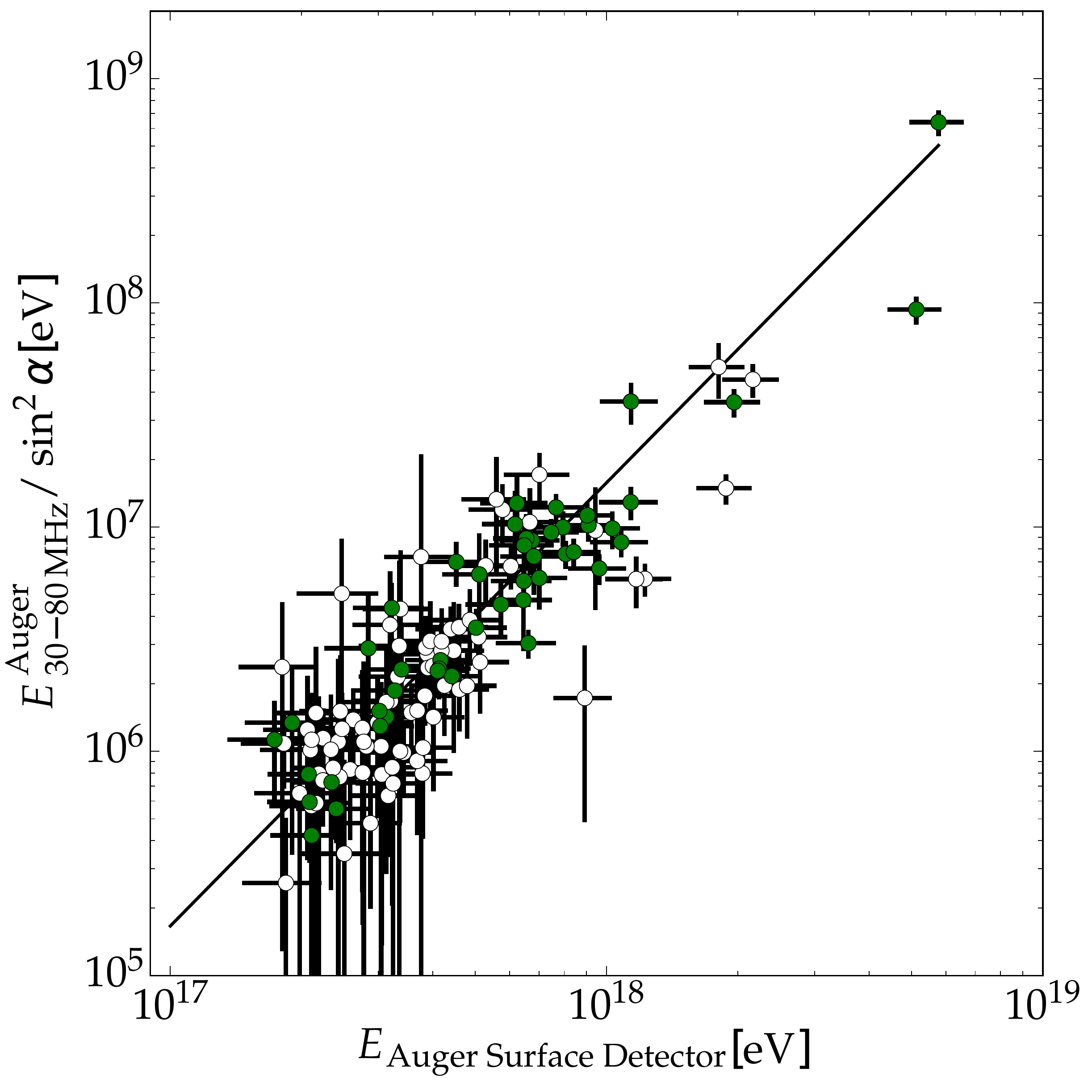}
\caption{Correlation between the radiation energy, normalized for 
incidence perpendicular to the geomagnetic field, and the cosmic ray energy
determined by the Auger surface detector. Open circles represent air showers with radio signals 
detected in three or four AERA detectors. Filled circles 
correspond to showers with five or more detected radio signals. Adapted 
from \citep{AERAEnergyPRD}.\label{fig:crosscalibration}}
\end{figure}

While the approaches discussed above used quantities 
such as the maximum amplitude at a characteristic distance or a fit parameter
of a two-dimensional lateral distribution function as estimators for the energy of the primary 
cosmic ray, AERA has recently published a result that has the major benefit 
of using an intuitive, well-defined, and universal quantity as an energy estimator: 
the total energy contained in the radio signal in the frequency band from 30 to 
80~MHz \citep{AERAEnergyPRD}. To determine this \emph{radiation energy}, the time-dependent 
electric field (in units of V/m) reconstructed from the AERA 
measurements at individual detector locations is squared to calculate the 
local Poynting flux. A time-integration over the detected pulse yields the 
energy fluence (in units of eV/m$^{2}$) measured at each individual 
radio detector. Using an adapted version of the two-dimensional LDF 
\citep{NellesLDF}, an LDF fit (see Fig.\ \ref{fig:2dldf}) and an integration over the shower plane 
is performed. This yields the total energy in the radio signal (in units of eV) in the 30 to 
80~MHz range. All known detector effects have been deconvolved from this 
\emph{radiation energy}. Since the contribution of the charge-excess 
effect to the radiation energy is minimal,
the radiation energy can be normalized with $\sin^{2}$ of the 
geomagnetic angle, yielding the radiation energy for air showers with 
perpendicular incidence to the geomagnetic field. This normalized 
radiation energy $E^{\mathrm{Auger}}_{30-80\,{\mathrm{MHz}}}/sin^{2}(\alpha)$ shows the 
expected quadratic correlation with the cosmic ray energy determined 
with the surface detector of the Pierre Auger Observatory, as is shown 
in Fig.\ \ref{fig:crosscalibration}. (Due to the radio signal coherence, 
amplitudes scale linearly with the cosmic ray energy, and the radiated 
energy scales quadratically.) From the power-law fit, the radiation 
energy for a cosmic ray shower with perpendicular incidence to the 
geomagnetic field at the Auger site can be read off. After a 
normalization with the strength of the geomagnetic field, this yields 
the following result:
\begin{figure}[!htb]
\centering
\includegraphics[width=0.5\textwidth]{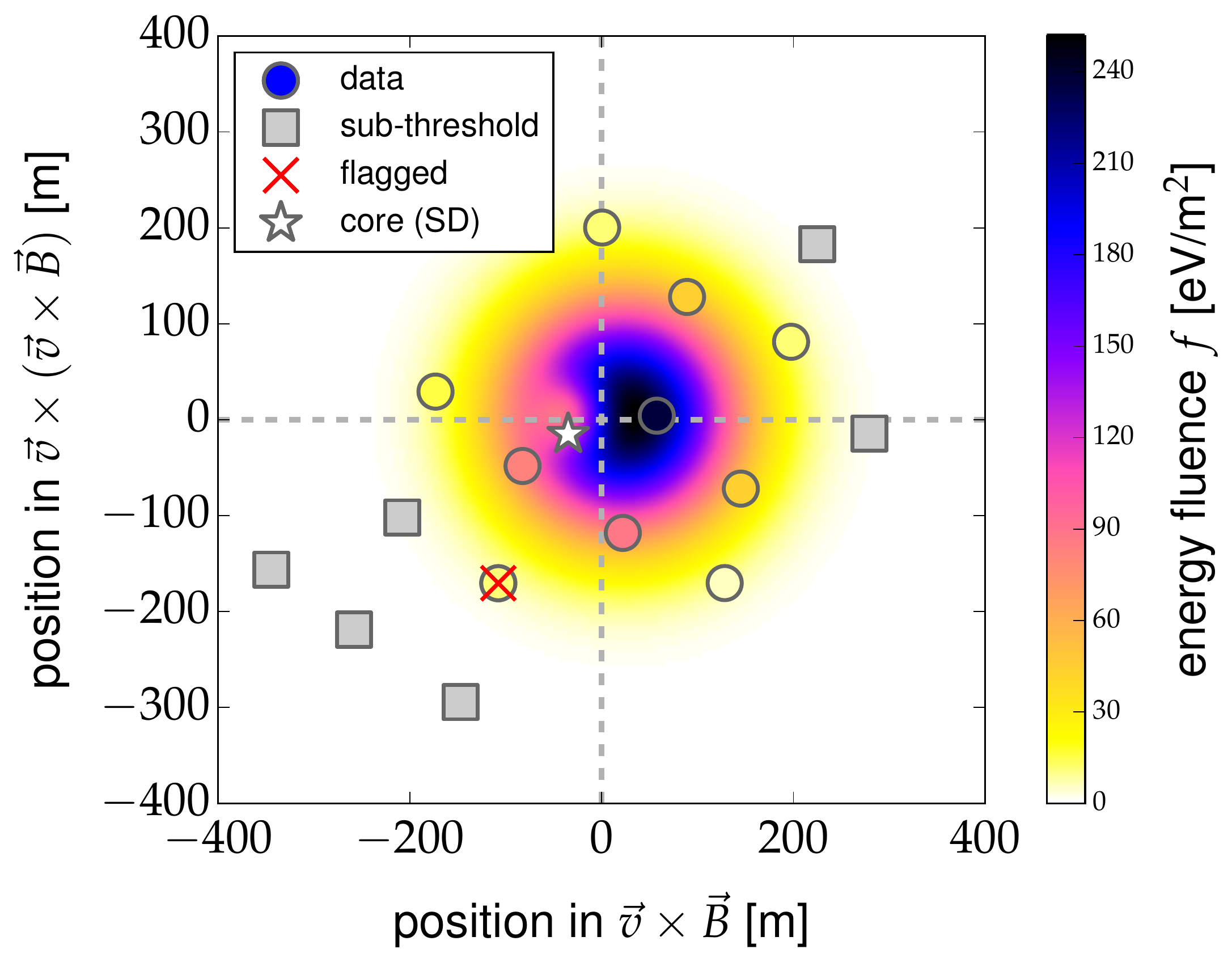}
\caption{Illustration of the energy fluence in the 
radio signals measured with individual AERA detectors and their fit with a two-dimensional lateral distribution 
function. Radio detectors with a detected signal 
(\emph{data}) as well as detectors with a signal below detection threshold 
(\emph{sub-threshold}) participate in the fit. Measurements exhibiting 
deviating polarisation characteristics are excluded (\emph{flagged}) 
to suppress transient radio-frequency interference.
The fit is performed in the shower plane, the x-axis being oriented along 
the direction of the Lorentz force for charged particles 
propagating along the shower axis $\vec{v}$ in the geomagnetic field 
$\vec{B}$. The best-fitting core position of the air shower is at the origin of the plot, 
slightly offset from the one reconstructed with the Auger surface 
detector (\emph{core (SD)}). The background-color illustrates the two-dimensional 
lateral distribution fit. Adapted from \citep{AERAEnergyPRD}.\label{fig:2dldf} }
\end{figure}

\begin{eqnarray}
 E_{30-80\,{\mathrm{MHz}}} &=& \left(15.8~ \pm 0.7\,\mathrm{(stat)} \pm 6.7 
\,\mathrm{(sys)}\right)\,\unit{MeV} \nonumber \\ &\times& \left(\sin\alpha 
\,\frac{E_\mathrm{CR}}{\unit[10^{18}]{eV}} \, \frac{B_\mathrm{Earth}}{\unit[0.24]{G}} \right)^2.
\end{eqnarray}
In other words, an extensive air shower with an energy of $10^{18}$~eV 
arriving perpendicular to a geomagnetic field with a strength of 
0.24~G radiates a total of 15.8~MeV in the form of radio signals in 
the frequency range from 30 to 80~MHz. This result should be directly 
comparable between different radio detectors (provided the air shower 
can evolve and radiate the bulk of its radio emission before it reaches 
the ground) and can thus be used for cross-calibration of detectors. 
The systematic uncertainty of the result is currently dominated by the 
uncertainty of the absolute energy scale of the Pierre Auger 
Observatory, which has been propagated from the fluorescence detector 
to the surface detector.

From the scatter of the energy reconstructed with the Auger 
surface detector and the radiation energy determined with AERA (Fig.\ 
\ref{fig:crosscalibration}), the resolution of the energy 
reconstructed from radio data has 
been determined. (The degree of correlation between the two quantities 
was estimated with a Monte Carlo simulation study.) For the high-quality subset of 
events measured in at least five AERA detectors, the radio energy 
resolution has been determined to 17\% \citep{AERAEnergyPRD}. This 
again illustrates the high resolution of radio energy measurements 
that has already been achieved today.

\subsection{Depth of shower maximum reconstruction} 
\label{sec:xmaxreconstruction}

In addition to the energy of the primary particle, the depth of shower 
maximum ($X_{\mathrm{max}}$) is a key quantity for the study of cosmic ray air 
showers. It is the parameter sensitive to the mass of the primary 
cosmic ray particle which is used in particular by  
fluorescence detector telescopes. Radio emission is sensitive to the 
depth of shower maximum, because the geometrical distance between the 
source and antenna directly shapes the radio emission arriving at the 
antenna, as discussed in some detail in section \ref{sec:distanceeffects}.

Sensitivity to \xmax in the radio signal was already presumed in the 
1970s \citep{AllanRefractive1971,HoughXmaxLDF} and has also been 
predicted by modern simulation studies \citep{HuegeUlrichEngel2008} (later confirmed on 
the basis of other calculations \citep{Konstantinov2009,deVriesEnergyXmax2010}).
In particular, the slope of the lateral distribution function provides information on the distance of the 
radio source: a far-away source (low \xmax value) produces a flatter LDF than 
a close-by source (high \xmax value).

\begin{figure}[!htb]
  \vspace{2mm}
  \centering
  \includegraphics[width=0.5\textwidth]{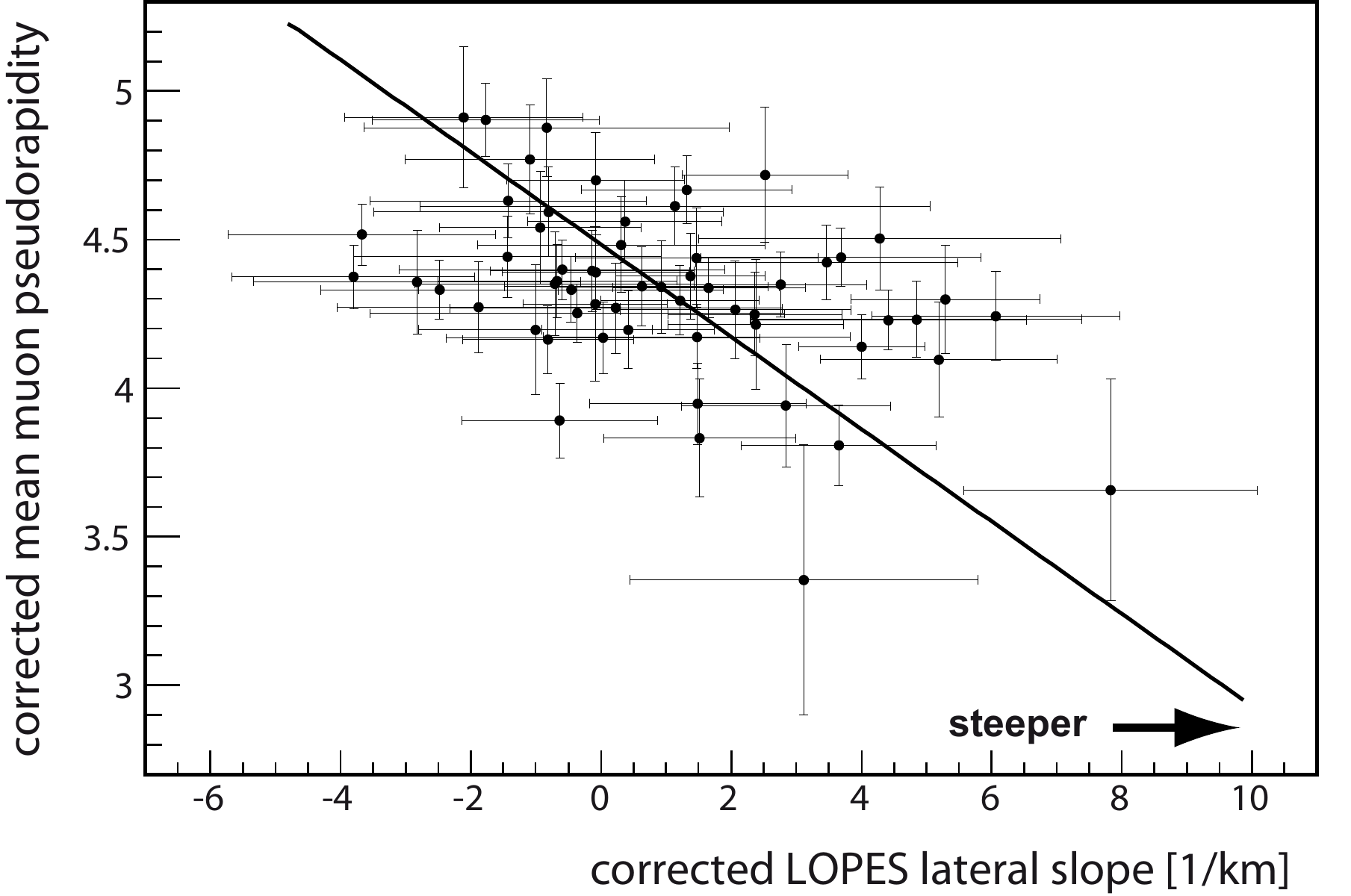}
  \caption{Correlation of the mean muon pseudorapidity as measured 
  with KASCADE-Grande and the slope of the lateral distribution 
  function as determined with LOPES. Higher muon pseudorapidities 
  (corresponding to larger production heights and thus showers 
  developing earlier in the atmosphere) are clearly associated 
  with flatter radio LDFs. Addapted from 
  \citep{ApelArteagaBaehren2012c}.}
  \label{fig:lopesmasses}
 \end{figure}

A first experimental proof that indeed the slope of the radio LDF probes the longitudinal air shower 
evolution was published by the LOPES experiment \citep{ApelArteagaBaehren2012c}. It could be shown 
that the mean muon pseudorapidity, a parameter related to the height 
of muon production and thus to the longitudinal air shower evolution, 
is correlated with the slope of the radio LDF (Fig.\ 
\ref{fig:lopesmasses}). 

\begin{figure*}[htb]
  \includegraphics[width=0.42\textwidth]{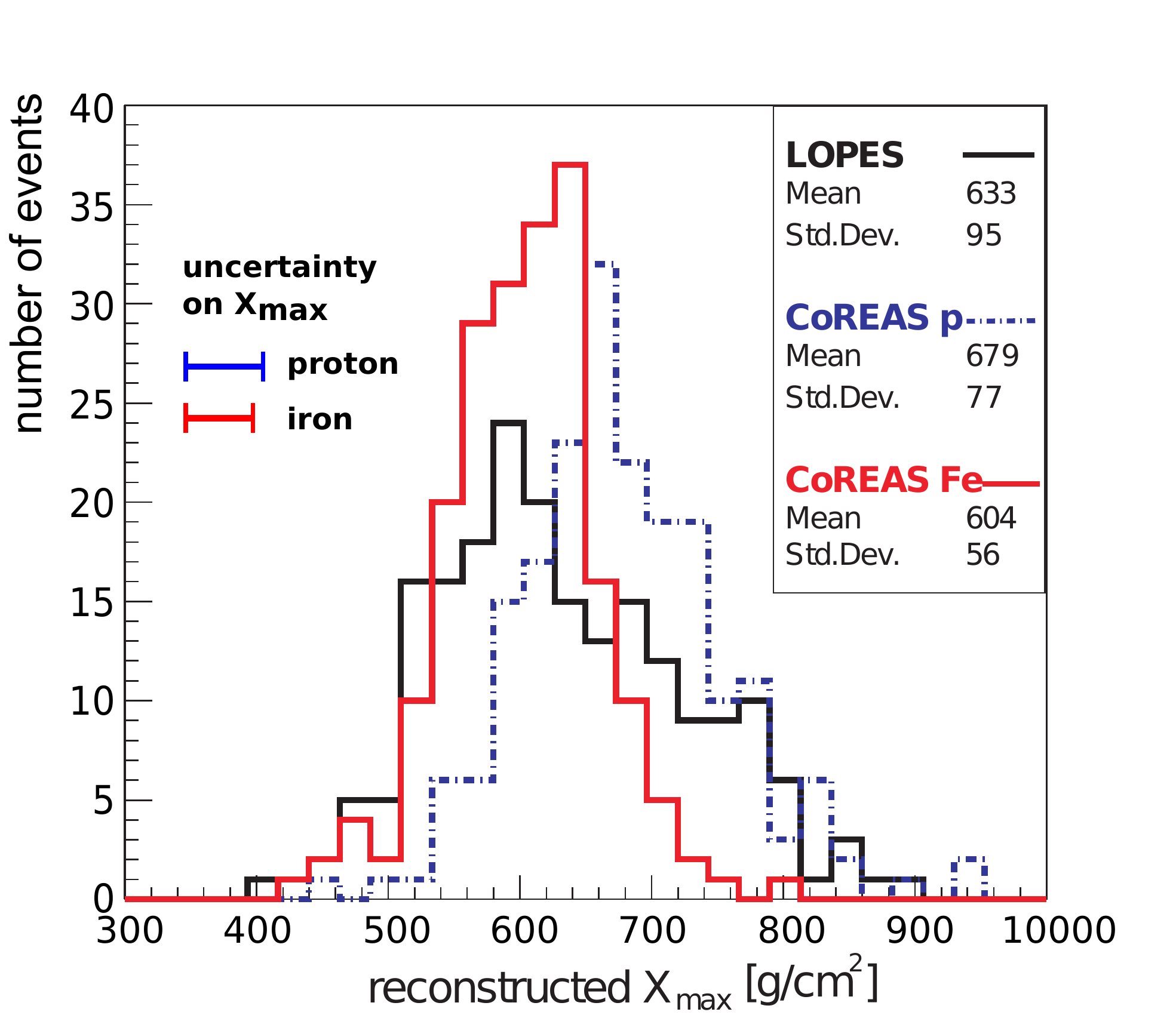}
  \includegraphics[width=0.59\textwidth,clip=true,trim=0cm 0cm 3cm 17.4cm]{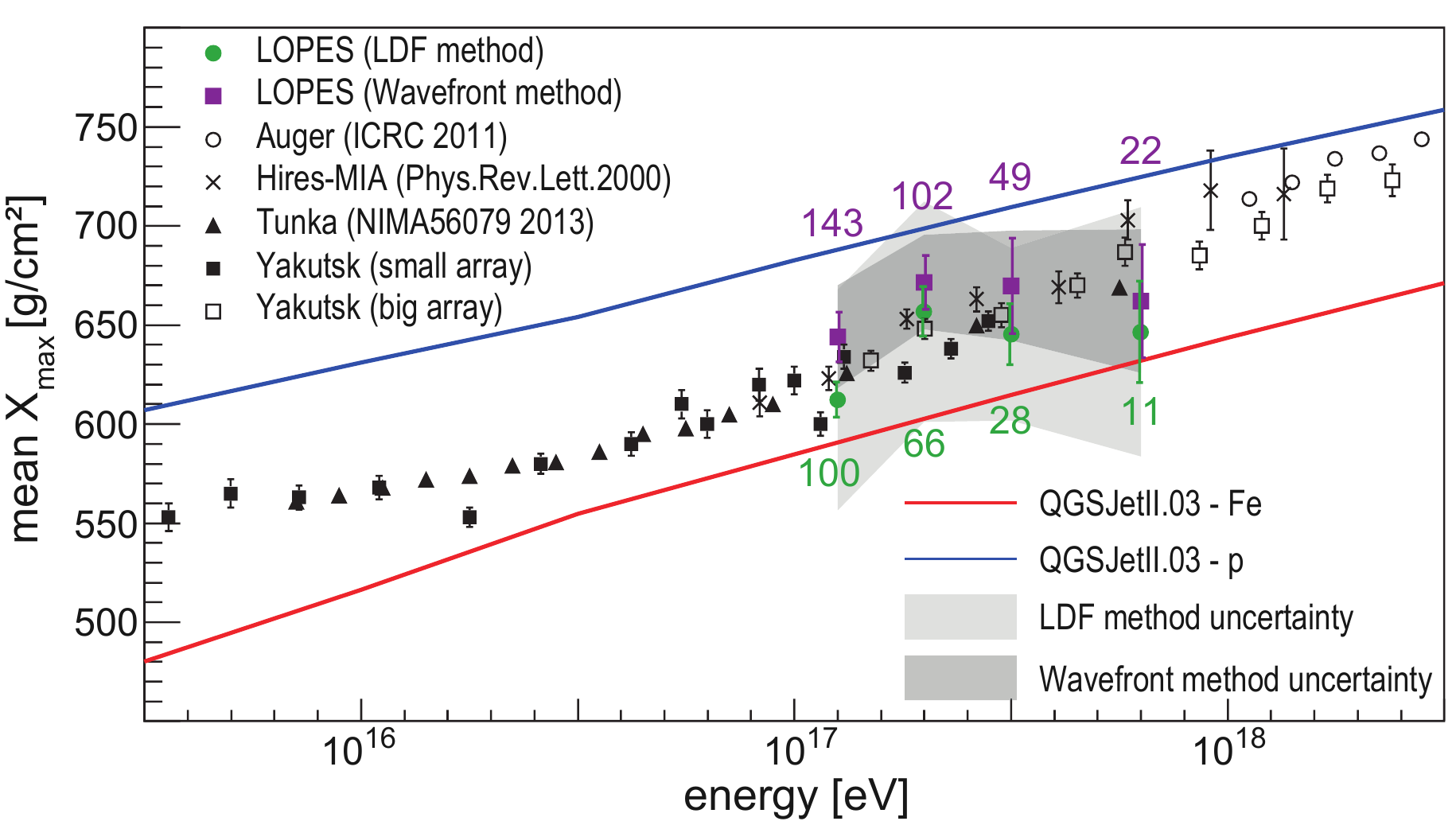}
  \caption{Left: Distribution of the \xmax values of air showers 
  determined from the slopes of the radio lateral distribution 
  functions measured with LOPES compared to the distributions expected 
  for CoREAS simulations of proton and iron primaries. Adapted from
  \citep{ApelArteagaBaehren2014}. Right: Mean 
  \xmax value as a function of energy as determined from the LDF and 
  wavefront \xmax analyses performed with LOPES. Adapted from
  \citep{SchroederLOPESECRS2014}.\label{fig:xmaxlopes}}
\end{figure*}

The next step in exploiting the radio LDF for \xmax determination has 
then been presented in \citep{ApelArteagaBaehren2014}. In this 
analysis, the authors first studied the relation 
between \xmax and the radio LDF slope on the basis of CoREAS 
simulations of LOPES events. This confirmed the previous predictions 
\citep{HuegeUlrichEngel2008} that the slope can be used to determine 
$X_{\mathrm{max}}$. The relation found on the basis of CoREAS simulations was then used to determine \xmax values 
from measured slope parameters (Fig.\ \ref{fig:xmaxlopes}). The 
analysis method achieved an uncertainty of $\approx 50$~g/cm${^2}$ and 
the overall systematic uncertainty of the result was $\approx 
90$~g/cm${^2}$, which is not competitive with fluorescence and 
Cherenkov light detectors which provide a resolution of $\sim 
20$~g/cm$^{2}$. Also, no independent measurement was available within 
LOPES to cross-check the validity of the determined \xmax values. Nevertheless,
this analysis can be seen as a proof of principle that such analyses 
are possible with radio detectors.

\begin{figure*}[htb]
  \includegraphics[width=0.32\textwidth]{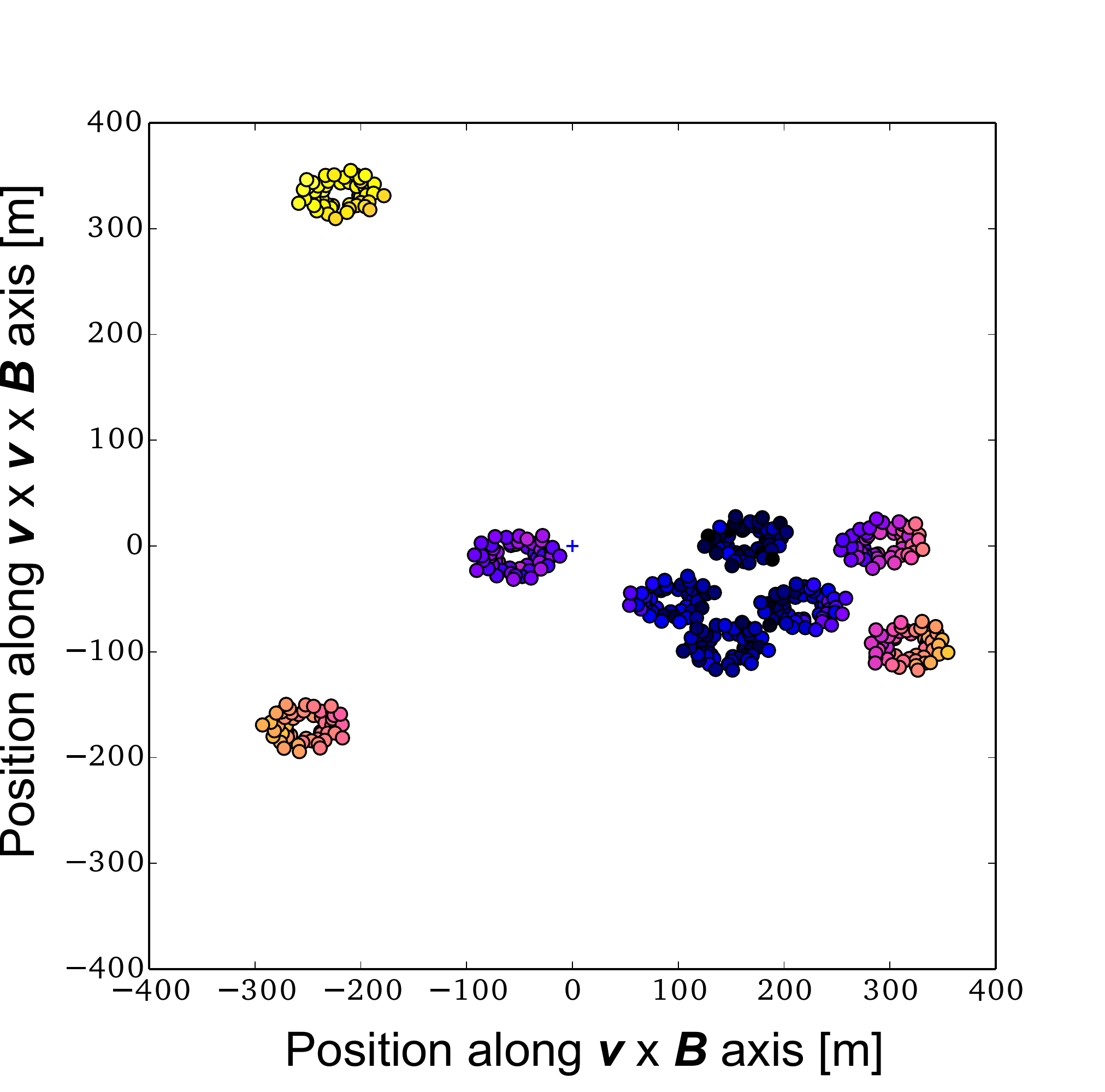}
  \includegraphics[clip=true,trim=50 0 45 0,width=0.35\textwidth]{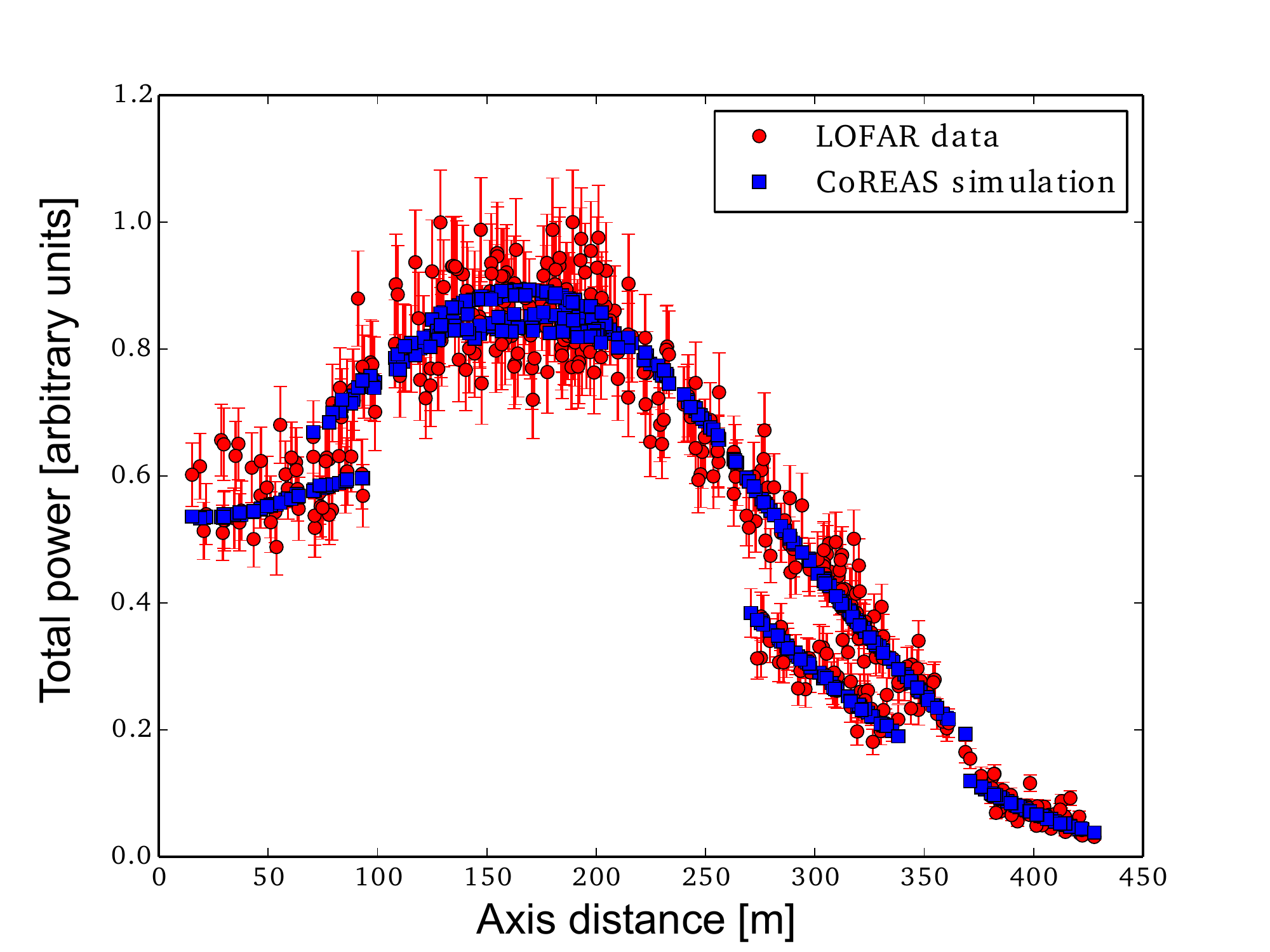}
  \includegraphics[width=0.31\textwidth]{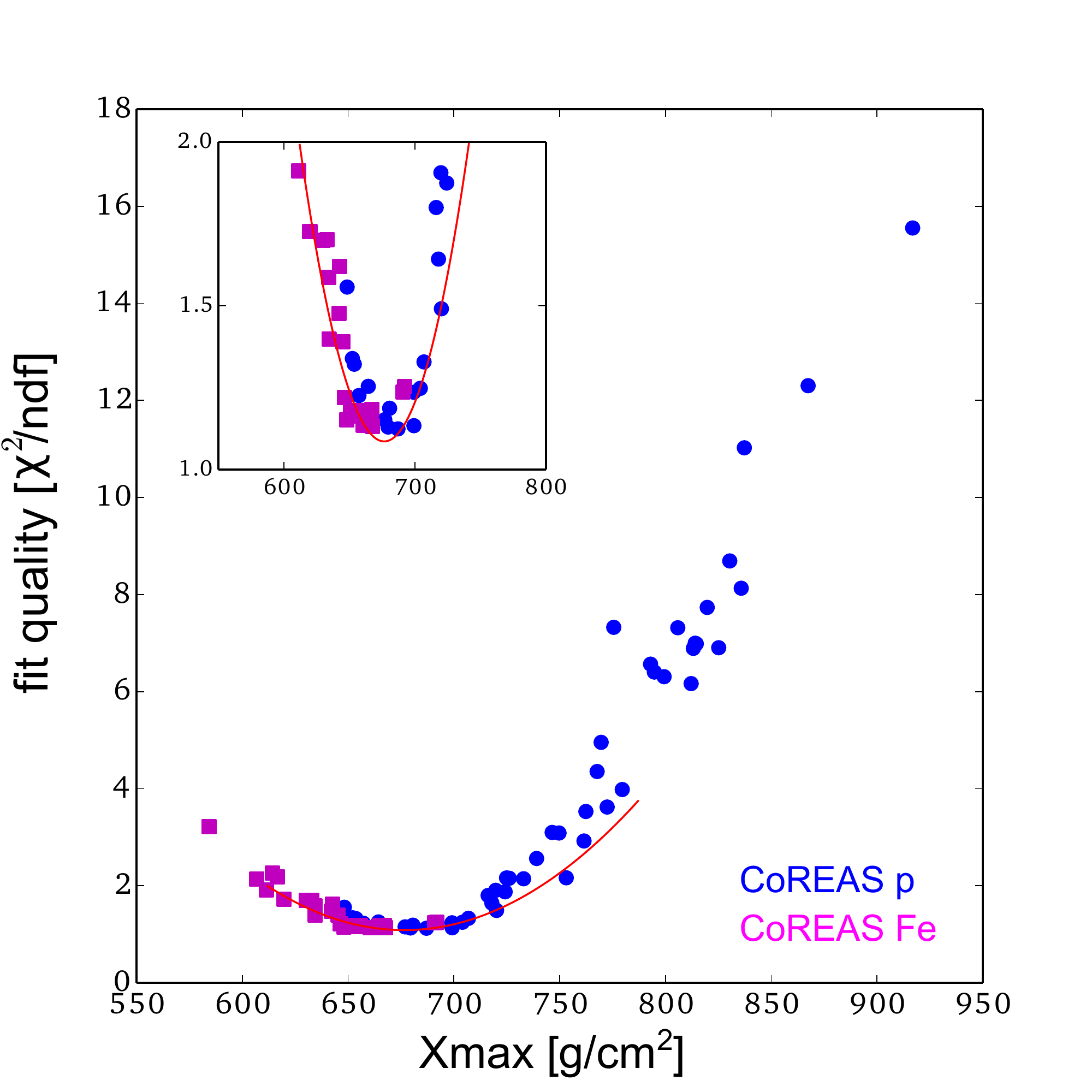}
  \caption{Left: Measured total power received at individual LOFAR 
  antennas (colored circles) in comparison with the two-dimensional 
  lateral distribution predicted by the best-fitting of a set of 
  CoREAS simulations (background-color) for a particular air shower 
  event. The depicted plane corresponds to the shower plane, defined 
  by the axes along the direction of the Lorentz force and the one 
  perpendicular to that. Middle: One-dimensional projection of the 
  two-dimensional lateral distribution. Right: Quality of the 
  agreement between the total power distribution measured with LOFAR 
  and the one predicted by different CoREAS simulations of the air 
  shower event. A clear correlation between the value of \xmax and the quality of the fit is 
  obvious. All diagrams adapted from \citep{LOFARXmaxMethod2014}.}\label{fig:xmaxlofar}
\end{figure*}

LOFAR has taken the approach of using the radio LDF for \xmax 
reconstruction to the next level. In their analysis 
\citep{LOFARXmaxMethod2014}, the complete complex two-dimensional distribution of the radio emission on the 
ground is used to identify the best-fitting out of a 
large number of CoREAS simulations for each given event. In addition to 
the LDF slope this approach also takes advantage of the 
asymmetries and the Cherenkov bump in the LDF. It turns out that the one parameter which governs the 
level of agreement between simulations and the LOFAR data is 
$X_{\mathrm{max}}$, which can be determined with a resolution of on average~17~
g/cm$^{2}$ (Fig. \ref{fig:xmaxlofar}). Again, an independent 
cross-check with \xmax information from an independent detector is not 
available in LOFAR, but the remarkable agreement between the 
simulations and data inspire confidence that the analysis is reliable 
and that radio detection can achieve an \xmax resolution competitive 
with other techniques.

\begin{figure}[!htb]
\centering
\includegraphics[width=0.48\textwidth]{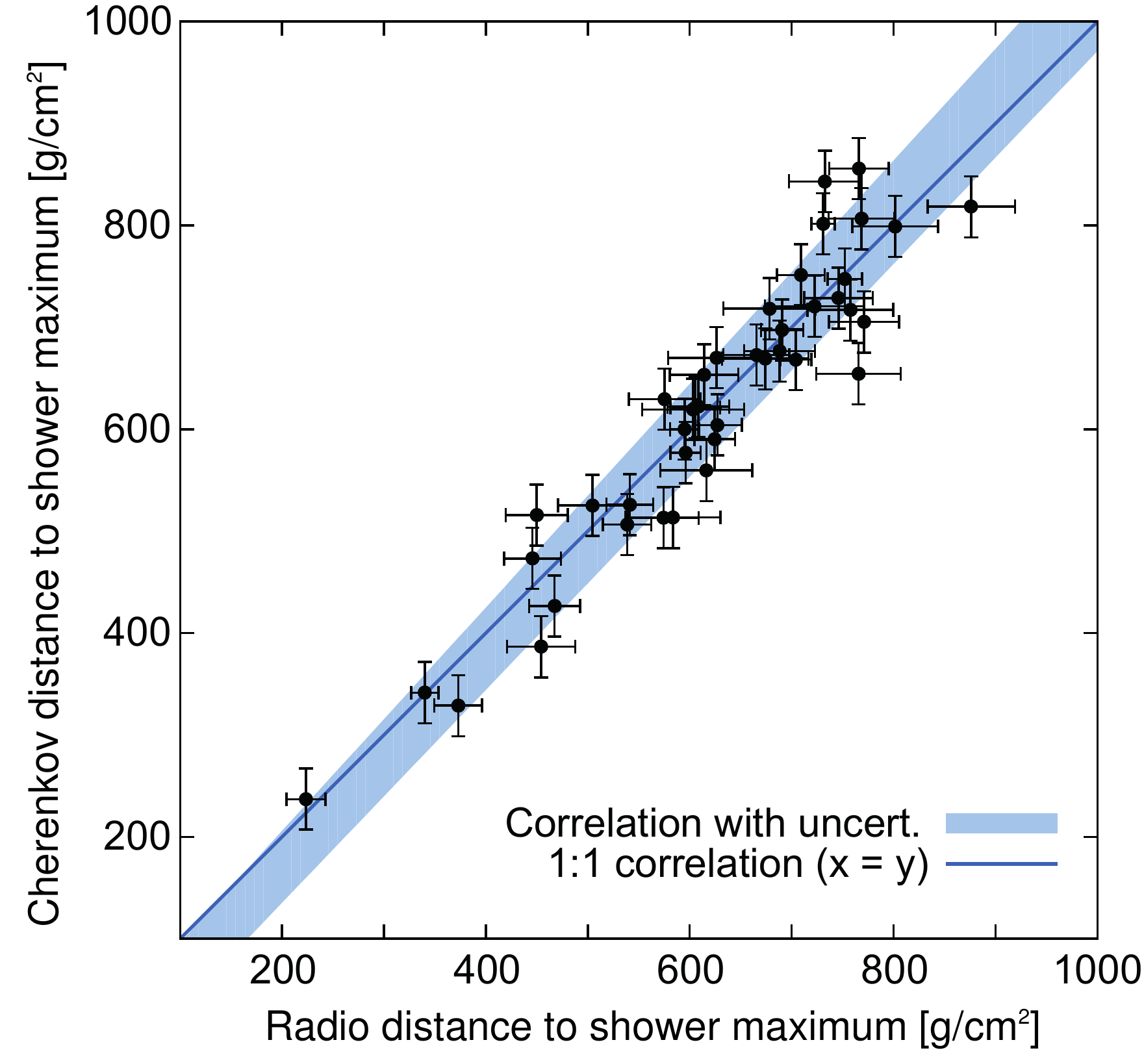}
\caption{Atmospheric depth between observer location and the depth of 
shower maximum as determined with the Tunka-Rex radio measurement and 
the Tunka-133 Cherenkov detectors. Adapted from \citep{TunkaRexCrossCalibration}.\label{fig:tunkarexxmax}}
\end{figure}

Finally, Tunka-Rex recently reported the first experimental comparison 
of an \xmax reconstruction using the slope of the radio LDF and 
the reconstruction with an independent detector, in this case the 
Cherenkov-light detectors of Tunka-133 \citep{TunkaRexCrossCalibration}. There 
is a very good correlation between the two reconstructions, as 
is shown in Fig.\ \ref{fig:tunkarexxmax}. The combined uncertainty of 
the two reconstructions currently amounts to $\sim 50$~g/cm$^{2}$, 
while the uncertainty of the Tunka-133 \xmax reconstruction alone is specified as 
28~g/cm$^{2}$. This first direct experimental proof of the \xmax sensitivity 
of radio measurements is another important milestone.

\begin{figure}[!htb]
  \includegraphics[clip=true,trim=0 0 0 430,width=0.55\textwidth]{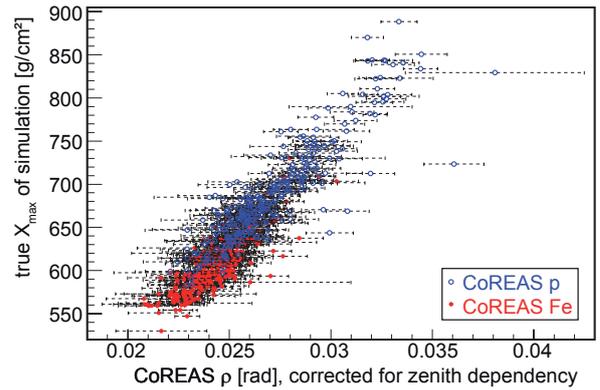}
  \caption{Correlation between the opening angle of the hyperbolic 
  wavefront $\rho$ and the depth of shower maximum as predicted by CoREAS 
  simulations for LOPES measurements. Adapted from \citep{LOPESWavefront2014}.}\label{fig:wavefrontxmax}
\end{figure}

The approaches discussed so far only used the radio LDF and thus the 
spatial distribution of signal amplitude or power, respectively. However, 
there are additional sensitivities of the radio signal to the source 
distance. As discussed in section \ref{sec:wavefront}, the radio 
wavefront is also sensitive to the distance of the radio source. A simulation study by the LOPES 
experiment \citep{LOPESWavefront2014} showed on the basis of CoREAS 
simulations of LOPES events that a clear correlation exists between 
the zenith-angle-corrected opening angle of the asymptotic cone of the 
hyperbolic wavefront and \xmax (Fig. \ref{fig:wavefrontxmax}). The 
method resolution in the absence of measurement uncertainties has been 
determined to be $\approx$~30~g/cm$^{2}$, but the overall
uncertainty in LOPES data was found to be 140~g/cm$^{2}$, probably limited by 
uncertainties in the determination of the pulse arrival times  
due to noise influence in the LOPES data. The method requires a high-quality timing calibration of the individual detector stations. 
The core position has significant impact in the analysis, which 
means that in principle it can also be determined in the course of a 
wavefront analysis (see section \ref{sec:LDFs}). As mentioned before, derived quantities such as the value of the opening 
angle depend on the exact definition of the signal arrival time 
(values determined with interferometric techniques are systematically 
different from those determined on the basis of the time of the pulse 
maximum, also remaining experimental characteristics in the 
deconvolved data can change the exact result). Therefore, comparisons between published results need to be 
performed with great care. The true potential lies in a combination of 
the LDF-based methods with a wavefront timing analysis (and polarisation information) to 
increase the accuracy of the radio-based \xmax measurement even 
further. The mean \xmax values as a function of energy determined with 
this approach are shown and compared with the values determined from 
the LDF analysis in Fig.\ \ref{fig:xmaxlopes}, right. Both methods 
agree within their (fairly large) systematic uncertainties.

A third possibility to access information on the source distance and 
thus \xmax is to study the radio pulse shape measured in individual 
antennas. Due to the geometrical time delays arising from the 
propagation of the radiating particles on the one hand and the radio 
emission on the other hand, the radio pulses become wider as the 
observer position moves away from the shower axis (with the exception 
of positions inside and near the Cherenkov ring where pulses are 
stretched and compressed, respectively). If this first-order effect can be corrected for, a second-order effect on 
the pulse width is given by the distance of the emission region from 
the ground.

Instead of measuring the pulse width, also the slope 
of the frequency spectrum can be used \citep{GrebeARENA2012}. (This 
relies only on the amplitude information, not the phases of the 
different frequency components.) Measurements of the frequency spectrum
of air shower radio emission have proven to be difficult, though, with the 
only published result from a ground-based array so far from the LOPES experiment 
\citep{NiglApelArteaga2008b}. One reason for this is that the air 
shower radio signal is localized in time, but spread in the frequency 
domain, leading to a lower signal-to-noise ratio in the determination 
of frequency components. Another reason is the required very good understanding of the 
frequency-dependent gain of the antenna used to measure the radio 
pulses. When trying to use frequency spectra in the determination of 
\xmax another complication is that the core position needs to be known 
rather precisely. On the other hand, the advantage of this method could be the use of single radio 
stations not necessarily requiring coincident detection in multiple 
radio detectors. Also, a combination with LDF and wavefront methods 
could be beneficial.

For the determination of \xmax from high-frequency emission please see 
the next subsection.

\subsection{High-frequency emission} \label{sec:highfreqresults}

\begin{figure*}[!htb]
  \centering
  \includegraphics[width=0.52\textwidth]{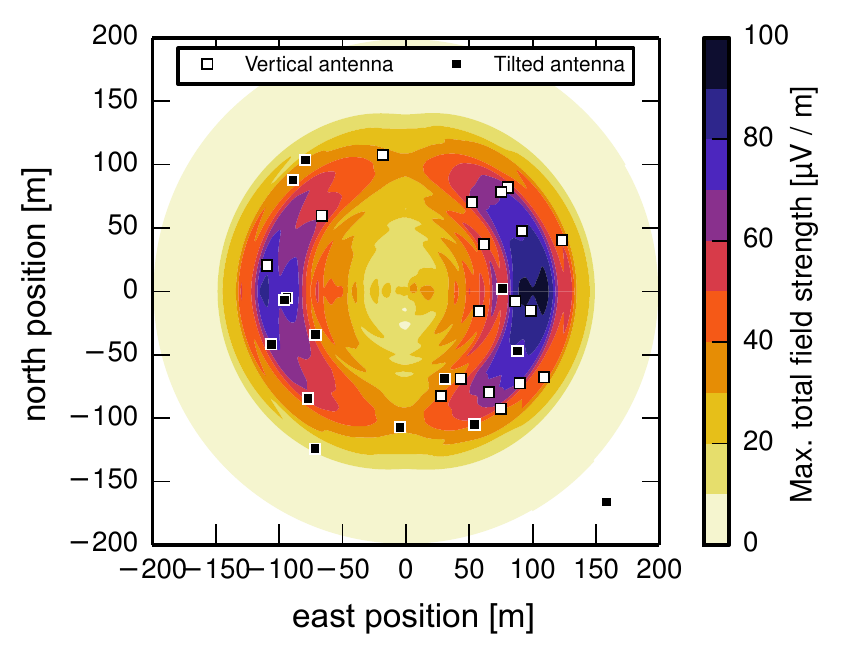}
  \hspace{3mm}
  \includegraphics[width=0.43\textwidth]{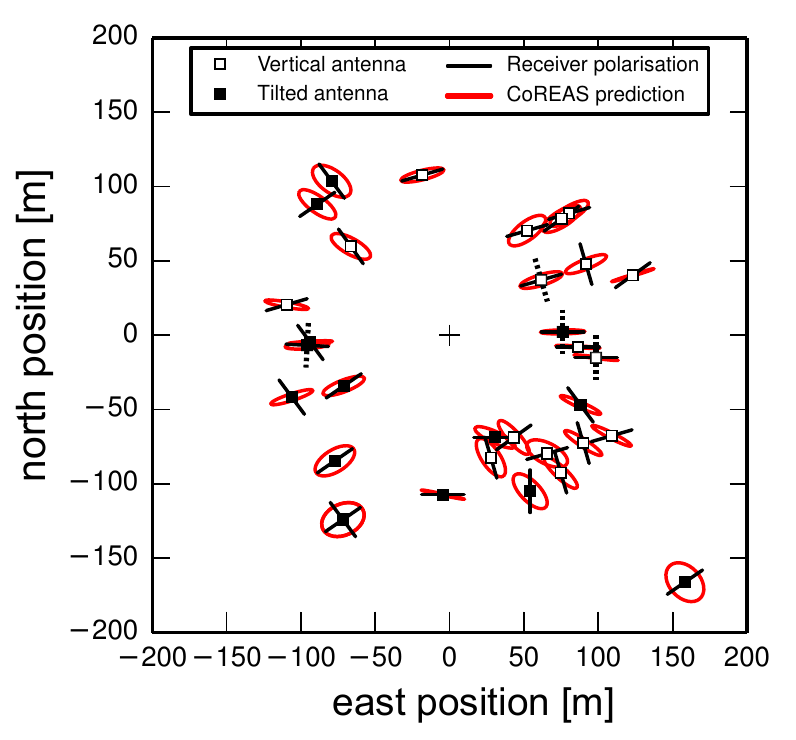}
  \caption{Left: Core positions of air showers measured in the 
  3.4-4.2~GHz band with CROME. All measured showers have their cores 
  at distances of $\sim 70-100$~m from the antenna. Their distribution 
  follows the prediction from CoREAS simulations, the  
  total field-strength of which is shown in the background-color. 
  Right: Signal polarisation measured with CROME in comparison to the 
  one predicted with CoREAS. Adapted from \citep{CROMEPRL}.}
  \label{fig:crome}
 \end{figure*}

Both EASIER \citep{GHzAtAuger} and CROME \citep{CROMEPRL} have reported successful detections of 
GHz radiation associated with air showers, and CROME has been 
able to study the characteristics of the radio emission in the 3.4 to 
4.2~GHz band in detail. There are clear indications that the radio emission is forward-beamed and 
polarised, which is not expected for ``molecular bremsstrahlung''. In 
fact, the CROME observations are compatible with the high-frequency 
radio emission expected from the interplay of geomagnetic and Askaryan 
radiation undergoing Cherenkov-like time-compression effects as described in 
section \ref{sec:cherenkov} and simulated here with CoREAS (Fig.\ 
\ref{fig:crome}). Even the changes in the emission pattern above 
2~GHz as predicted by CoREAS simulations, cf.\ section \ref{sec:additionalemissionaspects}, seem to be observed in the 
measurements (the lack of detected air showers along the north-south 
axis in Fig.\ \ref{fig:crome}, left).

\begin{figure}[!htb]
\centering
\includegraphics[width=0.48\textwidth,clip=true,trim=0cm 8cm 0cm 5cm]{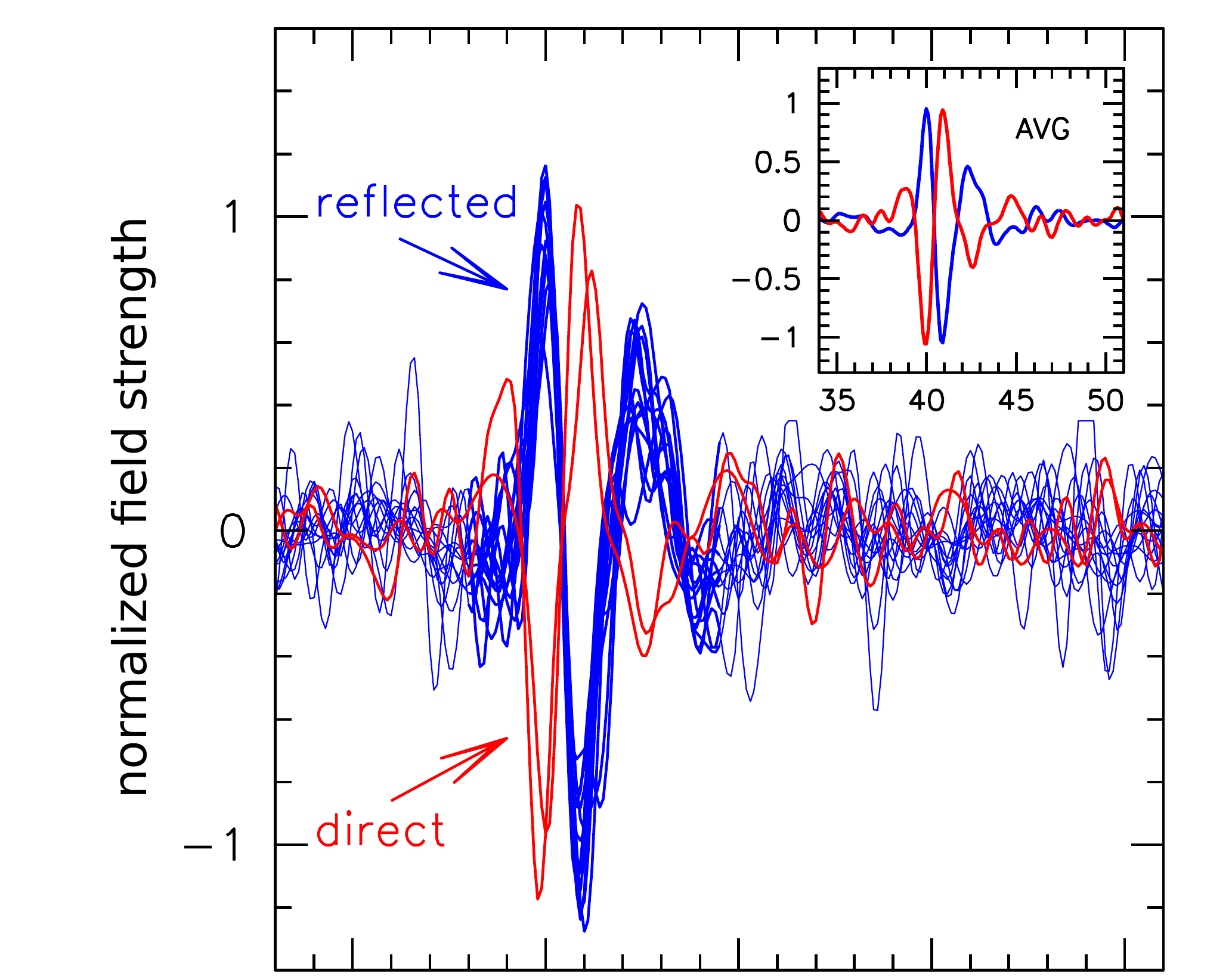}
\caption{Radio signals from 16 extensive air showers detected in the 
ANITA-I flight. 14 signals have been reflected off the antarctic ice, 
while 2 signals have been measured from Earth-skimming air showers. Adapted from 
\citep{HooverNamGorham2010}.\label{fig:anitaevents}}
\end{figure}

Further support to the validity of the simulations at high frequencies is given by the results from the 
ANITA-I experiment \citep{HooverNamGorham2010} which has observed 16 
impulsive radio signals in the frequency band from 200 to 1200~MHz. 
These pulses can be explained by geomagnetic and charge-excess radio 
emission along the Cherenkov angle, received either directly or reflected off the 
antarctic ice \citep{AlvarezMunizANITASims}. The fact that the radio 
pulses are very similar once normalized by amplitude (see Fig.\ 
\ref{fig:anitaevents}) illustrates that this high-frequency emission is 
indeed observed very near the Cherenkov angle. From the spectral index of the frequency 
spectrum, the off-axis angle (which must be close to the Cherenkov 
angle) can be determined, and once this is known the energy can be estimated on the basis of 
simulations \citep{ANITAEnergy}. The mean energy of the reflected ANITA 
events corresponds to 2.9~EeV, a value significantly lower than the previously published energy 
estimate of 15~EeV \citep{HooverNamGorham2010} based on assumptions 
for the radio emission not including refractive index effects and thus 
estimating a significantly wider radiation pattern. Note that such an 
energy analysis would not be possible at lower frequencies, as it 
would be essentially unknown at which off-axis angle a detected signal 
was recorded.

\begin{figure}[!htb]
\centering
\includegraphics[width=0.48\textwidth]{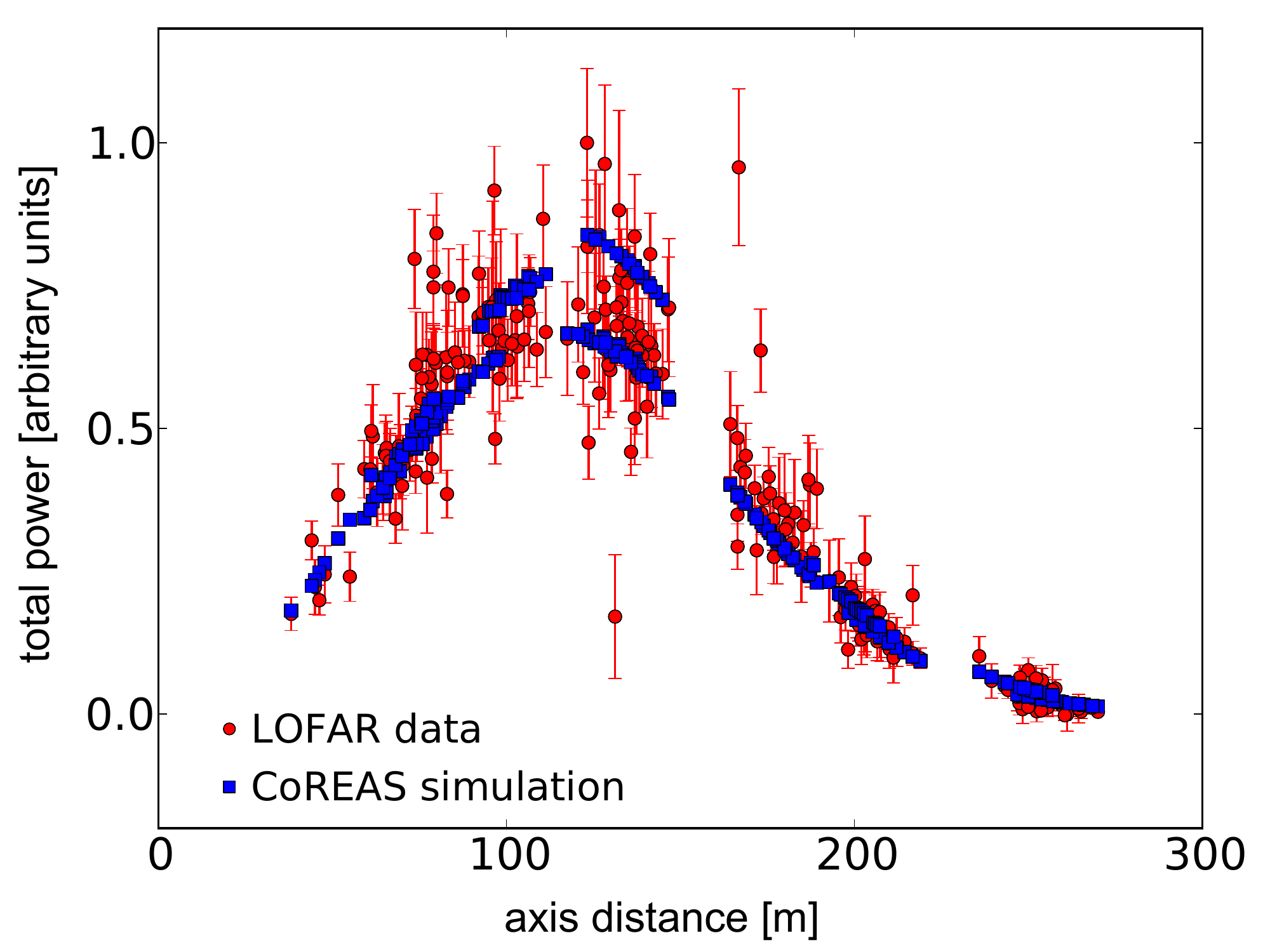}
\caption{A clear Cherenkov ring is visible in the measurements of 
extensive air showers with the LOFAR high-band antennas in the 
frequency range from 110 to 190~MHz. CoREAS simulations reproduce the 
measurements very precisely. Adapted from \citep{Nelles201511}.\label{fig:lofarhba}}
\end{figure}

Finally, LOFAR has observed air showers in the frequency range from 110 
to 190~MHz with their high-band mode. Again, these are in excellent 
agreement with CoREAS simulations, and clearly confirm the presence of a Cherenkov ring at higher 
frequencies (Fig.\ \ref{fig:lofarhba}).

An important fact to keep in mind is that high-frequency geomagnetic 
and charge-excess emission is only detectable at locations near the Cherenkov ring, where the radio 
emission from a large fraction of the shower evolution 
arrives simultaneously. To first order, this ring can be seen as the projection of a 
cone with an opening angle given by the Cherenkov angle starting from 
the shower maximum. Thus, the diameter of the Cherenkov ring is 
directly related to \xmax (and the atmospheric density at 
$X_{\mathrm{max}}$) and could also be used to determine the depth of shower 
maximum (see, e.g., \citep{deVriesCherenkovXmax}). However, successful detection requires a dense antenna 
spacing as only a very limited ring-like area is illuminated by the 
higher-frequency emission.

While there was great success in detecting and verifying high-frequency emission 
from the geomagnetic and charge-excess effects, time-compressed by the 
refractive index in the atmosphere, many searches for 
``molecular bremsstrahlung'' (see section 
\ref{sec:highfreqmeasurements}) were without success. Neither 
the air shower detectors CROME \citep{CROMEPRL}, EASIER, MIDAS and 
AMBER \citep{GHzAtAuger} nor the accelerator-based experiments AMY 
\citep{AMY}, MAYBE \citep{MAYBE} nor the Telescope Array Electron 
Light Source experiment \citep{ELSMBRIcrc2015} could find the emission at the level 
that was previously reported \citep{GorhamMBR}. Modern calculations 
also showed the emission to be much weaker than originally presumed 
(see ref.\ \citep{AlSamarai201526} and references therein). The 
prospects to use this isotropic radio emission from air showers for a 
``radio-fluorescence''-like detection approach thus seem very 
pessimistic today. Interestingly, one experiment reported 
somewhat forward-beamed microwave radiation at 11~GHz from a 95~keV electron 
beam in air \citep{ContiSartori}. The measured emission power scales 
linearly with the number of particles in the shower, and the authors 
interpret the radiation as arising from bremsstrahlung processes.
It is not entirely clear how this measurement relates to the negative searches for molecular 
bremsstrahlung reported above. It might not be at tension because of 
the more forward-beamed nature of the observed emission, in contrast 
to the presumed isotropy of molecular bremsstrahlung radiation.

\subsection{Influence of thunderstorms}

It has been known since the 1970s that the radio emission from 
extensive air showers can be strongly influenced by atmospheric electric fields 
\citep{Mandolesi19741431}. (In fact, the fear that radio measurements 
of extensive air showers could be unpredictable because they heavily 
rely on the state of unknown atmospheric electric fields was one of 
the reasons why activities in the field ceased in the 1970s.) Modern 
measurements with LOPES confirmed this influence 
\citep{BuitinkApelAsch2006}, finding amplified 
radio emission from air showers measured in thunderstorm conditions.
They also showed, however, that in fair weather and even rainy conditions the radio 
signal is unaffected. This means that, depending on the rate of 
thunderstorms at the location of a given experiment, reliable radio 
measurements are possible with duty cycles of more than $90-95$\%.

A simulation on the basis of (outdated) REAS2 simulations 
\citep{BuitinkHuegeFalcke2010} confirmed that 
air showers can be influenced significantly by strong electric fields 
and that also the radio emission can be strongly changed. This means 
that radio emission from air showers in thunderstorms carries 
information on the electric fields that they propagated through. 
LOPES, however, had too few antennas and was too imprecise to actually exploit 
this information practically.

\begin{figure}[!htb]
\centering
\includegraphics[width=\linewidth]{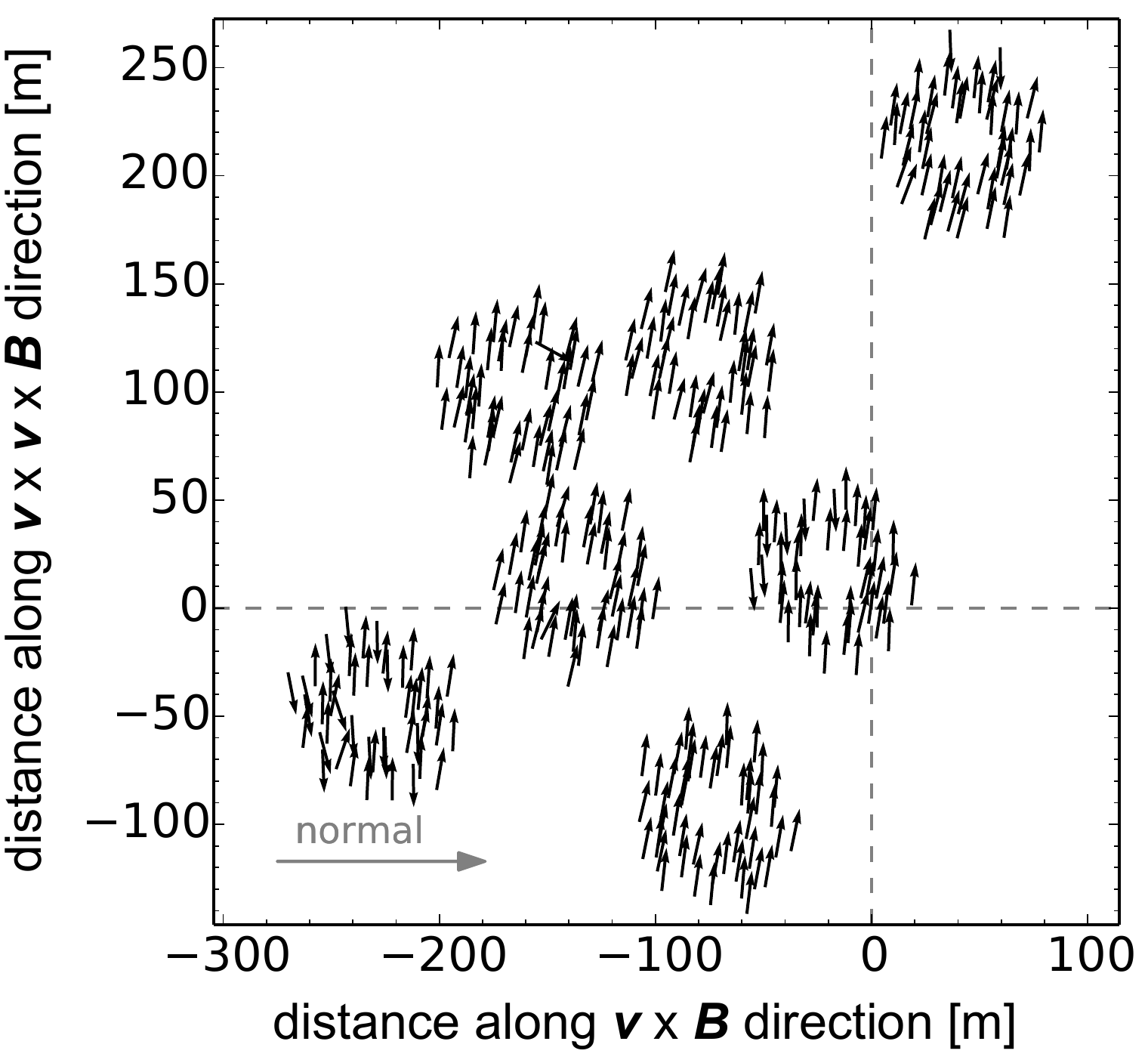}
\caption{LOFAR has measured air shower radio emission during 
thunderstorms which exhibit polarisation characteristics vastly 
different from the expectation of the geomagnetic emission mechanism 
(indicated by the arrow marked ``normal'') and its small modifications 
due to the charge-excess emission. Adapted from 
\citep{LOFARThunderstorms2015}.\label{fig:LOFARthunder}}
\end{figure}

\begin{figure}[!htb]
\centering
\includegraphics[width=0.5\textwidth]{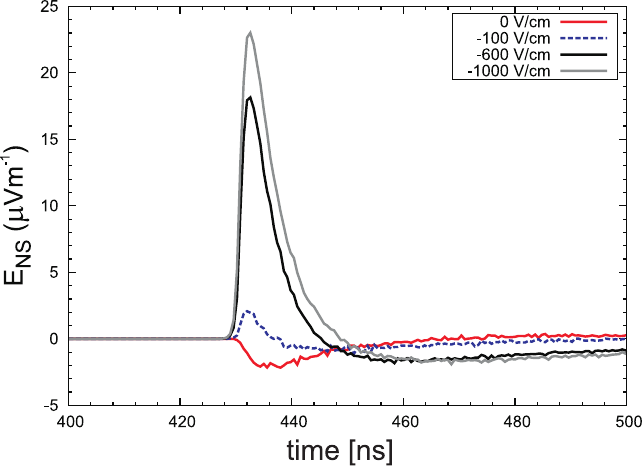}
\caption{CoREAS simulation of the north-south electric field component 
for an observer 250~m north of the core of a 30$^{\circ}$ inclined 
10$^{16}$ eV air shower in the presence of various vertical atmospheric 
electric fields. Fair-weather electric fields have values $\lesssim 3$~ 
V/cm. Electric field components perpendicular to the air shower axis influence 
the geomagnetic emission, while electric field components along the 
shower axis influence the charge-excess radiation.
Adapted from \citep{HuegeARENA2012b}.} \label{fig:efieldeffects}
\end{figure}

\begin{figure*}
\centering
\includegraphics[width=\textwidth]{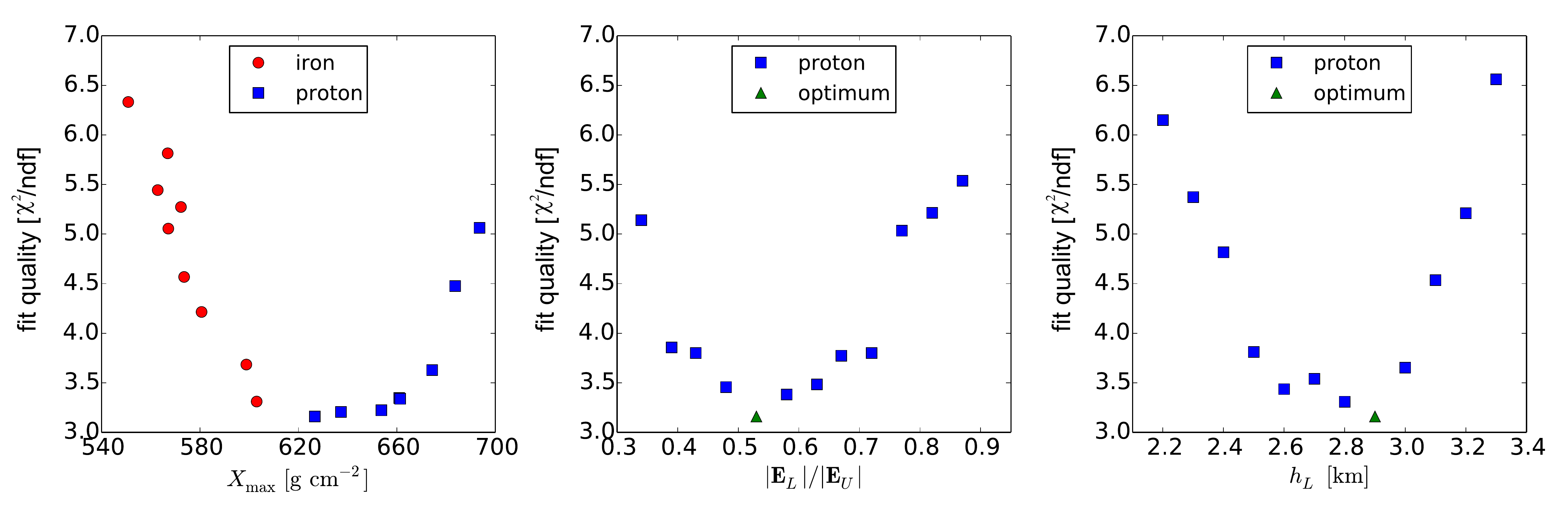}
\caption{Agreement between CoREAS simulations including atmospheric 
electric fields and a LOFAR air shower measurement recorded during 
thunderstorm conditions when varying various parameters of the 
simulations. Left: Variation of $X_{\mathrm{max}}$. Middle: Variation of the 
relative strength of the atmospheric electric field in the lower and 
upper layers. Right: Height at which the lower layer ends and upper 
layer starts. A specific set of these simulation parameters provided the best description 
of the measurement. Adapted from \citep{LOFARThunderstorms2015}.\label{fig:LOFARThunderchi2}}
\end{figure*}

This has changed very recently with a study performed by LOFAR 
\citep{LOFARThunderstorms2015}. LOFAR had measured several air showers 
which could not be reproduced with CoREAS simulations, in strong 
contrast to the majority of events that can be described very well. It 
turned out that many of these events were recorded during thunderstorm 
activity within 150~km of LOFAR. In particular, the polarisation 
(Fig.\ \ref{fig:LOFARthunder}) and amplitude distribution of these radio measurements were significantly 
different from those of air showers during fair weather. The authors 
used CoREAS, which is able to simulate the influence of electric 
fields on extensive air showers (Fig.\ \ref{fig:efieldeffects}) to 
probe possible atmospheric electric field configurations (Fig.\ 
\ref{fig:LOFARThunderchi2}) that rotate
the electric field vector towards the direction observed in the 
measurements. A good agreement with both the measured polarisation and 
amplitude distribution could be achieved with two oppositely oriented 
atmospheric electric field layers of different amplitudes at heights from ground to 2.9~km and 
from 2.9~km to 8~km. While it remains to be seen how reliable the 
atmospheric electric field information is that can be extracted from air 
shower measurements, the prospect of measuring atmospheric electric 
fields {\it in situ} using radio detection of cosmic ray showers is very 
promising and has received a lot of media attention.

In addition to probing electric fields in thunderclouds, radio 
detectors also have the potential to study possible connections between 
lightning initiation and extensive air showers, as they can 
record both the air shower radio pulse and the process of lightning 
initiation, which produces characteristic radio signals. Such 
connections have been long presumed \citep{Dwyer2014147}. In particular,
the scenario of a ``runaway breakdown'' \citep{Gurevich2013} has received
significant attention. Recent results indicating that indeed a combination of macroscopic ice 
particles in thunderclouds in combination with seed electrons from 
extensive air showers can initiate lightning at atmospheric electric
field strengths observed in nature \citep{DubinovaLightning}
provide strong motivation to intensify research in this direction.

\subsection{Determination of an energy spectrum of cosmic rays}

At this stage, the attentive reader might have wondered why no cosmic ray energy 
spectrum from radio measurements of extensive air showers has yet been discussed.
The reason is that for the determination of an energy spectrum,
the acceptance of the detector has to be known very accurately, and in 
radio detection, an accurate determination of the detector acceptance 
is rather challenging. First, the complication arises
that the detection threshold and thus the detector acceptance is strongly 
dependent on the air shower arrival direction, as the geomagnetic angle  
influences the radio emission strength. However, as we have discussed in 
great depth, the physics of the radio emission has by now been 
well-understood, so that the direction dependence can be modelled with 
confidence. Another problem arises from the 
time-variation of the radio background. The Galactic noise is 
well-known, including its time-variation, and its impact on the detection 
threshold can thus be taken into account reliably. If other sources 
of noise are significant, however, they have to be monitored in detail 
(continuous noise with a periodic trigger, transient noise with a 
pass-through trigger) so that their influence can be quantified.

\begin{figure}[!htb]
\centering
\includegraphics[width=0.48\textwidth]{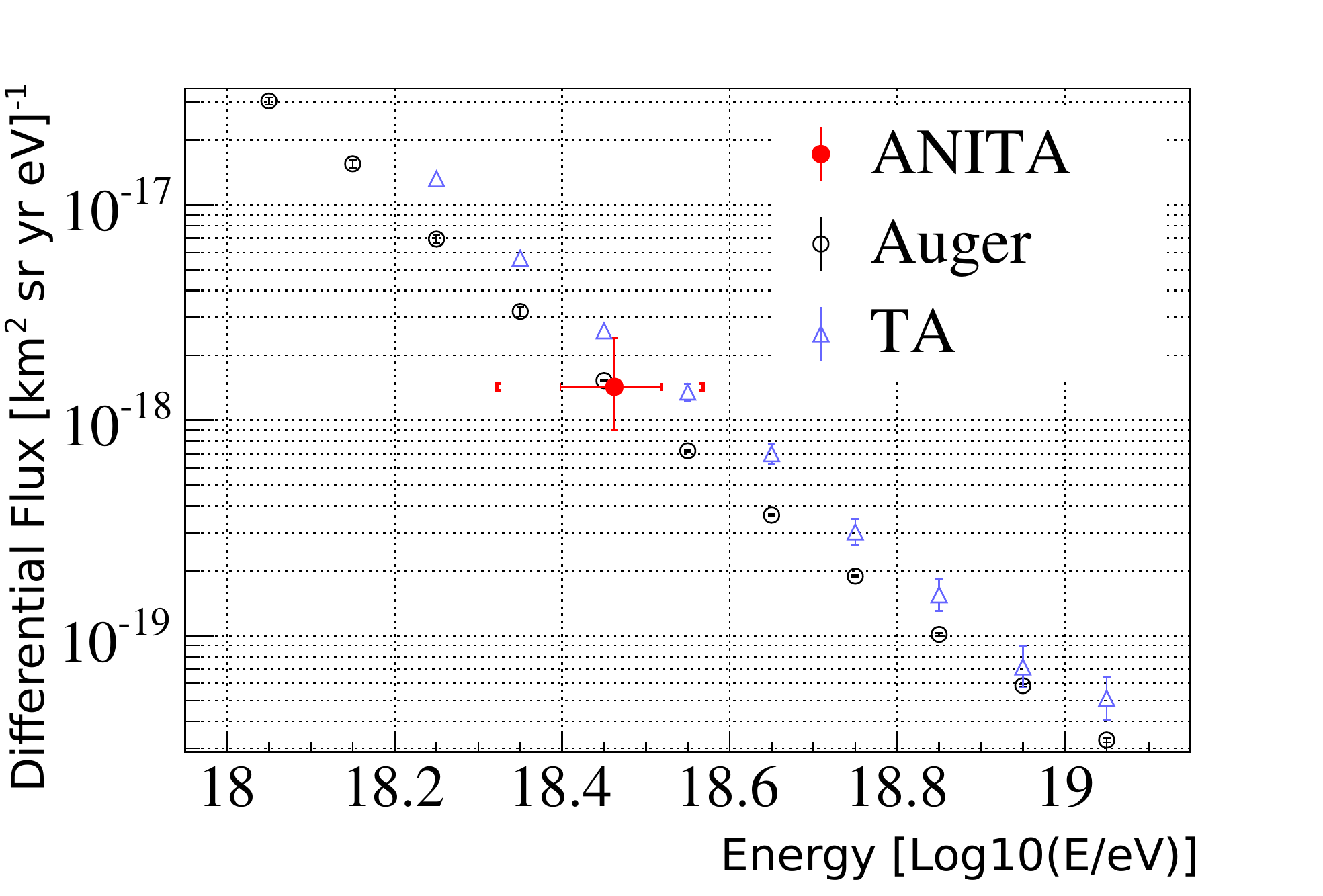}
\caption{The cosmic ray flux determined from the 14 events measured 
with the ANITA-I balloon flight. Adapted from \citep{ANITAEnergy}.\label{fig:anitaflux}}
\end{figure}

This discussion illustrates the difficulties in the determination of a cosmic ray 
energy spectrum from radio measurements. We stress, though, that while 
the determination of an energy spectrum requires significant effort, no 
principle problems are known to exist. Threshold-effects and the resulting 
detector acceptance can be determined precisely by Monte Carlo 
studies, which in fact very recently has been achieved by the ANITA 
collaboration for their first flight \citep{ANITAEnergy}. The 
determined flux at the mean event energy of 2.9~EeV is in good agreement 
with the one measured by the Pierre Auger Observatory and the 
Telescope Array, as is shown in Fig.\ \ref{fig:anitaflux}. This 
demonstrates that flux measurements with radio detectors are feasible, 
and should encourage other collaborations to perform similar analyses 
on their radio data.

\subsection{Measurements of the three-dimensional electric field vector}

When radio detection of extensive air showers with digital techniques 
began in the early 2000s, the detectors only sampled the electric field 
with single-polarised antennas at any given 
location (typically linearly polarised, except for CODALEMA-1 which used circularly polarised 
antennas). While the information gathered with this approach is of 
course valuable, it is nevertheless incomplete --- it is not possible 
to reconstruct the electric field at the measurement locations from 
this information. To do so, at least a second component of the 
electric field vector needs to be measured, which is why the 
experiments soon moved towards dual-polarised antennas. The typical 
scheme was to measure the north-south and east-west linear 
polarisations, i.e., the projection of the three-dimensional electric 
field vector onto the ground plane. As electromagnetic waves in the 
atmosphere are transverse waves and thus their electric field is 
polarised perpendicular to the propagation direction, this projection 
in combination with the arrival direction could be used to reconstruct 
the electric field vector.

This works well when the electric field vector is oriented such that 
the projected electric field is sizable, which is in particular 
fulfilled when the zenith angle of the air shower is small. For 
inclined air showers with zenith angles larger than 60$^{\circ}$, 
however, the electric field can have a sizable component in the 
vertical polarisation. This component would thus not be accessible by 
the horizontally aligned antennas, increasing the detection threshold 
significantly.

This reasoning prompted the LOPES-3D experiment with its tripole 
measurements of the electric field \citep{ApelArteagaBaehren2012a}. 
The results of the experiment, unfortunately, were somewhat 
inconclusive \citep{HuberARENA2014}. In general, anthropic noise was 
much more present in the vertical component of the electric field. 
This was expected, but in the end made it very difficult to exploit 
the additional vertical measurement with a clear benefit. One lesson 
learnt from LOPES-3D is that the added cost of instrumenting a vertical 
detection channel should only be undertaken after a survey of the 
transient noise environment in the vertical component at the 
designated detection site. If the noise is much stronger than in the 
horizontal components, instrumentation of a third channel might not be 
justified.


\section{Future directions}

Here, we discuss possible and proposed applications of the radio 
detection technique in future applications and try to assess their 
potential.

\subsection{Determination of the energy scale of cosmic rays} 
\label{sec:energyscale}

As discussed in section \ref{sec:energyreconstruction},
radio emission allows a calorimetric determination of the energy in 
the electromagnetic cascade of extensive air showers, comparable to 
fluorescence detection. To derive the energy of the primary cosmic particle, 
the fraction of energy not going into the electromagnetic cascade (and thus not leading to radio 
or fluorescence light emission) needs to be estimated for both 
techniques. While this fraction does depend on models of hadronic interactions, 
data-driven techniques have been developed \citep{ThePierreAuger:2013eja} wich reduce the systematic 
uncertainty arising from this correction to well below 5\% 
\citep{EnergyScaleICRC2013}. In contrast to fluorescence detection, 
however, radio signals do not undergo any significant scattering or 
absorption in the atmosphere. Also, they can be observed with relatively 
simple detection setups. Hence, there is high potential for using radio 
detectors as an excellent tool to cross-calibrate cosmic-ray experiments.
In particular, the determination of the energy in the radio signal by the 
Pierre Auger Collaboration \citep{AERAEnergyPRD} provides a 
transparent way to compare measurements at different detectors. It 
seems reasonable to deploy a small radio detector at 
every future cosmic ray detector for the sole reason of a precise
cross-calibration. 

In addition, as discussed in section \ref{sec:microscopicmodels}, the 
radio signal from extensive air showers can be calculated with 
first-principle calculations. The accurate description of the 
electromagnetic cascade of an air shower plus a formalism implementing 
a discretized calculation of classical electrodynamics suffice to 
calculate the radio emission on an absolute scale. This is markedly 
different from the complex situation of fluorescence detection, in 
which the ``fluorescence yield'' has been a limiting source of 
systematic uncertainty for a long time (today, the fluorescence yield plays only 
a minor role \cite{Ave:2012ifa}; signal propagation in the atmosphere, on the other hand, remains 
a challenging complication requiring very detailed atmospheric monitoring). The accurate absolute calibration of radio 
detectors will be the biggest challenge to overcome in this context, but there are 
many approaches that can be used to improve on the already good 
absolute calibration of today's experiments. For example, antennas can 
be optimized for independence from ground conditions, by design or by 
providing an appropriate ground shield (wire mesh, ...). External 
calibration sources can be improved further, and the cross-calibration 
with the Galactic noise, a well-understood, always-available 
calibration source, can be refined further. Galactic noise can even be 
used to monitor possible drifts of the absolute calibration 
effortlessly, although those are not expected to be important as radio detectors 
do not rely on hardware prone to aging effects such as 
photo-multiplier-tubes in the first place. First-principle calculations can thus be used 
in combination with a well-calibrated radio detector to independently check the 
absolute energy scale of cosmic ray detectors.

Both the cross-calibration and independent energy scale determination 
can be realized in the near future, and they will have a significant 
impact on cosmic ray physics beyond the field of radio detection.

\subsection{Large-scale measurements of inclined air showers} 
\label{sec:inclinedshowers}

From the beginning the hope was that radio detection could be applied 
on the largest experimental scales. This seemed particularly 
attractive as radio antennas can be built fairly cheaply (the SALLA 
antenna used in Tunka-Rex can be built for 500 USD including all 
analog electronics \citep{SchroederUHECR2014}) and electronics 
get cheaper with each new generation. Also, the near-100\% duty cycle 
of radio detectors marks a major advantage with respect to optical 
techniques such as fluorescence and Cherenkov-light detection with 
their typical duty cycles of $\sim 10$\%. However, it has become 
increasingly clear in recent years that the footprint of the radio 
emission is generally small, and stays small even as the energy of the 
primary particle is increased. The reason for this is of geometric 
nature, the forward-beamed emission is radiated in a narrow cone and 
thus typically subtends only regions with a few hundred metres in 
diameter. Radio detection arrays thus have to be fairly dense, with 
antenna spacings of order 300~metres or less to ensure coincident radio 
detection in several antennas. Instrumenting very large areas thus 
requires a very high number of radio detectors to be deployed. Even with cheap radio detectors, there 
is the problem of the cost of deployment and supporting infrastructure, 
in particular for power harvesting and communications. The current concepts can thus not 
easily be scaled to areas of hundreds of km$^{2}$. A radical rethinking of 
the design might still provide a solution, though (think pouring 
smartphone-like detectors out of an airplane for deployment ...).

The situation changes markably for inclined air showers, as was already 
illustrated in Fig.\ \ref{fig:inclined}. Showers with zenith angles 
larger than 70$^{\circ}$ illuminate areas of several km$^{2}$. 
The main reason is that the radio source is geometrically much 
further away, so that the same amount of energy is distributed over a 
much larger area on the ground. As long as the signal is still 
detectable above the Galactic noise level, air showers 
can thus be detected in coincidence with antennas on very sparse 
grids of more than a kilometre. The threshold for detecting pulses in 
the presence of Galactic noise is predicted from simulations to be 
around $~\sim 5 \cdot 10^{17}$~eV \citep{HuegeIcrc2013CoREAS}. 
Interferometric analysis approaches could lower this threshold 
considerably, since the radio pulses for very inclined air showers 
should in fact be very similar over many antennas (unlike for 
near-vertical air showers). The ``classical'' far-field interferometry 
as already exploited by LOPES is thus expected to work well for 
inclined air showers.

The intrinsic sensitivity for determination of \xmax will likely be weak 
for very inclined air showers: the geometrically further away the 
source, the smaller the relative changes in geometry. However, coincident detection of radio 
emission and particles can yield composition information on the basis of the ratio between the
electromagnetic and muonic cascade (particle detectors will register 
mostly muons for very inclined showers, the electromagnetic cascade 
has virtually died out when the shower reaches the ground). In fact, radio detection is the 
only technique allowing a measurement of the electromagnetic component 
of very inclined extensive air showers with a favourable acceptance. There is thus 
strong potential in using radio detection specifically for the 
detection of inclined air showers, a prospect that was actually 
already realized early on \citep{GoussetRavelRoy2004}.

The long-standing idea to detect near-horizontal neutrino-induced air 
showers with radio detection, unfortunately, does not seem very 
promising. The radio source will be close to the ground (this is the criterion with which 
a cosmic ray is excluded), but that means that the radio emission 
footprint will again be small and thus difficult to detect. The situation
may be more favorable if tau-lepton-induced air showers from Earth-skimming 
(upgoing) neutrinos or from neutrinos interacting in mountains are targeted,
as is the goal of the proposed GRAND project \citep{GRANDIcrc2015}.

In summary: While it currently seems unlikely that very-large-scale radio 
detectors focused on showers with zenith angles below 60$^{\circ}$ 
will be built in the near- to mid-term future, it seems a very 
realistic option to complement particle detectors such as those of the 
Pierre Auger Observatory with radio antennas focused on the detection 
of very inclined air showers for cosmic ray composition studies.

\subsection{Super-hybrid measurements with integrated detectors}

Radio detection has its strongest potential in combination with other 
detection techniques. This hybrid approach naturally solves the problem of triggering, 
and provides valuable additional information to reconstruct each 
individual air shower as well as possible. In particular, radio 
detection offers an excellent way to study the electromagnetic 
component of air showers (see previous subsection).

A large fraction of the cost involved in cosmic ray detection is in 
deployment, maintenance and ``infrastructure''. This is particularly 
true for radio detectors. Antennas can be built cheaply, also the cost for digital 
electronics is dropping continuously. The main cost of radio detectors is in fact incurred 
by power harvesting (solar panels are expensive, batteries age 
quickly and require regular replacement). High-bandwidth communications can also be problematic, 
although off-the-shelf components have recently become rather powerful.

The natural choice to decrease cost thus lies in integrating 
detectors. A ``one-does-all'' detector measuring particle content 
(preferably electromagnetic and muonic component separately), possibly 
fluorescence or Cherenkov light, and radio emission, seems like a very 
promising approach. First steps in such a direction have been followed 
in the framework of the AugerNext project \citep{HaungsIcrc2015} and 
with the TAXI prototype \citep{TAXI2014}. It seems very natural to 
proceed in this direction with increased efforts.

\subsection{Ultimate precision: the Square Kilometre Array}

As of 2020, the Square Kilometre Array will go into operation in 
western Australia. It will constitute the largest radio telescope ever 
built, with a rich programme in astrophysics and radio-astronomy, and its 
potential for cosmic ray detection was already considered over a 
decade ago \citep{FalckeGorhamProtheroe2004}.

\begin{figure*}
\centering
\includegraphics[width=0.48\textwidth]{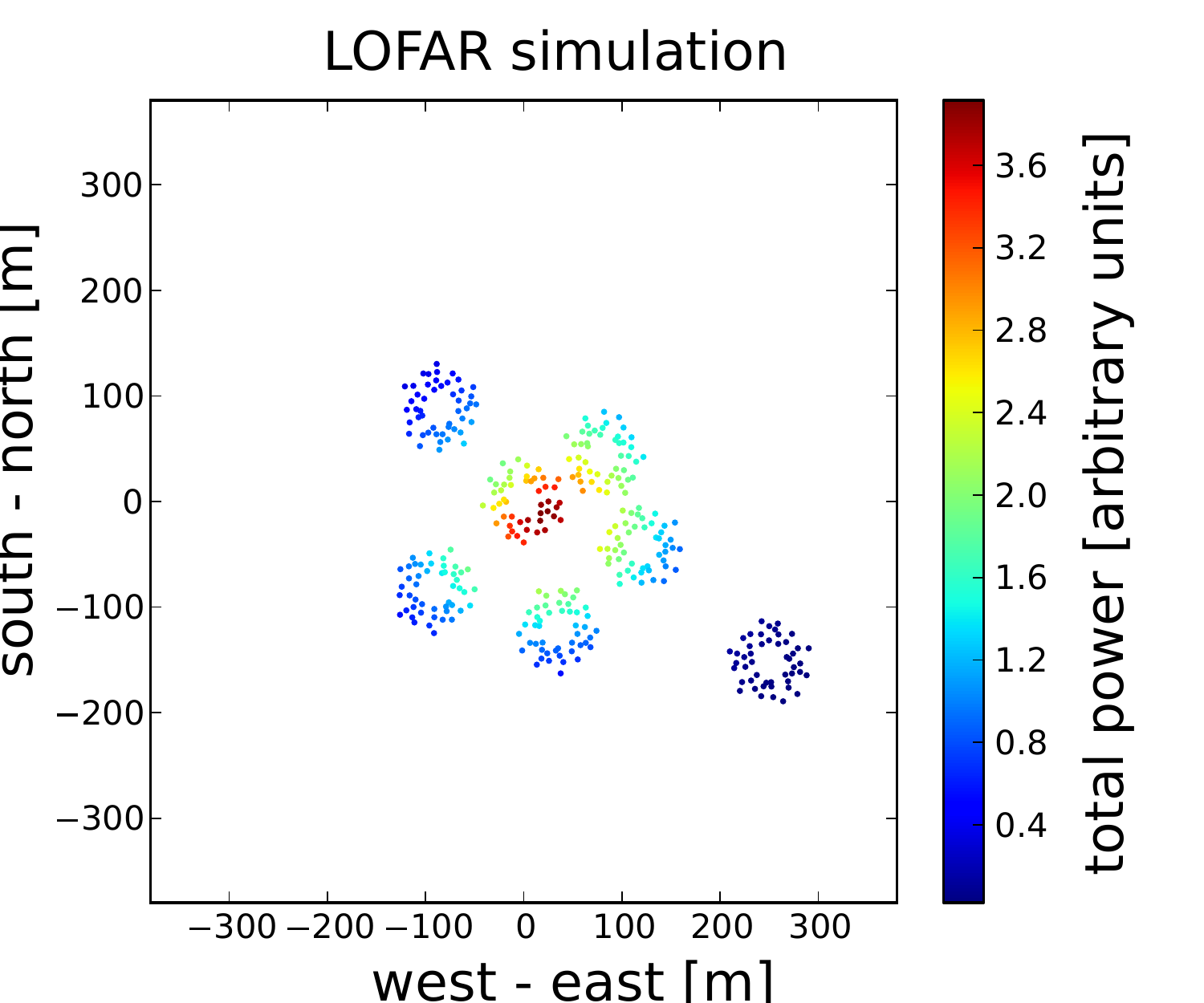}
\includegraphics[width=0.48\textwidth]{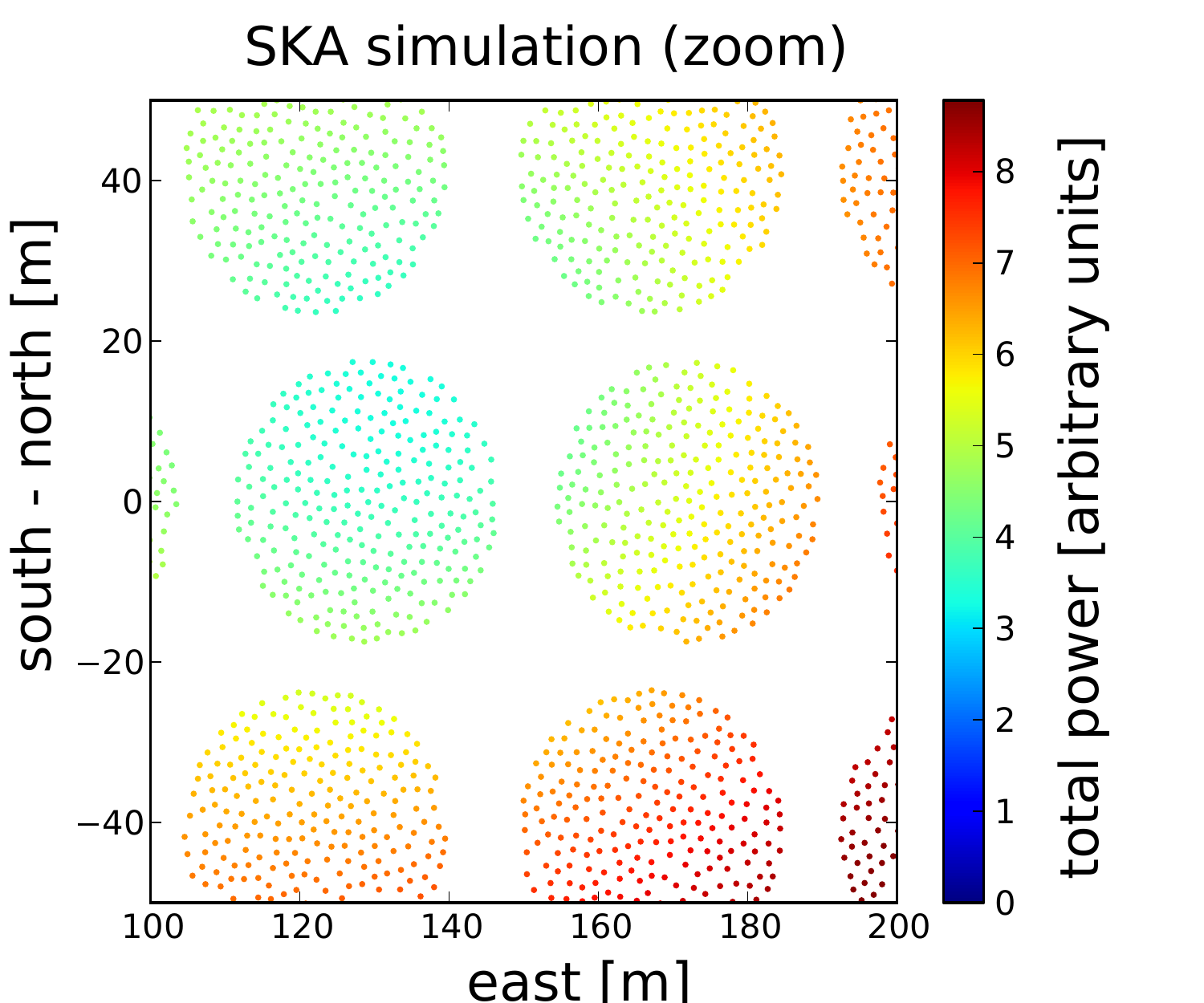}
\includegraphics[width=1.00\textwidth]{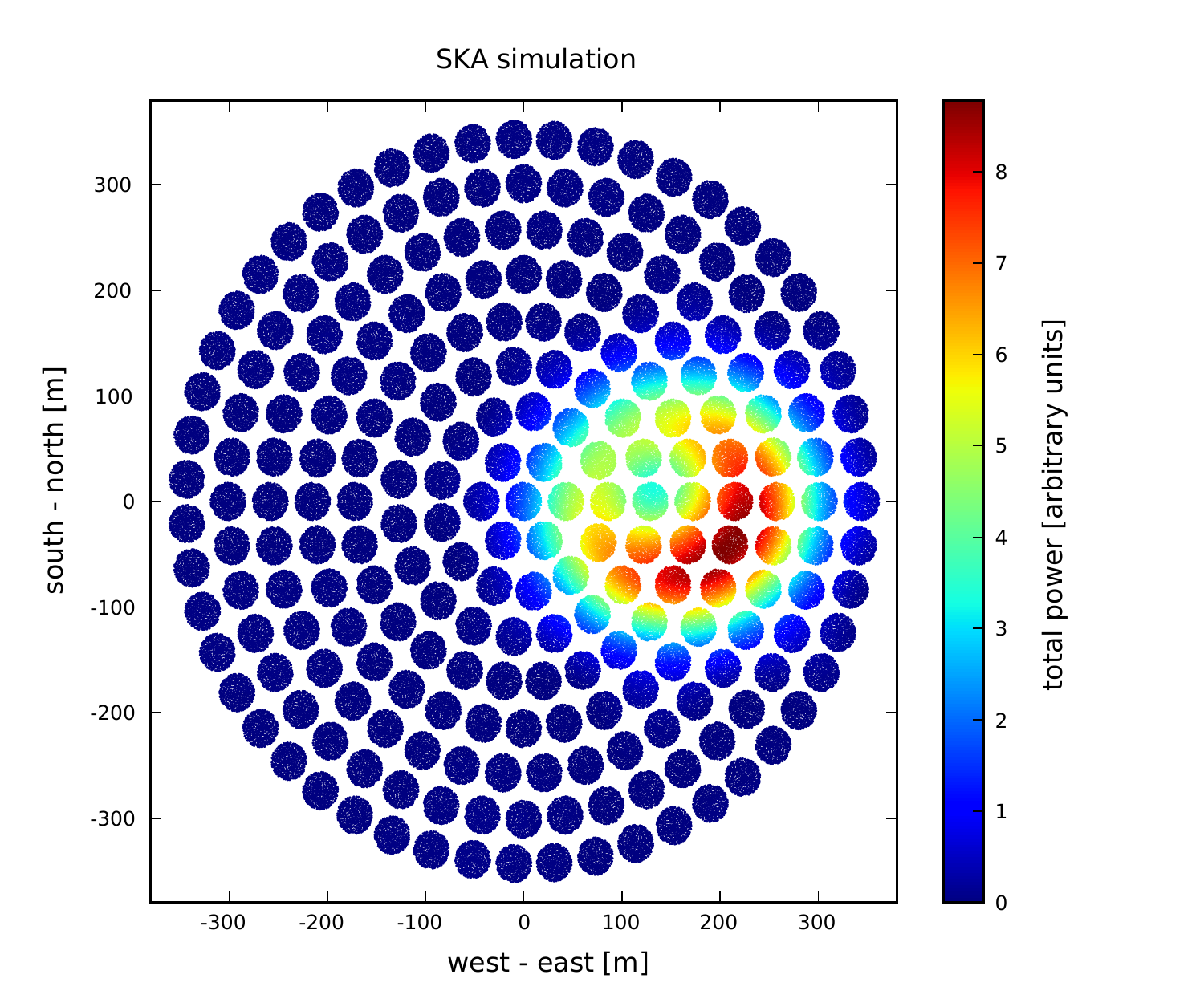}
\caption{CoREAS simulation of the radio emission footprint of an air shower 
with 30$^{\circ}$ zenith angle and an energy of $10^{18}$~eV sampled 
with LOFAR (top-left) and SKA-low (bottom, zoomed-in at top-right). Each point
represents a measurement with an individual dual-polarised antenna. Even for an ideal core position,
LOFAR only achieves an incomplete sampling of the radio 
signal. The SKA-low sampling, on the other hand, is extremely homogeneous 
and detailed, irrespective of the core position within the antenna 
array. The appearance of a Cherenkov ring in the SKA-low measurement is 
due to the measurement of higher-frequency components up to 
350~MHz. Adapted from \citep{HuegeSKAIcrc2015}. \label{fig:lofarvsska}}
\end{figure*}

If equipped with a suitable particle detector array for triggering 
purposes and adequate buffering capabilities at individual antennas, 
the dense core of the low-frequency part (50-350~MHz) of SKA can be used for detection 
of extensive air showers in the energy range of $\gtrsim 10^{16}$~eV 
to $\gtrsim 10^{18}$~eV. An overwhelming number of 60,000 antennas 
will be deployed in a circular region with 750~m diameter, with a very 
homogeneous spacing. The results of LOFAR, in particular with respect to the 
reconstruction of the depth of shower maximum, were already impressive, 
but the level of detail measured with SKA-low will beat the one of LOFAR by far, as is obvious from Fig.\ 
\ref{fig:lofarvsska}. The current expectation is that the 
average resolution of SKA-low on measurements of \xmax will be below 
10~g/cm$^{2}$ and thus significantly better than that of any other 
detector existing today. SKA-low can therefore be used for precision 
studies of the mass composition in the transition from 
Galactic to extragalactic cosmic rays, for the study of interaction 
and air shower physics at very high energies and in the extreme 
forward regime, and for studies of thunderstorm 
and lightning physics, including possible connections to cosmic rays 
\citep{HuegeSKA2014}.

Furthermore, with the detailed sampling of each individual air shower achieved by 
SKA-low, also ``tomographic imaging'' with near-field interferometry 
seems a very promising prospect. In principle, it should be possible to do a three-dimensional 
``tomography'' of the extended source region of air shower radio emission. In 
other words, one should be able to make a three-dimensional image of the electromagnetic component of 
the air shower.  It will be a major challenge to devise this analysis technique.
However, there is very high potential to extract much more 
information from the radio signals than just \xmax and energy.

A complementary observation mode of the SKA for cosmic particles of the highest 
energies exists in detecting cosmic-ray showers (and also neutrino showers) 
via radio emission generated in the lunar regolith as opposed to the Earth's atmosphere 
\citep{BraySKA2014,JamesSKAIcrc2015}.

\subsection{Balloon-borne and satellite-based detection}

When the ANITA balloon-borne experiment, conceived to measure radio 
pulses from neutrinos interacting in the antarctic ice, measured more 
than a dozen radio pulses, it was quickly realized that these were 
caused by radio emission from cosmic rays, reflected off the ice (or 
transmitted directly in an Earth-skimming geometry). It was unclear 
for some time what the energy of the measured cosmic rays was. If the 
16 detected events were of very high energy, then balloon-borne cosmic 
ray detection could be a way to achieve very large exposures and hence 
collect statistics of cosmic rays at the very highest energies. 
Projects like EVA \citep{EVABalloon} and SWORD \citep{SWORD} have consequently been 
proposed.

\begin{figure}[!htb]
\includegraphics[width=0.48\textwidth]{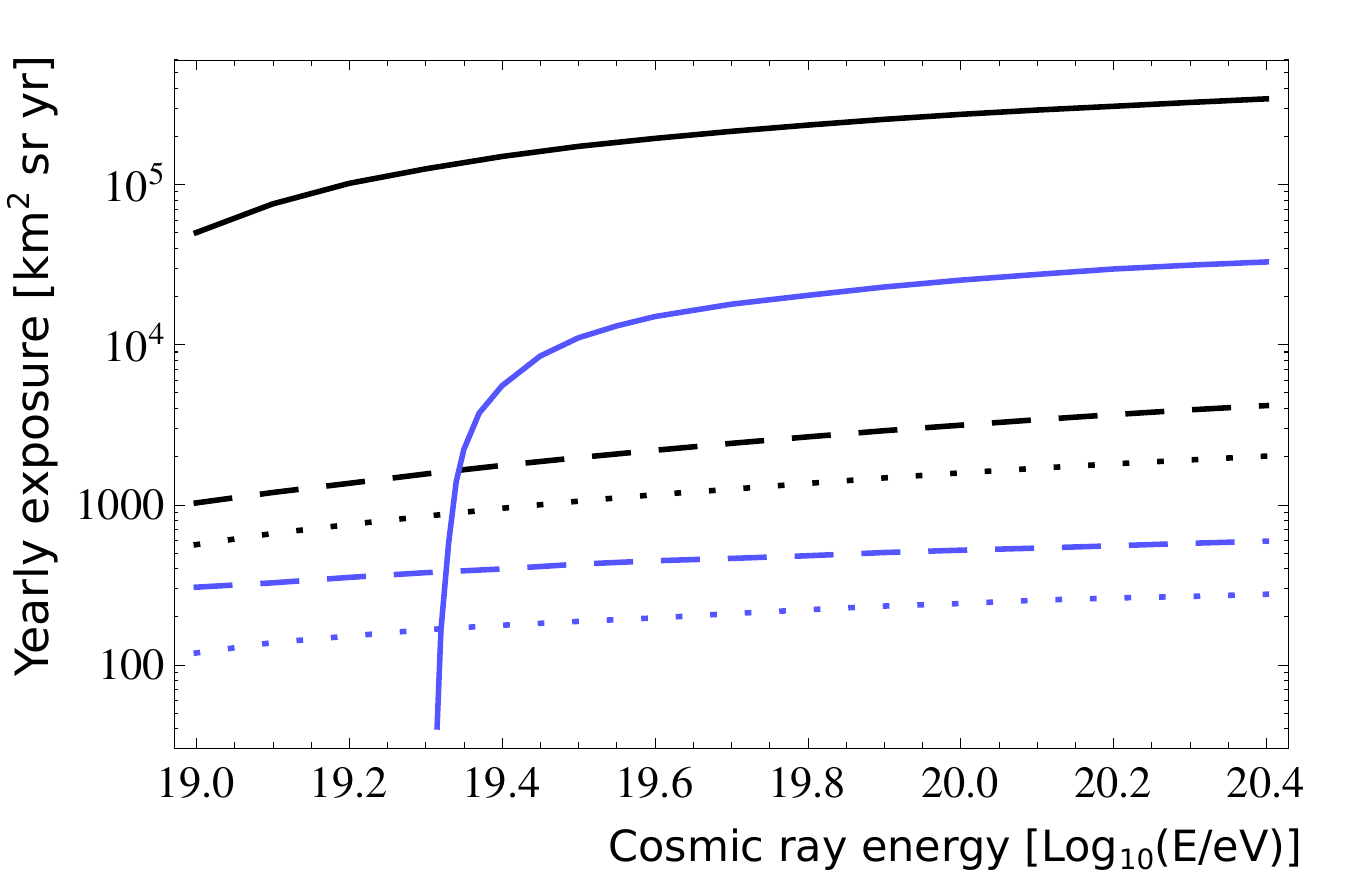}
\caption{Yearly exposure (acceptance) as a function of energy for a 
radio detector on a mountain (dotted line), a balloon (dashed line) and 
a satellite (solid line). The angular distributions of radio signals 
underlying these predictions are based on a SWORD model (black lines) and 
on CoREAS simulations (blue lines). For comparison, the exposure collected by the 
Surface Detector array of the Pierre Auger Observatory for air showers 
up to 55 degrees zenith angle in the period from Jan 2004 to Dec 2014
amounts to 42,500 km$^{2}$ ~sr~y \citep{ValinoIcrc2015}. Adapted from \citep{Motloch201440}.
\label{fig:balloonaperture}}
\end{figure}

A recent study by Motloch et al.\ \citep{Motloch201440}, however, 
showed on the basis of CoREAS simulations that the radio emission is 
beamed in such a narrow cone around the air shower axis, that the 
effective acceptance reached by radio detectors on balloons or even 
satellites (let alone mountains \citep{TarogeIcrc2015}) in the end is not as large as one had 
hoped, see Fig.\ \ref{fig:balloonaperture}. This is not even remedied by 
measuring at frequencies below 100~MHz where the emission is not as 
focused on a Cherenkov ring as it is at higher frequencies.

This pessimistic view is confirmed by a recent publication of the ANITA collaboration 
\citep{ANITAEnergy} in which the average energy of the ANITA cosmic ray events has 
been determined at 2.9~EeV, significantly lower than the 15~EeV 
\citep{HooverNamGorham2010} estimated originally on the basis of an 
incomplete understanding of the radio emission properties.

Based on today's knowledge of the physics of radio emission from 
extensive air showers, balloon- or satellite-based observations thus 
do not seem as promising as originally thought.

\subsection{Low-frequency radio emission} \label{sec:lowfrequency}

Current experiments for air shower radio detection focus on the 
frequency range between the short-wave band and the FM band (typically 
30-80~MHz). At lower frequencies, atmospheric noise quickly rises, as 
signals (e.g., lightning radio pulses) from very far-away are still measurable due 
to their reflection between the Earth's surface and the ionosphere. 
Radio emission from air showers at low MHz (and even kHz) frequencies 
have been performed before the renaissance of air shower radio 
detection, see ref. \citep{MHzIndians} and references therein. Successful detections were reported and attributed 
mostly to transition radiation of the air shower cascade entering the 
ground, but these activities did not gather significant momentum and were not 
independently confirmed. Today, two ideas, however might make 
low-frequency measurements an interesting topic to investigate once 
more.

First, there is a frequency window from $\sim 1-5$~MHz 
where, during day, the atmospheric noise is low, cf.\ Fig.\ \ref{fig:noise}. The reason is that 
the ``D-layer'' in the ionosphere absorbs radio emission at these 
frequencies during daytime. Simulations predict that radio 
signals from extensive air showers should be measurable over large 
areas at these frequencies, i.e., relatively sparse 
antenna arrays could be used for such measurements. However, timing 
information would be very coarse at such low frequencies, and the duty 
cycle would be limited to less than 50\%.

Second, it has been discussed \citep{RevenuSuddenDeath} that the near-instantaneous 
stopping of the charged particles in an air shower when hitting the 
ground should lead to strong low-frequency radio emission that can be measured 
kilometres away from the impact point. While it is clear that 
this radio emission must occur (a large number of charged particles is 
decelerated very quickly), calculations presented so far have been somewhat 
incomplete. They described the emission from the fast deceleration of 
particles (``sudden death''), yet did not take into account the fact 
that the deceleration actually takes place in the Earth, that the 
signal has to be propagated through the Earth/air boundary and that 
the transmission of the signals along the Earth will quickly dampen at 
least those components of the electromagnetic waves polarised parallel 
to the Earth's surface. A consistent treatment of all these effects is 
imperative to judge the potential of this technique. By now, also 
experimental activities trying to measure this effect have been 
started in the EXTASIS project \citep{DallierIcrc2015}.


\section{Conclusions}\label{sec:conclusion}

Radio detection of cosmic ray air showers has undergone an impressive 
decade of progress. \emph{The} major breakthrough of the past years has been 
achieved with a detailed understanding of the radio emission physics, 
culminating in Monte Carlo simulations on the basis of first 
principles such as CoREAS and ZHAireS which can successfully explain every measurement made so far. Unlike ten 
years ago, this means that new experiments and analysis procedures can 
now be developed in a targeted fashion on a solid theoretical 
foundation. The field has clearly left the pioneering phase where many 
basic questions were unanswered. While the previous decade was focused 
on unerstanding the physics and detection techniques for air shower 
radio emission themselves, the next decade will clearly focus on studying 
\emph{cosmic rays} using radio techniques.

Many of the promises initially made by radio enthusiasts could 
be fulfilled. Indeed, radio detection can measure the energy of cosmic ray particles with 
excellent resolution (17\% having been achieved experimentally, 
with potential to go below 10\%). The hoped-for sensitivity on \xmax and hence 
the mass of cosmic rays has been demonstrated convincingly with 
simulations, and measurements with dense antenna arrays such as LOFAR 
have already achieved a resolution of $\sim 17$~g/cm$^{2}$ or better, 
i.e., at a level competitive with today's fluorescence and Cherenkov-light detectors.
Furthermore, Tunka-Rex measurements have brought the experimental 
proof that \xmax values derived with radio measurements are indeed in good 
agreement with those measured by other detectors, and similar endeavors 
are currently being followed within AERA.

Difficulties arose in the use of radio detectors as an autonomous 
detection technique not using any input from other detectors. 
Self-triggering of radio signals remains very challenging. Real-time 
interferometric triggering strategies could help. Yet self-triggering does 
not seem a goal of high priority, as the strength of radio 
detection does not lie in its isolated application but in its 
combination with other detection techniques, in particular particle 
detectors. The combination of these two techniques, both with near-100\% 
duty cycle, can deliver very detailed information on individual air 
showers. This is especially true if the particle detectors provide 
a dedicated measurement of the muonic component of air showers, while
the radio detectors provide accurate information on the electromagnetic component,
including its longitudinal evolution with atmospheric depth. Such 
``radio-hybrid'' measurements can thus contribute significantly to our 
understanding of air shower physics, for example in testing hadronic 
interaction models.

As the footprint of the radio emission for non-inclined air showers is 
fairly limited, instrumentation of areas larger than a few dozen of 
km$^{2}$ still requires the development of concepts that can scale to 
thousands of individual detectors at moderate cost. The main challenge 
here lies in deployment and ``infrastructure'', in particular in power 
harvesting. But technology is progressing fast, and this potential 
should be borne in mind. For inclined air showers, on the other hand, already
today's concepts could be used to instrument hundreds of 
km$^{2}$ and thus measure cosmic rays at energies well beyond 
$10^{18}$~eV up to the highest energies, as the radio detector
spacing can be of order a km or larger. As no other detection technique can measure the 
electromagnetic component of air showers for very inclined air showers with a 
reasonable acceptance, this seems like a very promising area of 
application for large-scale radio detection, especially if detectors 
are integrated to share infrastructure. First activities in this 
direction are being followed in the frame of AERA at the Pierre Auger 
Observatory; in fact, a combination of radio detectors and the 
Auger surface detector array could at some point possibly extend the precision measurements 
targeted with the AugerPrime upgrade to zenith angles
$\gtrsim 65^{\circ}$.

Dense radio detectors such as LOFAR and the upcoming SKA can access 
the energy range from $\gtrsim 10^{16}$~eV to $\gtrsim 10^{18}$~eV. 
This is the region where transitions from Galactic to 
extragalactic cosmic rays already seem to take place. With the high duty 
cycle and good \xmax resolution achievable with radio detectors, the mass composition in this 
energy range could be probed with high precision. With sufficient 
statistics, composition-sensitive anisotropy studies could be 
performed, and with those dense radio detectors could make a major 
contribution in studying the physics of cosmic rays in the transition region. Very 
significant potential also still lies in an improved analysis of the radio data, in particular in ``imaging'' approaches such 
as near-field interferometry applied to dense radio detectors. We are 
only using a fraction of the information in the radio signal so far, and imaging analyses could allow detailed insights in the 
physics of air showers that we have not even thought of today.

The biggest impact that radio detection is likely to have already in the near
future, however, lies in the accurate cross-calibration of the energy 
scale of cosmic ray detectors worldwide, as well as in the independent determination of the 
absolute energy scale of cosmic rays on the basis of first-principle 
calculations. Radio detection is uniquely well-suited for this goal, 
as the emission is well-predictable, the signal is not absorbed or 
scattered in the atmosphere, and the techniques have been developed to 
precisely calibrate radio detectors. It might well be that every 
cosmic ray detector will soon be equipped with a small radio antenna 
array, for the sole purpose of an accurate absolute energy calibration.

\section*{Acknowledgements}

I would like to thank J.\ Knapp for his encouragement and valuable advice in 
writing this review article, C.\ Grupen, A.\ Haungs and F.\ Schr\"oder 
for their very detailed and constructive criticism on the manuscript and A.\ 
Zilles for compiling the map with an overview of existing experiments. 
I would also like to thank all colleagues in the field of radio detection of cosmic 
ray air showers for a decade of fruitful collaboration in a truly 
pioneering spirit.






\end{document}